\documentclass[11pt, a4paper]{article}

\usepackage[utf8]{inputenc}

\usepackage{graphicx}

\usepackage[labelfont=bf]{caption}

\usepackage{authblk}

\usepackage{lscape}

\usepackage{setspace}

\usepackage[dvipsnames]{xcolor}

\usepackage[colorlinks=true,urlcolor=purple,linkcolor=purple,citecolor=purple]{hyperref}
\usepackage[authoryear]{natbib}

\usepackage[labelsep=period]{caption}

\usepackage{mathpazo}

\usepackage{lscape}

\usepackage[margin=1in]{geometry} 

\usepackage{soul}
\usepackage[dvipsnames]{xcolor}
\definecolor{blue}{RGB}{140,173,206}
\definecolor{orange}{RGB}{233,179,116}

\DeclareRobustCommand{\hlblue}[1]{{\sethlcolor{blue}\hl{#1}}}
\DeclareRobustCommand{\hlorange}[1]{{\sethlcolor{orange}\hl{#1}}}

\usepackage{tikz}

\usepackage{verbatim}

\usepackage{booktabs}
\usepackage{longtable}
\usepackage{threeparttable}
\usepackage{threeparttablex}
\usepackage[capposition=top]{floatrow}
\usepackage{tabularx}
\usepackage{tabu}
\usepackage{etoolbox}
\usepackage{adjustbox}
\usepackage{amssymb}
\renewcommand{\arraystretch}{1.18}
\usepackage[section]{placeins}

\usepackage{booktabs}

\usepackage[T1]{fontenc}

\usepackage[bottom]{footmisc}

\title{Investigating individual writing style as a contributor to gender gaps in science and technology\thanks{Correspondence to: rfunk@umn.edu. We thank Dr. Rebecca Levitan and Dr. Martin Chodorow for guidance and  feedback on an earlier version of this work. We released an open-source Python package to compute linguistic features used in this paper, available here: \url{https://pypi.org/project/stylometer}. Financial support from the National Science Foundation (grants 1829168 and 1932596) is gratefully acknowledged. Any errors are those of the authors alone.}}
\author[1]{Kara Kedrick}
\author[2]{Ekaterina Levitskaya}
\author[3]{Russell J. Funk}

\affil[1]{Carnegie Mellon University,       kkedrick@andrew.cmu.edu}
\affil[2]{Coleridge Initiative, ekaterina.levitskaya@coleridgeinitiative.org}
\affil[3]{University of Minnesota,        rfunk@umn.edu}

\date{} 

\begin{document}

\maketitle

\begin{abstract}
Gender gaps in how scientific work is evaluated are well documented, but their sources remain debated. We ask whether an overlooked factor---the linguistic style of the writing itself---is gendered and consequential. Drawing on a framework that distinguishes informational features (which emphasize facts) from involved features (which emphasize relationships), we analyze single-authored abstracts of academic papers and patents across all fields of science and technology. Women's writing is systematically more involved than men's---richer in relational, audience-oriented features and higher in the balance of involved to informational language---a difference that holds across scientific fields, in collaborative as well as single-authored work, and in a large open-access biomedical corpus, throughout the full text of papers, not only their abstracts. This stylistic signature also shapes how work is received---papers whose abstracts are more involved are cited more by women and less by men, and the association persists even when papers are compared against their most content-similar alternatives, indicating a gendered signal independent of topic. That a near-costless feature of writing---word choice rather than substance---leaves a systematic, gendered trace on who cites scientific work points to a subtle channel through which evaluation bias may persist, in tension with the universalist ideal that ideas be judged independently of their author.
\end{abstract}

\pagebreak

\section{Introduction}

\noindent In his classic essay, ``The Normative Structure of Science,'' sociologist Robert K. Merton identified universalism as a foundational principle of the scientific enterprise, one that distinguishes science from other competing systems of knowing. According to Merton's formulation \cite[p. 270]{merton1973}, universalism holds that the evaluation of scientific contributions ``is not to depend on the personal or social attributes of their protagonist; his race, nationality, religion, class, and personal qualities are as such irrelevant.'' The value of universalism is manifested perhaps most concretely in the practice of double-blind peer review, wherein the identities of both those making scientific claims and those evaluating them are obscured from one another \citep{bornmann2011}. 

While scholars have long observed that adherence to the principle of universalism is far from universal \citep{mulkay1976, cole1992, longandfox1995}, the growing availability of large-scale databases is creating opportunities for unprecedented insight into processes of scientific visibility and evaluation \citep{teplitskiy2018,dondio2019,lane2021}, including the barriers that inhibit objective assessments. 
Recent literature in particular has raised considerable concern about the role of gender in scientific evaluation, grant awarding, and hiring practices \citep{mossracusin2012, reuben2014, oliveira2019, card2020}. 
There is growing evidence to suggest that the contributions of women are disseminated, recognized, and valued less than those of men \citep{joshi2014, sarsons2017, vasarhelyi2021, ni2021}, thereby undermining the imperative of universalism, and ultimately weakening the scientific enterprise by suppressing diverse perspectives. 
Specifically, prior work has shown that relative to men, women's contributions are awarded less grant funding from prestigious sponsors \citep{oliveira2019, kolev2020}, are less likely to be granted patent protection \citep{jensen2018}, and also receive fewer citations from other academic papers \citep{lariviere2013}. 
These gaps appear to be driven at least in part by homophily, as men tend to cite other men over women, even in fields with higher female representation, such as the Social Sciences \citep{potthoff2017, dion2018, ghiasi2018}. There is also some evidence that men tend to self-cite at a much higher rate than women \citep{king2017}, although findings are mixed and likely vary across fields \citep{azoulay2020}. 

Yet much remains unknown about the role of gender differences in citation behavior. Prior scholarship has highlighted differences in field, topic choice, or research quality as playing an important role in explaining observed gaps between men and women \citep{leahey2007, key2019}. While these factors are undoubtedly important, studies have also shown that even when accounting for field, topic choice, and quality, men's and women's research is evaluated differently \citep{hengel2017, hengel2020}. Studies further report gender gaps even in the presence of double-blind review \citep{kolev2020, mahajan2020}, suggesting that conscious or unconscious bias based on the identity of the author is likely not the determinative factor \citep{vanderlee2015, hospido2019, card2021, card2020_gender}.

In this article, we aim to contribute to the understanding of gender differences in how scientific papers and patents are cited, using a linguistic lens. Although prior work suggests that this discrepancy may arise from explicit ways in which women communicate about their work, such as underselling their accomplishments \citep{lerchenmueller2019} or expressing a lack of confidence more readily than men \citep{Exley2024}, we examine how gender differences may manifest more broadly in writing style. Specifically, we draw on a prior framework, developed by \cite{biber1988}, which contrasts linguistic features characteristic of a high information density (``informational'' features) with linguistic features associated with more affective, interactional content (``involved'' features). This framework allows us to take a systematic approach to feature selection and also to examine features which have been overlooked as indicators of variation in the previous literature on gender gaps, such as the use of determiners or pronouns. \cite{biber1988} found in a corpus of epistolary writing that women used more ``involved'' features (e.g. increased use of pronouns) for the purpose of building a relationship with the audience, while men used more ``informational'' features (e.g. increased use of cardinal numbers in texts) for the purpose of conveying the factual information. Using this framework, we conduct a large-scale analysis of gender differences in scientific and technical writing, leveraging two distinctive document types---academic articles and patented inventions---which collectively span all major fields of science and technology. 

Several noteworthy findings emerge from our study. First, even in such restricted genres as academic and invention abstracts, works with a single female author use more ``involved'' features than those with a single male author---a pattern that holds across all scientific fields and most patent subcategories. This pattern generally persists in collaboratively authored work, the full text of papers, and the rigidly formatted text of patent claims. This difference in writing style is the largest in the Social Sciences and Arts \& Humanities and smallest in the Physical Sciences and Technology, whose stricter writing conventions leave less room for gendered stylistic variation to surface. In the patent data, the gender of the lawyer is more predictive of writing style than that of the inventor, underscoring both the role of attorneys in crafting patent text and the breadth of the pattern across individuals with very different professional training. Second, this stylistic signature is associated with how work is received---papers with more involved abstracts are cited more by women and less by men, and the association persists even when papers are compared against their most content-similar non-citers, indicating a gendered signal independent of topic rather than shared subject matter. The force of this result lies less in the size of any single coefficient than in the asymmetry between its cause and its consequence---a near-costless feature of writing, word choice rather than substance, leaves a systematic, content-independent trace on who cites a paper, pointing to a subtle channel through which evaluation bias may persist.


Our study builds on, but also extends, a small but growing stream of research in the Science of Science on gender and language (c.f., \citeauthor{fortunato2018}, \citeyear{fortunato2018}).  Prior work in this area has yielded mixed findings regarding gender differences in scientific writing style. Some studies report notable differences between men's and women's writing \citep{lerchenmueller2019, kolev2020, kim2022}, while other findings suggest that differences are minimal or statistically insignificant, particularly in longer text formats \citep{horbach2022, franco2021,lerchenmueller2019}. Additional research suggests that academic writing styles among men and women share similarities \citep{francis2001,rubin1992}, and that gender-based variation may be more pronounced in less constrained tasks \citep{Newman2008}. However, the choice of features in much of this work has often not been explicitly grounded in a linguistic framework, thereby complicating the interpretation of the results and making comparisons across studies difficult. In addition, existing research has been done within particular fields and on particular kinds of text (e.g., scientific papers), and therefore little is known about whether the patterns observed thus far will generalize across diverse fields and document types (such as patent data), which have diverse norms.

Our study also adds to the linguistics literature. First, previous research on gender differences in language utilization has largely been done on informal or conversational text, typically focusing on speech patterns, and using corpora like emails \citep{thomson2001}, essays \citep{mulac2006}, social media \citep{garimella2016}, and blogs \citep{pennebaker2011}. There has been far less research on gender differences in formal writing, and the work that has been done has relied on relatively small samples \citep{argamon2003, biber1988}. Second, we study a new source of text data---patented inventions---that has not been studied in previous linguistic analyses. As a form of written professional legal discourse, patents are restricted in terms of stylistic expression \citep{bhatia1993}, and therefore finding stylistic differences by gender is especially notable.

\section{Data and methods}

\subsection{Conceptual framework and feature selection}

We draw on a previously established linguistic framework, known as ``Informational versus Involved Production,'' outlined by \cite{biber1988}, which characterizes the joint usage of certain linguistic features based on their communicative function. While the linguistics community has developed a number of potentially applicable frameworks, we chose Biber's because it was created using a diverse set of corpora, including non-fiction, and was intended to capture universal features of English language; therefore, we anticipate that it would likely capture useful variation across a broad range of scientific and technological fields, encompassing potentially diverse norms of composition. We specifically chose the ``informational''/``involved'' dimension of Biber's framework, as studies note that men tend to focus more on relaying factual information (``informational''), while women tend to focus on building a relationship with an audience in their communications (``involved''; \citeauthor{argamon2003}, \citeyear{argamon2003}). Based on a corpus of e-mail communication, speech, and blogs, \cite{pennebaker2011} stipulated that ``males categorize their worlds by counting, naming, and organizing the objects they confront'' (informational aspect), while women ``personalize their topics'' (involved aspect).

Biber outlines two factors that influence the choice of individuals to use a specific linguistic mode---the production circumstances of real-time constraints (speech or conversation) versus editing possibilities (academic writing), and the primary purpose of the writer (informational versus involved). With respect to the production circumstances, given the type of corpus that we are exploring in this study (formal written text), we can hold these constant, as both papers and patents are produced in settings where editing is possible and likely (i.e., the authors are under no real-time constraints, unlike the case of a conversation). With respect to the primary purpose of the writer (informational versus involved), \cite{argamon2003} noted that formal written text is intended for a ``broad unseen audience.'' \cite{biber_1998} found that letters written by women and addressed to women were more involved and letters written by men and addressed to men were less involved. He also found that, overall, letters written by women are more involved than letters written by men. Even though in formal written text there is no specific addressee, and academic papers and patents are intended for a ``broad unseen audience'', our goal is to understand if we are still able to observe the variation by the gender of the writers. 

In addition to Biber's framework, the specific set of ``involved'' and ``informational'' features used in our study was informed by the follow-up work of \cite{pennebaker2011} and \cite{argamon2003}. We chose three distinctive features for ``involved'' and ``informational'' writing modes (Figure \ref{fig:gender_schematic}). Pronouns, questions, and non-phrasal coordination (``and'' connector) are highly correlated as linguistic features and are reflective of the ``involved'' writing style. The use of pronouns indicates interaction between the writer and the reader---the writer assumes that the reader knows what they are referring to with a pronoun, as the reader is following the writer's story in a text; pronouns ``present things in a relational way'' \cite[p. 7]{argamon2003}. We consider questions to be an important involved feature as well, as they point to the presence of a hypothetical addressee in a formal written text, and are therefore a marker of interaction \citep{biber1986, marckworthbaker1974}. Non-phrasal coordination (``and'') is used ``to string clauses together in a loose, logically unspecified manner, instead of integrating the information into fewer units'' \cite[p. 106]{biber1988}, and thus is more associated with a narrative type of communication (``involved'' aspect). For ``informational'' writing mode indicators, we use past tense, determiners, and cardinal numbers. We use past tense as an ``informational'' feature, to indicate a ``disconnection'' with the audience, absence of interaction in the narrative. The other two ``informational'' features---determiners and cardinal numbers---are taken from \cite{argamon2003} who used the Balanced Winnow algorithm to identify features that are strong indicators of either male or female writing. Argamon and colleagues note that, based on their examination of texts, ``the use of determiners reflects that male writers are mentioning classes of things, in contrast to female writers, who are personalizing their messages'' \cite[p. 12]{argamon2003}. The greater use of cardinal numbers (quantification) is also noted as a strong male indicator in both the work of \cite{pennebaker2011} and \cite{argamon2003}.

Based on this framework and prior work, we hypothesize that (1) female authors will use involved features at a higher rate whereas male authors will use informational features at a higher rate, resulting in a higher Involved--Informational Ratio among female authors compared to male authors; and (2) papers (and, more weakly, patents) with higher rates of involved features will tend to be cited more by female authors, while those with higher rates of informational features will tend to be cited more by male authors.

Although Biber's original framework encompasses more linguistic features associated with the involved and informational dimensions, our selection of these six features reflects the specific constraints of our corpus. Other features excluded from our analysis such as nouns, longer and more specialized words, and present tense are already predominantly used in formal scientific and technical text \citep{biber1988}, and therefore offer little discriminating power in our corpus. Similarly, while private verbs (e.g., \textit{think}, \textit{feel}), emphatics, and amplifiers are involved features in Biber's framework, characterized as markers of heightened interpersonal interaction and personal feeling \citep[][pp. 105--106]{biber1988}, they are primarily associated with interactive and affective discourse rather than formal written genres. We therefore expect them to appear too infrequently in our corpus of scientific and technical writing to serve as reliable indicators of stylistic variation between male and female authors. Finally, several other features were excluded because they cannot be identified reliably at scale without complex syntactic parsing beyond the part-of-speech tagging and string matching that formed the basis of our computational pipeline; these include that-deletions, participial clauses, and WH-clauses. The six features we focus on are therefore both theoretically central to the involved/informational distinction and sufficiently prevalent within this genre to serve as meaningful indicators of writing style. Future work should examine whether expanding the feature set to include additional features from Biber's framework meaningfully alters the observed patterns.

\subsection{Corpus}

As noted previously, we draw on two different corpora: (1) abstracts of academic papers from Web of Science\footnote{Web of Science: \url{https://clarivate.com/webofsciencegroup/solutions/web-of-science/}} (``WoS data'') and (2) abstracts of patents from PatentsView (``PatentsView data'')\footnote{PatentsView: \url{http://www.patentsview.org/}}. Both databases have been widely used in prior research in the Science of Science, and provide generally representative coverage of scientific papers and patented inventions, respectively \citep{funk2017, birkle2020, jaffe2002,kedrick2024cp}.

For the WoS data, we excluded document types that are not typically meant for the publication of scientific findings (e.g., book reviews, letters, music scores, and bibliographies). We further limited our attention to articles that included full-text abstracts, and that were written in English\footnote{The WoS data do not include accurate information on the language of paper abstracts. We identified English language abstracts using a commercial machine translation API.}. For the PatentsView data, we limited our attention to utility patents, which encompass roughly 90\% of the patents granted by the United States Patent and Trademark Office (USPTO), and have therefore been the central focus of prior work \citep{jaffe2002}. All patents granted by the USPTO are written in English, and more than 90\% (going back to 1976) include full-text abstracts. Patents without abstracts were excluded from our analysis.

For the purposes of our study, we further narrow our focus to single-authored papers and patents (i.e., we exclude papers and patents written by teams). Our motivation for doing so was to allow for a clearer identification of the relationship between gender and writing style, which would be more complex in the case of teams, particularly those of mixed gender. Note that in the patent data, we also narrowed our sample to include only those inventors with a single lawyer, in order to more cleanly control for gender. This process reduced our sample from 34,210,804 to 2,189,643 papers and from 3,840,570 to 2,087,588 patents.  In addition, we require that abstracts have more than 100 words, in order to normalize frequency counts per 100 words, which is consistent with prior research in linguistics \citep{biber1988}. This threshold helps ensure that rare linguistic features can be measured reliably. This is especially relevant for the question feature. As shown in Table~\ref{table:2}, questions occur at a mean rate of only 0.03 per 100 words in our corpus. As a result, word-level normalization alone may not adequately distinguish genuine absence of a feature from the effects of limited text length. Among the 1,714,804 single-authored male papers and 474,839 single-authored female papers, approximately 298,065 (17.4\%) and 53,970 (11.4\%), respectively, were excluded because their abstracts contained fewer than 100 tokens. The reduction was larger for patents. Among the 1,964,816 single-inventor male patents and 122,772 single-inventor female patents, approximately 630,626 (32.1\%) and 43,052 (35.1\%), respectively, were excluded because their abstracts contained fewer than 100 words. In total, we use 512,940 single-authored papers and 32,106 single-inventor patents in the analysis; paper models with field fixed effects drop 420 papers that lack a field classification, leaving 512,520. Additional details on the procedure used to narrow down our samples from the population of articles and patents, including the number of observations lost at each stage, are given in Figure \ref{fig:Corpus_Size} in the Supplementary Materials.

Women are underrepresented in the population of both single authors and inventors. For perspective, in our data subset, roughly 21.7\% of authors and 5.9\% of inventors are women.\footnote{This is an estimate for academic papers with one author and for patents with one inventor, where the gender of the author/inventor was known/assigned.} Moreover, men and women from our sample of single-authored papers tend to publish in different areas, and their relative representation changes over time. Therefore, to help facilitate comparisons, we subset the WoS and PatentsView data by building a matched sample of papers and patents written by male and female authors and inventors. To do so, we identified, for each academic paper written by a single female author in a particular WoS subject area (e.g.,
``Sociology'', ``Cell Biology'', ``Mathematics'') and in a particular year, a corresponding academic paper written by a single male author in the same subject area and year. We follow a similar procedure for patents, using National Bureau of Economic Research (NBER) patent subcategories to proxy for the field of technology, and the grant year of the patent as the year of publication.

\begin{figure}[tp]
\centering
\includegraphics[width=\textwidth]{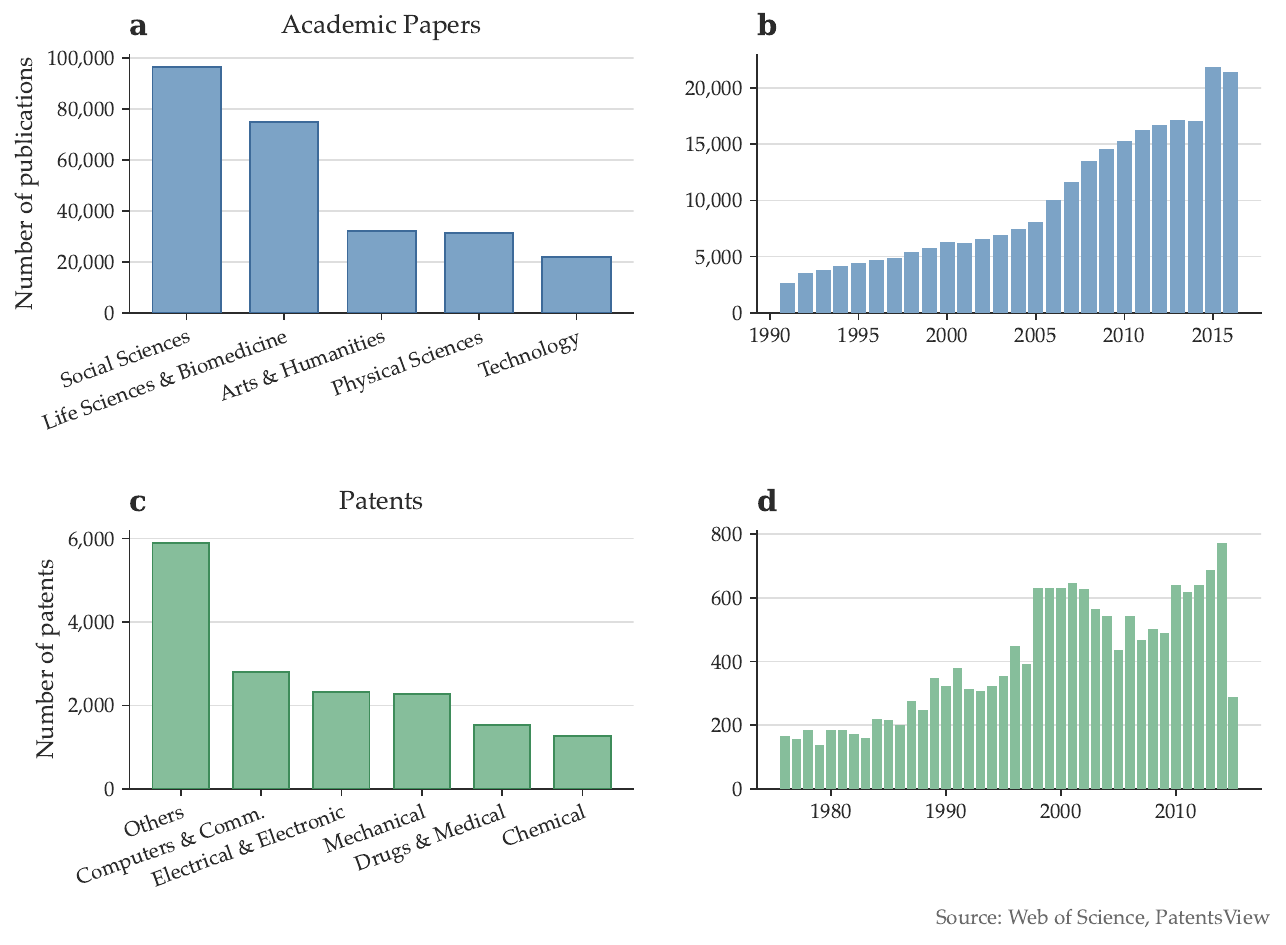}
\caption{\textbf{Composition of the matched analytical sample, by field and over time.} Number of single-authored works in the gender-balanced analytical sample, counting works authored or invented by women (because the sample is matched one-to-one on field and year, the counts for men are identical): \textbf{(a)} academic papers by field; \textbf{(b)} academic papers by publication year; \textbf{(c)} patents by NBER category; \textbf{(d)} patents by grant year. The field composition of the full source corpora is shown in Figure \ref{fig:academic papers_all_field}.}
\label{Figure 1}
\end{figure}

For the WoS data, we subset the corpus to years starting in 1991, as abstracts are not well represented in earlier periods (Figure \ref{Figure 1}). After matching, the largest share of abstracts in our final analytical sample are from the Social Sciences (about 38\%), followed by Life Sciences and Biomedicine. (Figure \ref{fig:academic papers_all_field}(a) shows the field distribution in the full Web of Science corpus.) This is consistent with prior work, as women are well represented in the Social Sciences and biomedical domains \citep{ceci2014, su2015}. For the PatentsView data, our analysis begins in 1976, which is the first year in which machine readable patent records are available from the USPTO. After matching, the largest share of patents come from the ``Others'' NBER category (about 37\%), which includes ``Miscellaneous'' patent types. (Figure \ref{fig:academic papers_all_field}(b) shows the category distribution in the full PatentsView corpus.) 

Gender is assigned using author and inventor names. For papers, author gender is coded based on first and (when available) middle names, as reported in the Web of Science database, using the genderize.io application programming interface (API). For patents, we assign inventor gender by using a coding provided by the United States Patent and Trademark Office, as included with the PatentsView data. More details about the gender assignment can be found in the Gender coding section of the Supplementary Materials. 

\subsection{Linguistic measures}
For each paper and patent in our sample, we constructed a series of measures tracking the utilization of informational and involved features. We use pronouns, questions, and non-phrasal coordination (``and'' connector) as indicators of the ``involved'' style of writing, and past tense, determiners, and cardinal numbers as indicators of ``informational'' style of writing.  Figure \ref{fig:gender_schematic} gives an overview of the specific features of interest; more details about the computation of these features are included in the Table \ref{table_features} in the Supplementary Materials.\footnote{We also released an open-source Python package to compute these linguistic features, available here: \url{https://pypi.org/project/stylometer}.}

We define three separate measures---involved rate, informational rate, and Involved--Informational Ratio. Because our corpus includes academic writing and invention abstracts, there is already a high utilization of informational features by both female and male authors; therefore, in addition to involved and informational rates, we are also interested in seeing the relative rate at which writers use involved features in relation to informational features (Involved--Informational Ratio). We define the involved and informational rates as the sum of three involved (or informational) linguistic features divided by the total number of tokens in a given abstract, and normalized to units of 100 words \citep{biber1988}. We define the Involved--Informational Ratio as the involved rate divided by the informational rate.  

\begin{equation}
\textrm{Involved Rate} = \frac
{N_{pron}+N_{and}+N_{q}}{N_{tokens}}\times 100
\end{equation}

\begin{equation}
\textrm{Informational Rate} = \frac{N_{det}+N_{past}+N_{num}}{N_{tokens}}  \times 100
\end{equation}

\begin{equation}
\textrm{Inv.-Inf. Rate} = \frac{\textrm{Involved Rate}}{\textrm{Informational Rate}}
\end{equation}

To better understand the relationship between author gender and writing style, we estimated a series of linear regression models, predicting the utilization of informational and involved features in paper and patent abstracts. For models of both papers and patents, predictor variables included the gender of the author (or inventor), the year of academic paper (or patent grant), and the principal scientific (or technological) field. For models of patents, we also included as a predictor the gender of the lawyer, as patent attorneys often take part in writing and editing the patent text. 

\subsection{Citation measures}
The possibility that men and women write in systematically different ways naturally raises questions about the relationship between writing styles and the reception of intellectual products. Specifically, differences in how men and women present their research in written text may help account for some of the disparities in rates of citation documented in prior work \citep{caplar2017,potthoff2017, dion2018, dworkin2020, zurn2020}. 

To evaluate this possibility, we conducted a series of analyses in which we decomposed the number of citations made by papers and patents to those in our sample, according to the gender of the citing author or inventor. 
For each sample paper and patent, we computed two related measures: (1) citations from citing works whose first author or inventor is female (male), and (2) citations from citing works whose last author or inventor is female (male). For each measure, we excluded self-citations, i.e., those made by the author of the sample paper to his or her own work. In addition, to facilitate interpretation, we transform each measure into a rate per 100 citations. Subsequently, we estimated a series of regression models, wherein the four measures are our outcomes of interest, and the predictors are the gender of the author or inventor (and lawyer for patents) and the rates of informational and involved features. For papers and patents cited 0 times, we impute a rate of 0, and add an imputation indicator as a covariate. We control for the field and year of academic paper/patent.

To assess whether writing style influences citation patterns beyond content similarity, we constructed a content-based risk set of potential citers for each single-authored paper and patent. This dataset included both observed citations and up to ten counterfactual citations---papers and patents that were highly similar in content (based on SBERT embeddings) but did not cite the work. We identified potential counterfactual citers by using FAISS to retrieve the top 100 nearest neighbors (based on cosine similarity between SBERT embeddings) within the same field and publication year as the observed citers. This design enabled us to test whether involved and informational rates predict citation patterns when we control for semantic similarity. We examined gender differences in citation patterns using regression models that estimated the likelihood that citations to the focal single-authored papers and patents came from citing works with (1) a female (male) first author or inventor or (2) a female (male) last author or inventor. The citing works included multi-authored documents. The dependent variable was coded as 1 if the citation was observed and the citing work met the gender criterion, and 0 otherwise. All models included fixed effects for field and publication year, and we clustered standard errors by paper or patent to account for variation in the number of observed and counterfactual citations. For additional details, see the Controlling for content similarity section in the Supplementary Materials.

\subsection{Writing examples}
Before presenting our quantitative results, we first highlight several examples from our data that illustrate the differences between the informational and involved styles. Beginning with papers, consider the following excerpts from two abstracts, one written by a female author (1) and the other written by a male (2). Both papers were published in the same field (Business \& Economics), in the same journal (\emph{Financial Management Journal}), during the same time period (1990s).

\begin{quote}
\textbf{(1)} \emph{Why do companies in the United Kingdom pay scrip (i.e., in stock) dividends? Are tax savings the sole motive for this option? If so, are all tax-loss companies offering this option? Or is this option driven by other motives, such as cash savings, signaling, and agency conflicts? In this study, I provide some insights into the motivation for scrip-dividend payment by comparing the operational performance and other characteristics of all companies that distributed scrip dividends with those of a control group of non-scrip-paying, but otherwise similar, firms. Scrip dividends are offered by an increasing number of companies in the United Kingdom as an option whereby shareholders are able to choose between receiving dividends in cash or the equivalent in the form of shares (scrip). Companies stress tax savings when they offer this option because, unlike cash dividends, the scrip option is not subject to a payment of the advanced corporation tax (ACT)}.
\end{quote}

\begin{quote}
\textbf{(2)} \emph{The rich variety of securities innovations in recent years continues to intrigue academicians and practitioners alike. The array of innovative securities includes a variety of equity-linked debt securities. On January 28, 1991, the Republic of Austria publicly offered \$100 million principal amount of a new equity-linked debt security called stock index growth notes (``SIGNs'') in the United States. The SIGNs were scheduled to mature in approximately 5.5 years. They would make no payments of interest prior to maturity. If the value of the S\&P 500 is below 336.69 on the maturity date, the holder will receive \$10, the initial offering price. If the value exceeds 336.69, the holder will get \$10 plus \$10 multiplied by the percentage appreciation in the S\&P 500 above 336.69. SIGNS may thus be characterized as a package consisting of (i) a 5.5-year triple-A-rated zero coupon note plus (ii) a 5.5-year European call option, or warrant, on the S\&P 500 with a strike price of 336.69}.
\end{quote}

Although both abstracts address similar topics, they are presented in dramatically different ways. Abstract (2), written by the male author, is striking for its heavy utilization of cardinal numbers (e.g., ``\$100 million'', ``5.5 years'', ``above 336.69''), which are mostly avoided by the female author. Abstract (1), by contrast, stands out for its greater utilization of ``connector words'' (e.g., ``and'') and questions, the latter of which are not used at all in Abstract (2). The female abstract reads a lot like a story, building a relationship with the reader through the use of pronouns, present tense, asking questions, and by stringing the clauses together with ``and'' in a loose, storytelling-like manner; the male abstract is focused on relaying the factual information, via the use of numbers and determiners, and past tense. Figure \ref{fig:comparison} shows the full text of both abstracts, zoomed out, with informational and involved features highlighted in different colors (blue for informational, orange for involved). 

This general pattern is consistent across papers and patents in our larger sample. Figure \ref{fig:academic papers_patents_bars} shows the female share of abstracts, with 95\% confidence intervals, across bins of the Involved--Informational Ratio for paper and patent abstracts, respectively. In both, the female share generally rises with the ratio---for academic papers it moves from just below gender parity (50\%) at the lowest ratios to well above parity at higher ratios, with confidence intervals that exclude parity; for patents the same upward shift is present but weaker and less precisely estimated. Note in the patents, the number of bins is smaller because we observe a narrower range of Involved--Informational Ratio values for patents, perhaps reflecting the more constrained style of the genre.

\section{Results}

\subsection{Gender differences in writing style}
Table \ref{table:2} shows the breakdown of informational and involved features for abstracts from academic papers and patents, respectively. The results there suggest that, in absolute terms, informational features are used significantly more often in both text sources, by both men and women. This distribution seems plausible given the nature of our corpus (i.e., abstracts from scientific articles and patents are generally concerned with communicating information), and is also consistent with prior work; \cite{biber1988}, for example, situated academic prose closer to the informational dimension, and \cite{argamon2003} noted the ``greater quantification inherent to most non-fiction genres.'' Table \ref{table:Features_academic papers_and_patents} also shows the mean and standard deviation of involved features, informational features, and the Involved--Informational Ratio in the academic papers and patents samples.  

The results of our regression analysis are shown in Tables \ref{table:coefficients_academic papers_rate} (papers) and \ref{table:coefficients_patents_rate} (patents)\footnote{We evaluated our regression models for multicollinearity; all variance inflation factors were below the conventional threshold of 10, suggesting little multicollinearity among the covariates.}. Beginning with papers, we find, as expected, a statistically significant association between author gender and the Involved--Informational Ratio of paper abstracts ($\beta=0.0209; p<0.001$, Model 2); relative to men, women tend to use a higher ratio of involved to informational features in their scientific writing. Holding all other variables at their means, the predicted Involved--Informational Ratio (based on Model 2 of Table \ref{table:coefficients_academic papers_rate}) is about 6.7\% greater for female than male authors. Models 4 and 6, also of Table \ref{table:coefficients_academic papers_rate}, help to unpack this finding by estimating separate regressions predicting the utilization of involved and informational features, respectively. We find that relative to men, women use involved features at a significantly greater rate ($\beta=0.23; p<0.001$, Model 4), and informational features at a significantly lower rate ($\beta=-0.18; p<0.001$, Model 6) in their writing. This suggests that the differences we detect between men and women in the Involved--Informational Ratio are likely driven by both a higher rate of involved feature use and a lower rate of informational feature use among female authors.

While women's higher involved rate and Involved--Informational Ratio hold across all fields of science and technology for single-authored abstracts, we find heterogeneity in the magnitude of the effect. Figure \ref{fig:Coefficients} shows the predicted gender differences by field for our subset of single-authored abstracts. The differences are greater in the Social Sciences and Arts \& Humanities and less pronounced in Physical Sciences and Technology---the predicted gender gap in the Involved--Informational Ratio is roughly six times larger in the Social Sciences ($\beta=0.028$) than in the Physical Sciences ($\beta=0.005$; Table~\ref{detailed_papers})---which may be explained by the fact that the former fields generally permit greater freedom of stylistic expression and higher use of involved features than the latter fields, which have more restricted stylistic rules \citep{argamon2003}. For patents, the involved rate is higher for female inventors across most NBER categories. Full regression results for each field are presented in Table \ref{detailed_papers} (for academic papers) and in Table \ref{detailed_patents} (for patents) in the Supplementary Materials. Note that our sample only includes single-authored papers and patents that have at least 100 words in their abstracts. While these exclusions substantially narrow the scope of the corpus, focusing on single-authored work was a deliberate choice intended to isolate writing style free from the confounding influence of the gender of coauthors. 

Turning next to patents, we find similar patterns. There is a statistically significant association between inventor gender and the Involved--Informational Ratio of patent abstracts ($\beta=0.0034; p<0.01$, Model 2 of Table \ref{table:coefficients_patents_rate}); relative to men, women tend to use a higher ratio of involved to informational features in their technical writing. Once again holding all other variables at their means, the predicted Involved--Informational Ratio (based on Model 2 of Table \ref{table:coefficients_patents_rate}) is about 2.3\% greater for female than male inventors. In Model 2 of Table \ref{table:coefficients_patents_rate}, we evaluate the relationship between the gender of the patent lawyer and the Involved--Informational Ratio of the patent. Here, we find results that are similar to those we observed for inventor gender, with a statistically significant association between lawyer gender and Involved--Informational Ratio ($\beta=0.0076, p<0.01$, Model 2). The magnitude of the association, however, is larger than what we observed for inventor gender; holding all other variables at their means, the predicted Involved--Informational Ratio (based on Model 2 of Table \ref{table:coefficients_patents_rate}) is about 5.1\% greater for patents with female than male attorneys. 

Following our analytical strategy with papers, in Models 4 and 6 (Table \ref{table:coefficients_patents_rate}), we run separate regressions predicting the rate of utilization for involved and informational features, respectively. Here, the findings are a bit more complex. We find that women use both involved ($\beta=0.14, p<0.001$, Model 4) and informational ($\beta=0.41, p<0.001$, Model 6) features at a significantly greater rate in their writing than men. Female lawyers use involved features in their writing at a rate that is statistically indistinguishable from men (Model 4), but statistically fewer informational ones ($\beta=-0.62, p<0.001$, Model 6). 

\subsection{Writing style and the gender of citing authors}
Next, we turn to our analysis of gender differences in citations. To simplify the presentation, in this section, we only report models predicting citation as a function of the rate of informational and involved feature utilization (i.e., we do not report the results of models that use the Involved--Informational Ratio as a predictor). We found substantially similar results, however, using the ratio measure.

We present a set of nested regression models to assess whether linguistic variables mediate the relationship between author gender and citation counts. We compare a base model that includes author gender and relevant controls (field and year fixed effects, citation-zero flag) to a full model that adds our two key linguistic style measures. These results are shown in Table \ref{table:RegressionsCitationsByGenderOfCitingAuthorPapersSeparateMeasuresV4Unstandardized} (unstandardized coefficients for papers) and Table \ref{table:RegressionsCitationsByGenderOfCitingInventorPatentsSeparateMeasuresV4Unstandardized} (unstandardized coefficients for patents). See Supplementary Tables \ref{table:RegressionsCitationsByGenderOfCitingAuthorPapersSeparateMeasuresV4Standardized} and \ref{table:RegressionsCitationsByGenderOfCitingInventorPatentsSeparateMeasuresV4Standardized} for analyses with standardized coefficients. We also provide a visualization of our findings in Figure \ref{fig:citation_analysis}.

Overall, we observe significant gender homophily in citation patterns. Papers written by a single female author receive a greater share of citations from papers with a female first author ($\beta=4.40, p<0.001$, Model 1; $\beta=4.33, p<0.001$, Model 2) and papers with a female last author ($\beta=3.65; p<0.001$, Model 5; $\beta=3.57, p<0.001$, Model 6). In relative terms, women's single-authored papers draw roughly 15\% more of their citations from female-first-authored work than comparable papers by men---about 4.4 percentage points on a base of 29\%.\footnote{The 29\% and 50\% citing-share figures are the mean shares of a paper's citations that come from female-first- and male-first-authored papers, respectively, in our matched sample.} The opposite pattern holds for single-authored papers written by men.

Taking into account the linguistic features, we see a consistent and statistically significant pattern across the outcome measures---papers with higher rates of utilization of involved features in their abstracts tend to be cited at a higher rate by papers with a female first author ($\beta=0.25, p<0.001$, Model 2) and papers with a female last author ($\beta=0.29, p<0.001$, Model 6). More substantively, we can interpret the coefficient estimates as indicating that for each unit increase in the rate of involved features used, the rate of citation by papers with a female first and last author is predicted to increase by 0.25 and 0.29, respectively. When we consider the use of informational features, however, the pattern flips ($\beta=-0.08, p<0.001$, Model 2 for citation by papers with a female first author; $\beta=-0.06, p<0.001$, Model 6 for citation by papers with a female last author). Papers with higher rates of utilization of involved features are cited at a lower rate by papers with a male first author ($\beta=-0.18, p<0.001$, Model 4) and papers with a male last author ($\beta=-0.30, p<0.001$, Model 8). We find that the opposite holds for utilization of informational features, although only in models predicting citation by papers with a male first author ($\beta=0.03, p<0.01$, Model 4). 

Turning to patents, we find a similar directional pattern to that observed for papers, but the significant associations are both smaller in magnitude and fewer in number. While patents that use more informational features are less likely to be cited by patents with a female first ($\beta=-0.07, p<0.001$, Model 2) or last ($\beta=-0.07, p<0.001$, Model 6) inventor, none of the remaining coefficients tracking the utilization of informational or involved features reach conventional significance; only the involved rate for citations by male first inventors is marginally significant ($\beta=0.20, p<0.10$, Model 4). These results are not surprising, considering that patent abstracts, as a type of text, allow much less stylistic expression than academic papers do.

Including the linguistic measures attenuates the author-gender coefficient---the effect of being a female first author declines from 4.40 to 4.33, with similar reductions across the other models---and produces a small but consistent gain in explained variance ($R^2$ rises from 0.3080 to 0.3084 in the female-first-author model; Table \ref{table:RegressionsCitationsByGenderOfCitingAuthorPapersSeparateMeasuresV4Unstandardized}), consistent with writing style partially mediating the gendered citation gap. We do not take writing style to be a primary driver of citation patterns---the per-paper association is modest. The magnitude of that association, however, is best weighed against how little the wording must change. A one standard deviation increase in involved language is associated with a 0.52 percentage-point increase in the share of citations from female-first-authored papers (Table \ref{table:RegressionsCitationsByGenderOfCitingAuthorPapersSeparateMeasuresV4Standardized}, Model 2); and because one unit of the involved rate is a single involved word per 100, moving from the 10th to the 90th percentile of involved language\footnote{The 10th and 90th percentiles of the involved rate are 2.2 and 7.5 involved features per 100 tokens, respectively---a spread of about 5.3, or roughly eight words across a typical 150-word abstract.}---about eight words in a typical 150-word abstract and a rounding error within a full paper---shifts the gender composition of a paper's citers by about 30\% as much as the author's own gender does. That so slight a difference in wording, demanding no change in substance or rigor, leaves a systematic and gendered trace on who cites a paper is, to us, the substantive point---one whose bearing on cumulative advantage does not require the per-paper effect to be large.

Finally, we note that these patterns align with our theoretical expectations---male authors appear less likely to cite involved language and more likely to cite informational language, while female authors exhibit the opposite tendency. Because male-first-authored papers supply the larger share of citations in our data---roughly 50\%, versus 29\% from female-first-authored papers---the involved features more characteristic of women's writing are precisely those the field's majority citing audience rewards least. These findings suggest that linguistic style may function not only as a marker of epistemic stance, but also as a gendered cue that interacts with how work is cited. Taken together, the nested models offer greater transparency into the structure of these relationships and help clarify the mechanisms by which gendered citation disparities may arise.

\subsection{Robustness and extensions}
Because our central design choices---focusing on single-authored abstracts and constructing a matched, gender-balanced sample---could in principle account for the patterns reported above, we subjected the core findings to a series of robustness checks and extensions. In every case, the gender differences in writing style persist, and the citation relationship continues to hold for papers under content matching.

\textbf{Collaborative authorship.} We focus on single-authored work to isolate individual writing style, free from the confounding influence of coauthors of potentially different genders. A natural question, however, is whether the patterns generalize to the collaboratively authored papers and multi-inventor patents that make up the majority of contemporary science and technology. We therefore re-estimated our models on collaboratively authored works, measuring gender by either the first or the last author (Supplementary Tables~\ref{tab:c_pap_m}--\ref{tab:c_pat_m}). The same gender differences generally appear in both cases---the writing style of collaborative teams with female as first (or last) author are associated with a higher involved rate and Involved--Informational Ratio. As expected when stylistic signal is shared across coauthors, the differences are generally attenuated relative to single-authored work, but the involved-rate and ratio differences remain in the same direction. For example, the coefficient for involved rate decreased from $\beta = 0.23$ in the single-authored paper model to $\beta = 0.11$ in the first-author model (both $p < 0.001$). The disciplinary pattern is likewise preserved---the gender gap is larger in the Social Sciences and smaller in the Physical Sciences (Supplementary Table~\ref{tab:disc_pap}).

\textbf{The full, unmatched sample.} Our main models are estimated on a matched, gender-balanced sample, which equalizes the field and year composition of men's and women's work but departs from the underlying population. To confirm that our findings are not an artifact of this design, we re-estimated the same regressions on the full, unmatched sample of both single- and multi-authored works. The involved-rate and ratio results are substantively unchanged---for single-authored papers the involved-rate coefficient is $\beta=0.28$ in the full sample versus $\beta=0.23$ in the matched sample (both $p<0.001$)---indicating that the matching procedure is not driving the observed differences (Supplementary Tables~\ref{tab:c_pap_f}--\ref{tab:c_pat_f}).

\textbf{Patent claims.} Patents are structured into distinct components---typically an abstract, a description of the technology, and a set of claims---each subject to different constraints by patent examiners. Because claims are tightly constrained by legal and technical requirements and are expected to carry substantially more factual content, they offer a hard test of whether stylistic differences survive in the most rigidly formatted patent text. Using a matched sample of 7,616 patents for which both components were available in PatentsView, we compared abstracts and claims directly (Supplementary Section~\ref{sec:claims_supp}, Tables~\ref{tab:claims_compare}--\ref{tab:claims_reg}). Claims exhibit slightly higher involved and informational rates than abstracts (involved: 3.01 versus 3.18 per 100 tokens, $t=-8.52$, $p<.001$; informational: 21.48 versus 21.84, $t=-8.09$, $p<.001$), but both rates rise by similar proportions, so the Involved--Informational Ratio is statistically indistinguishable between the two components ($t=0.38$, $p=.705$). Replicating the regression analysis on claims yields the same gender pattern observed for abstracts---the predicted ratio remains greater for female than male inventors ($\beta=0.0044$, $p<.01$), and female inventors write at higher involved ($\beta=0.1153$, $p<.001$) and informational ($\beta=0.1777$, $p<.01$) rates. For patent lawyers, none of the three associations reach significance in the claims models, in contrast to the significant effects in abstracts, suggesting that the stylistic imprint of lawyers is more constrained by the rigid legal structure of claims.

\textbf{Full text.} Because our main analysis is based on abstracts, we examined whether the same differences appear in the full text of papers. We matched our sample papers by DOI to the PubMed Central Open Access Subset---the largest corpus of full-text scientific articles available for systematic text analysis---retrieved the full text, and parsed it into its constituent sections, computing the same measures for the abstract, each section, and the full body. The gender difference is not confined to abstracts---for single-authored papers, women use significantly more involved features ($\beta=0.26$ in the full body) and a higher Involved--Informational Ratio ($\beta=0.0195$; both $p<0.001$) in the full body \emph{and} in every individual section---introduction, methods, results, and discussion (Supplementary Table~\ref{tab:ft_solo}). The effect is largest for single-authored work, as expected when the writing reflects a single individual, and is present, if smaller, for collaboratively authored papers measured by either the first or last author (Supplementary Tables~\ref{tab:ft_collab}--\ref{tab:ft_collab_last}).

\textbf{Content similarity.} Papers and patents are often cited because their content relates to the citing work, raising the possibility that the citation patterns above reflect topical similarity rather than writing style. To address this, we constructed a content-based risk set of potential citers---for each focal work, we used SBERT embeddings and a nearest-neighbor search to assemble up to ten content-matched non-citing works, drawn from each focal work's 100 nearest neighbors---highly similar in content---that were ``at risk'' of citing it but ultimately did not (Supplementary Tables~\ref{table:risk_papers_citations}--\ref{table:risk_patents_citations}). Even after this adjustment, involved and informational rates remain significant predictors of the gender of citing authors---for academic papers, a higher involved rate is associated with more citations from female first ($\beta=0.0015$) and last ($\beta=0.0019$) authors and fewer from male first ($\beta=-0.0045$) and last ($\beta=-0.0053$) authors (all $p<0.001$). For patents, the informational-rate associations are directionally consistent, but the involved-rate associations are weak and do not reproduce the paper pattern. The findings are robust to using 50 or 150 nearest neighbors in place of 100 (Supplementary Tables~\ref{table:risk_papers_citations_k50}--\ref{table:risk_patents_citations_k150}). As detailed in the Supplementary Materials, this analysis is best viewed as complementary to the rate-based models reported in the main text.

\section{Discussion}

The norm of universalism in science and technology stipulates that the evaluation of scientific and technological contributions should be based on impersonal criteria \citep{merton1973}. Nevertheless, growing evidence suggests that gender gaps in scholarly assessment remain both prevalent and widespread \citep{mossracusin2012, reuben2014, oliveira2019, card2020}. In this study, we examined variation in writing style by gender in single-author scientific and technical abstracts, and considered the role this variation plays in the citation patterns. To do so, we developed an approach motivated by a larger linguistic framework, examining differences between ``involved'' and ``informational'' writing styles.

Based on our sample of single-authored academic papers and patents, we find differences in the writing styles of female and male authors, with female writers using involved features more frequently (features associated with more affective, interactional content)---a pattern that holds across all scientific fields and most patent subcategories, is more pronounced in the Social Sciences and Arts \& Humanities than in the Physical Sciences (likely because the former permit more stylistic expression), and generally persists in collaboratively authored work, across the full text of papers, and in the rigidly formatted text of patent claims. In the single-author patent abstracts, the gender of the lawyer is more predictive of this pattern than that of the inventor, suggesting that attorneys exert substantial influence over the editing of patent abstracts. The same signature is also associated with how work is received---papers with higher involved rates receive relatively more citations from women and fewer from men---an association that survives matching each paper to its most content-similar non-citers, indicating that it reflects style rather than shared topic. The per-paper coefficient is small (a one standard deviation increase in involved language corresponds to a 0.52 percentage-point increase in the share of citations from female-first-authored papers), but its significance lies in its character rather than its size---a near-costless difference in wording leaves a systematic, content-independent, gendered imprint on citation, attenuating the author-gender coefficient by roughly 1.6--2.1\% in the female-author models and offering a channel through which stylistic conventions could quietly translate into unequal recognition.

Unlike academic articles---where authors have less limitations on the references they cite---patents include citations from three actors with distinct incentives---inventors reference technical lineage, lawyers reference prior work to defend the breadth of the claim, and examiners insert references contextualizing the invention in a way that may challenge its novelty. Our analysis only focuses on the final list of references, without accounting for the source of each citation. A more fine-grained approach would involve distinguishing citations added by inventors, lawyers, and examiners to examine how their contributions shape overall citation patterns. It is possible that gender influences citation practices at all levels---for example, female citers may be more likely to reference patents with lower informational rate, as suggested by Table \ref{table:RegressionsCitationsByGenderOfCitingInventorPatentsSeparateMeasuresV4Unstandardized}. Alternatively, examiners---who are tasked with assessing the factual content of an invention---may be less influenced by stylistic or personal features of the text, resulting in citation patterns that are less sensitive to involved or informational rates. Future work should aim to disentangle these factors and examine how gender and role-specific responsibilities affect citation behavior.

Our work is not without limitations. Importantly, our study inferred the gender of authors and inventors based on their names, which offers only a rough proxy. Moreover, our gender assignments were binary, corresponding to sex-based categories (female and male), which oversimplifies the true diversity of gender identities among authors and inventors. We also limited our attention to single-authored papers and patents and patents with a single lawyer. While this approach helps to make the interpretation of our results clearer, these patterns are not confined to single-authored work---they generally persist among collaboratively authored papers and multi-inventor patents, whether gender is measured by the first or the last contributor (Supplementary Tables~\ref{tab:c_pap_m}--\ref{tab:c_pat_m}). Further work on analyzing writing styles of mixed-gender teams could help provide a more nuanced understanding of the observed pattern. Additionally, our study is based on observational data, and therefore our results should not be interpreted as causal. As such, future behavioral studies, involving experimental assessment of the response of reviewers to highly involved and highly informational texts based on the reviewer's gender, would complement our corpus study of academic papers and patents.

Finally, our primary analysis is limited to abstracts of papers and patents, making it unclear whether the same patterns would hold in full‑length texts. Research that examines full-text funding applications reports only minimal stylistic variation between men and women \citep{horbach2022}, and another study found gender differences in abstracts but not in introductions \citep{lerchenmueller2019}. In our own data, however, a robustness analysis on the full text of open-access papers finds the same gender differences throughout the body and in every individual section, indicating that the pattern is not confined to abstracts; its generalizability beyond this open-access subset, and to the full text of patents, remains uncertain. Even so, abstracts play a pivotal role in the dissemination of research and significantly influence readers' decisions on whether to read the entire article. Therefore, gender‑related differences at the abstract level are still likely to exert a meaningful influence on citation behavior, even if those differences diminish in the body of the document. 

Notwithstanding these limitations, our study makes several contributions. First, it advances the Science of Science. We introduce a systematic, theoretically grounded set of linguistic features---motivated by the involved/informational framework established in linguistics as indicative of differences in male and female communication---for measuring writing style at scale. More consequentially, we move beyond documenting that style is gendered to show that it \emph{matters}---writing style predicts the gender of a paper's citers even when papers are compared against their most content-similar non-citers, identifying a content-independent linguistic channel through which gender gaps in evaluation may operate.
 
Second, our study adds new insights to the linguistics literature. Specifically, we use a larger sample of formal written text and utilize a new source of data (patent abstracts) which to our knowledge has not been used for the purposes of the analysis of gender differences in writing before. We also released an open-source Python package to compute the linguistic features used in this paper for future use of researchers, available here: \url{https://pypi.org/project/stylometer}.

Finally, our study also has several natural policy implications. In particular, our findings suggest that writing in science and technology carries personal, gendered markers and could therefore constitute a source of bias during evaluation. Some journals have already started implementing policies aimed at reducing stylistic bias, for example by discouraging the use of wording that emphasizes the novelty of the work, a phenomenon known to be used more frequently by men \citep{lerchenmueller2019}. Our measure can provide a tool for identifying the extent to which scientific writing relies on involved versus informational language, potentially giving editors an additional consideration when selecting reviewers to evaluate a paper. In addition, editors could use our measures to evaluate whether a manuscript is unusually high or low in involved or informational content. Editors could then encourage authors to revise such language to make their writing more accessible and appealing to a broader audience. Authors themselves may also benefit from greater awareness of these linguistic dimensions when communicating their work and may choose to adopt more neutral language. However, establishing rigid guidelines for communication styles in science and technology may not be the right solution. Instead, promoting universalism in scientific evaluation may require scientific institutions to develop processes that better account for diversity in communication styles while maintaining focus on the underlying quality of the research. We suggest that, in order to mitigate bias in evaluation processes, the diversity among the contributors should be matched by the same diversity among the evaluators. For example, previous researchers noted the need to increase the number of female reviewers \citep{kolev2020}, as well as to encourage citation diversity \citep{zurn2020} and continue raising awareness of language bias in established institutions, structures, and processes, including academia \citep{clements2021}. Such measures could help to uphold the norm of universalism in scientific contributions.

\section{Acknowledgements}
We thank Dr. Rebecca Levitan and Dr. Martin Chodorow for guidance and feedback for this work. The views expressed here are the authors' and do not reflect those of the Coleridge Initiative. 

\section{Funding}
This work was supported by the National Science Foundation under grants 1829168 (RJF) and 1932596 (RJF).

\section{Competing interests}
The authors declare that they have no competing interests.

\section{Data and materials availability}
The Python 3, MySQL 8, RStudio 1.2, and Stata 16 code used to analyze and visualize the data for the current study will be available from the corresponding author. PatentsView data used in the study are publicly available directly from the publisher. Data from the Web of Science are available from Clarivate Analytics, but restrictions apply to the availability of these data, which were used under license for the current study, and so are not publicly available. These data are however still available from the authors upon reasonable request and with permission from the publisher.

\clearpage
\bibliographystyle{apalike}
\bibliography{references}

@book{bhatia1993,
  author    = {Bhatia, Vijay K.},
  title     = {Analysing Genre: Language Use in Professional Settings},
  publisher = {Longman},
  address   = {London},
  year      = {1993}
}

@article{tekles2022,
  author    = {Tekles, Alexander and Auspurg, Katrin and Bornmann, Lutz},
  title     = {Same-gender citations do not indicate a substantial gender 
               homophily bias},
  journal   = {PLOS ONE},
  volume    = {17},
  number    = {9},
  pages     = {e0274810},
  year      = {2022},
  doi       = {10.1371/journal.pone.0274810}
}

@inproceedings{ghiasi2018,
  author    = {Ghiasi, Gita and Mongeon, Philippe and Sugimoto, Cassidy R. 
               and Lariviére, Vincent},
  title     = {Gender homophily in citations},
  booktitle = {23rd International Conference on Science and Technology 
               Indicators},
  pages     = {1519--1525},
  year      = {2018},
  publisher = {Centre for Science and Technology Studies (CWTS)},
  url       = {https://hdl.handle.net/1887/65291}
}

@article{Exley2024,
Author = {Exley, Christine L. and Nielsen, Kirby},
Title = {The Gender Gap in Confidence: Expected but Not Accounted For},
Journal = {American Economic Review},
Volume = {114},
Number = {3},
Year = {2024},
Month = {March},
Pages = {851–85},
DOI = {10.1257/aer.20221413},
URL = {https://www.aeaweb.org/articles?id=10.1257/aer.20221413}}

@article{kedrick2024cp,
	author = {Kedrick, Kara and Levitskaya, Ekaterina and Funk, Russell J.},
	date = {2024/10/01},
	doi = {10.1038/s41562-024-01957-x},
	id = {Kedrick2024},
	isbn = {2397-3374},
	journal = {Nature Human Behaviour},
	number = {10},
	pages = {1915--1923},
	title = {Conceptual structure and the growth of scientific knowledge},
	url = {https://doi.org/10.1038/s41562-024-01957-x},
	volume = {8},
	year = {2024}}

@misc{reimers2019sbert,
      title={Sentence-BERT: Sentence Embeddings using Siamese BERT-Networks}, 
      author={Nils Reimers and Iryna Gurevych},
      year={2019},
      eprint={1908.10084},
      archivePrefix={arXiv},
      primaryClass={cs.CL},
      url={https://arxiv.org/abs/1908.10084}, 
}

@article{douze2024faiss,
      title={The Faiss library},
      author={Matthijs Douze and Alexandr Guzhva and Chengqi Deng and Jeff Johnson and Gergely Szilvasy and Pierre-Emmanuel Mazaré and Maria Lomeli and Lucas Hosseini and Hervé Jégou},
      year={2024},
      eprint={2401.08281},
      archivePrefix={arXiv},
      primaryClass={cs.LG}
}

@article{rubin1992,
	author = {Rubin, Donald L. and Greene, Kathryn},
	doi = {https://doi.org/10.58680/rte199215447},
	issn = {1943-2348},
	journal = {Research in the Teaching of English},
	number = {1},
	pages = {7-40},
	publisher = {NCTE},
	title = {Gender-Typical Style in Written Language},
	type = {Journal Article},
	url = {https://publicationsncte.org/content/journals/10.58680/rte199215447},
	volume = {26},
	year = {1992}}

@article{francis2001,
	author = {Becky Francis and Jocelyn Robson and Barbara Read and},
	doi = {10.1080/03075070120076282},
	eprint = {https://doi.org/10.1080/03075070120076282},
	journal = {Studies in Higher Education},
	number = {3},
	pages = {313--326},
	publisher = {SRHE Website},
	title = {An Analysis of Undergraduate Writing Styles in the Context of Gender and Achievement},
	url = {https://doi.org/10.1080/03075070120076282},
	volume = {26},
	year = {2001}}

@article{franco2021,
	author = {Marina Christ Franco and Danielle B. Rice and Helena Silveira Schuch and Odir Antonio Dellagostin and Maximiliano S{\'e}rgio Cenci and David Moher},
	doi = {https://doi.org/10.1016/j.jclinepi.2021.01.018},
	issn = {0895-4356},
	journal = {Journal of Clinical Epidemiology},
	pages = {37-43},
	title = {The impact of gender on scientific writing: An observational study of grant proposals},
	url = {https://www.sciencedirect.com/science/article/pii/S0895435621000299},
	volume = {136},
	year = {2021}}

@article{Newman2008,
	author = {Matthew L. Newman and Carla J. Groom and Lori D. Handelman and James W. Pennebaker and},
	doi = {10.1080/01638530802073712},
	eprint = {https://doi.org/10.1080/01638530802073712},
	journal = {Discourse Processes},
	number = {3},
	pages = {211--236},
	publisher = {Routledge},
	title = {Gender Differences in Language Use: An Analysis of 14,000 Text Samples},
	url = {https://doi.org/10.1080/01638530802073712},
	volume = {45},
	year = {2008}}

@article{horbach2022,
	author = {Serge P.J.M. Horbach and Jesper W. Schneider and Maxime Sainte-Marie},
	doi = {https://doi.org/10.1016/j.joi.2022.101332},
	issn = {1751-1577},
	journal = {Journal of Informetrics},
	number = {4},
	pages = {101332},
	title = {Ungendered writing: Writing styles are unlikely to account for gender differences in funding rates in the natural and technical sciences},
	url = {https://www.sciencedirect.com/science/article/pii/S1751157722000840},
	volume = {16},
	year = {2022}}

@article {azoulay2020,
author = {Pierre Azoulay and Freda B. Lynn },
title = {Self-Citation, Cumulative Advantage, and Gender Inequality in Science},
journal = {Sociological Science},
volume = {7},
number = {7},
issn = {2330-6696},
url = {http://dx.doi.org/10.15195/v7.a7},
doi = {10.15195/v7.a7},
pages = {152--186},
year = {2020},
}

@book{merton1973,
  title={The sociology of science : theoretical and empirical investigations},
  author={Merton, R.K. and Storer, N.W.},
  isbn={},
  lccn={},
  series={},
  url={},
  year={1973},
  publisher={Chicago: University of Chicago Press}
}

@article{bornmann2011,
author = {Bornmann, Lutz},
title = {Scientific peer review},
journal = {Annual Review of Information Science and Technology},
volume = {45},
number = {1},
pages = {197-245},
doi = {https://doi.org/10.1002/aris.2011.1440450112},
url = {https://asistdl.onlinelibrary.wiley.com/doi/abs/10.1002/aris.2011.1440450112},
eprint = {https://asistdl.onlinelibrary.wiley.com/doi/pdf/10.1002/aris.2011.1440450112},
year = {2011}
}

@article{mulkay1976,
author = {Michael J. Mulkay},
title ={Norms and ideology in science},
journal = {Social Science Information},
volume = {15},
number = {4-5},
pages = {637-656},
year = {1976},
doi = {10.1177/053901847601500406},

URL = { 
        https://doi.org/10.1177/053901847601500406
    
},
eprint = { 
        https://doi.org/10.1177/053901847601500406
    
}}

@article{longandfox1995,
author = {Long, J. Scott and Fox, Mary Frank},
title = {Scientific Careers: Universalism and Particularism},
journal = {Annual Review of Sociology},
volume = {21},
number = {1},
pages = {45-71},
year = {1995},
doi = {10.1146/annurev.so.21.080195.000401},

URL = { 
        https://doi.org/10.1146/annurev.so.21.080195.000401
    
},
eprint = { 
        https://doi.org/10.1146/annurev.so.21.080195.000401
    
}}

@book{cole1992,
  title={Making Science: Between Nature and Society},
  author={Cole, S.},
  isbn={9780674543478},
  lccn={lc91042192},
  url={https://books.google.com/books?id=AusBqjmZM9AC},
  year={1992},
  publisher={Harvard University Press}
}

@article{teplitskiy2018,
title = {The sociology of scientific validity: How professional networks shape judgement in peer review},
journal = {Research Policy},
volume = {47},
number = {9},
pages = {1825-1841},
year = {2018},
issn = {0048-7333},
doi = {https://doi.org/10.1016/j.respol.2018.06.014},
url = {https://www.sciencedirect.com/science/article/pii/S0048733318301598},
author = {Misha Teplitskiy and Daniel Acuna and Aïda Elamrani-Raoult and Konrad Körding and James Evans}
}

@article{dondio2019,
title = {The “invisible hand” of peer review: The implications of author-referee networks on peer review in a scholarly journal},
journal = {Journal of Informetrics},
volume = {13},
number = {2},
pages = {708-716},
year = {2019},
issn = {1751-1577},
doi = {https://doi.org/10.1016/j.joi.2019.03.018},
url = {https://www.sciencedirect.com/science/article/pii/S1751157718304206},
author = {Pierpaolo Dondio and Niccolò Casnici and Francisco Grimaldo and Nigel Gilbert and Flaminio Squazzoni}
}

@article{lane2021,
author = {Lane, Jacqueline N. and Teplitskiy, Misha and Gray, Gary and Ranu, Hardeep and Menietti, Michael and Guinan, Eva C. and Lakhani, Karim R.},
title = {Conservatism Gets Funded? A Field Experiment on the Role of Negative Information in Novel Project Evaluation},
journal = {Management Science},
volume = {0},
number = {0},
pages = {null},
year = {2021},
doi = {10.1287/mnsc.2021.4107},

URL = { 
        https://doi.org/10.1287/mnsc.2021.4107
    
},
eprint = { 
        https://doi.org/10.1287/mnsc.2021.4107
    
}}

@article {mossracusin2012,
	author = {Moss-Racusin, Corinne A. and Dovidio, John F. and Brescoll, Victoria L. and Graham, Mark J. and Handelsman, Jo},
	title = {Science faculty{\textquoteright}s subtle gender biases favor male students},
	volume = {109},
	number = {41},
	pages = {16474--16479},
	year = {2012},
	doi = {10.1073/pnas.1211286109},
	publisher = {National Academy of Sciences},
	issn = {0027-8424},
	URL = {https://www.pnas.org/content/109/41/16474},
	eprint = {https://www.pnas.org/content/109/41/16474.full.pdf},
	journal = {Proceedings of the National Academy of Sciences}
}

@article {reuben2014,
	author = {Reuben, Ernesto and Sapienza, Paola and Zingales, Luigi},
	title = {How stereotypes impair women{\textquoteright}s careers in science},
	volume = {111},
	number = {12},
	pages = {4403--4408},
	year = {2014},
	doi = {10.1073/pnas.1314788111},
	publisher = {National Academy of Sciences},
	issn = {0027-8424},
	URL = {https://www.pnas.org/content/111/12/4403},
	eprint = {https://www.pnas.org/content/111/12/4403.full.pdf},
	journal = {Proceedings of the National Academy of Sciences}
}

@article{oliveira2019,
title = "Comparison of National Institutes of Health Grant Amounts to First-Time Male and Female Principal Investigators",
author = "Oliveira, {Diego F.M.} and Yifang Ma and Woodruff, {Teresa K} and Brian Uzzi",
year = "2019",
month = mar,
day = "5",
doi = "10.1001/jama.2018.21944",
language = "English (US)",
volume = "321",
pages = "898--900",
journal = "JAMA - Journal of the American Medical Association",
issn = "0098-7484",
publisher = "American Medical Association",
number = "9",
}

@article{card2020,
    author = {Card, David and DellaVigna, Stefano and Funk, Patricia and Iriberri, Nagore},
    title = "{Are Referees and Editors in Economics Gender Neutral?*}",
    journal = {The Quarterly Journal of Economics},
    volume = {135},
    number = {1},
    pages = {269-327},
    year = {2020},
    month = {2},
    issn = {0033-5533},
    doi = {10.1093/qje/qjz035},
    url = {https://doi.org/10.1093/qje/qjz035},
    eprint = {https://academic.oup.com/qje/article-pdf/135/1/269/32526130/qjz035.pdf},
}

@techreport{card2021,
 title = "Gender Differences in Peer Recognition by Economists",
 author = "Card, David and DellaVigna, Stefano and Funk, Patricia and Iriberri, Nagore",
 institution = "National Bureau of Economic Research",
 type = "Working Paper",
 series = "Working Paper Series",
 number = "28942",
 year = "2021",
 month = "June",
 doi = {10.3386/w28942},
 URL = "http://www.nber.org/papers/w28942"
}

@techreport{card2020_gender,
 title = "Gender neutrality in economics: The role of editors and referees",
 author = "Card, David and DellaVigna, Stefano and Funk, Patricia and Iriberri, Nagore",
 institution = "VOX, CEPR Policy Portal",
 number = "28942",
 year = "2020"
}

@article{mahajan2020,
  doi = {10.3171/2020.6.jns20902},
  url = {https://doi.org/10.3171/2020.6.jns20902},
  year = {2020},
  publisher = {Journal of Neurosurgery Publishing Group ({JNSPG})},
  volume = {135},
  number = {2},
  pages = {352--360},
  author = {Uma V. Mahajan and Harsh Wadhwa and Parastou Fatemi and Samantha Xu and Judy Shan and Deborah L. Benzil and Corinna C. Zygourakis},
  title = {Does double-blind peer review impact gender authorship trends? An evaluation of two leading neurosurgical journals from 2010 to 2019},
  journal = {Journal of Neurosurgery}
}

@TechReport{toole2021,
  author={Andrew A. Toole and Michelle J. Saksena and Charles A. W. Degrazia and Katherine P. Black and Francesco Lissoni and Ernest Miguelez and Gianluca Tarasconi},
  title={{Progress and Potential: 2020 update on U.S. women inventor-patentees}},
  year=2021,
  month=Jan,
  institution={HAL},
  type={Working Papers},
  url={https://ideas.repec.org/p/hal/wpaper/hal-03098153.html},
  number={hal-03098153},
  doi={},
}

@article{joshi2014,
author = {Aparna Joshi},
title ={By Whom and When Is Women’s Expertise Recognized? The Interactive Effects of Gender and Education in Science and Engineering Teams},
journal = {Administrative Science Quarterly},
volume = {59},
number = {2},
pages = {202-239},
year = {2014},
doi = {10.1177/0001839214528331},

URL = { 
        https://doi.org/10.1177/0001839214528331
    
},
eprint = { 
        https://doi.org/10.1177/0001839214528331
    
}}

@article{sarsons2017,
Author = {Sarsons, Heather},
Title = {Recognition for Group Work: Gender Differences in Academia},
Journal = {American Economic Review},
Volume = {107},
Number = {5},
Year = {2017},
Month = {May},
Pages = {141-45},
DOI = {10.1257/aer.p20171126},
URL = {https://www.aeaweb.org/articles?id=10.1257/aer.p20171126}}

@article{kolev2020,
Author = {Kolev, Julian and Fuentes-Medel, Yuly and Murray, Fiona},
Title = {Gender Differences in Scientific Communication and Their Impact on Grant Funding Decisions},
Journal = {AEA Papers and Proceedings},
Volume = {110},
Year = {2020},
Month = {May},
Pages = {245-49},
DOI = {10.1257/pandp.20201043},
URL = {https://www.aeaweb.org/articles?id=10.1257/pandp.20201043}}

@article{jensen2018,
author = {Jensen, Kyle and Kovacs, Balazs and Sorenson, Olav},
year = {2018},
month = {04},
pages = {307-309},
title = {Gender differences in obtaining and maintaining patent rights},
volume = {36},
journal = {Nature Biotechnology},
doi = {10.1038/nbt.4120}
}

@article{lariviere2013,
author={Larivi{\`e}re, Vincent
and Ni, Chaoqun
and Gingras, Yves
and Cronin, Blaise
and Sugimoto, Cassidy R.},
title={Bibliometrics: Global gender disparities in science},
journal={Nature},
year={2013},
month={Dec},
day={01},
volume={504},
number={7479},
pages={211-213},
issn={1476-4687},
doi={10.1038/504211a},
url={https://doi.org/10.1038/504211a}
}

@article{potthoff2017,
author={Potthoff, Matthias
and Zimmermann, Fabian},
title={Is there a gender-based fragmentation of communication science? An investigation of the reasons for the apparent gender homophily in citations},
journal={Scientometrics},
year={2017},
month={Aug},
day={01},
volume={112},
number={2},
pages={1047-1063},
issn={1588-2861},
doi={10.1007/s11192-017-2392-0},
url={https://doi.org/10.1007/s11192-017-2392-0}
}

@article{dion2018, title={Gendered Citation Patterns across Political Science and Social Science Methodology Fields}, volume={26}, DOI={10.1017/pan.2018.12}, number={3}, journal={Political Analysis}, publisher={Cambridge University Press}, author={Dion, Michelle L. and Sumner, Jane Lawrence and Mitchell, Sara McLaughlin}, year={2018}, pages={312–327}}

@article{leahey2007,
author = {Erin Leahey},
title ={Not by Productivity Alone: How Visibility and Specialization Contribute to Academic Earnings},
journal = {American Sociological Review},
volume = {72},
number = {4},
pages = {533-561},
year = {2007},
doi = {10.1177/000312240707200403},

URL = { 
        https://doi.org/10.1177/000312240707200403
    
},
eprint = { 
        https://doi.org/10.1177/000312240707200403
    
}}

@article{key2019, title={You Research Like a Girl: Gendered Research Agendas and Their Implications}, volume={52}, DOI={10.1017/S1049096519000945}, number={4}, journal={PS: Political Science \& Politics}, publisher={Cambridge University Press}, author={Key, Ellen M. and Sumner, Jane Lawrence}, year={2019}, pages={663–668}}

@article {vanderlee2015,
	author = {van der Lee, Romy and Ellemers, Naomi},
	title = {Gender contributes to personal research funding success in The Netherlands},
	volume = {112},
	number = {40},
	pages = {12349--12353},
	year = {2015},
	doi = {10.1073/pnas.1510159112},
	publisher = {National Academy of Sciences},
	issn = {0027-8424},
	URL = {https://www.pnas.org/content/112/40/12349},
	eprint = {https://www.pnas.org/content/112/40/12349.full.pdf},
	journal = {Proceedings of the National Academy of Sciences}
}

@article{hospido2019,
  title={Gender Gaps in the Evaluation of Research: Evidence from Submissions to Economics Conferences},
  author={Laura Hospido and Carlos Sanz},
  journal={Labor: Demographics \& Economics of the Family eJournal},
  year={2019}
}

@TechReport{hengel2017,
  author={Hengel, E.},
  title={{Publishing while Female. Are women held to higher standards? Evidence from peer review}},
  year=2017,
  month=Dec,
  institution={Faculty of Economics, University of Cambridge},
  type={Cambridge Working Papers in Economics},
  url={https://ideas.repec.org/p/cam/camdae/1753.html},
  number={1753},
  doi={},
}

@misc{hengel2020,
  title={Gender and equality at top economics journals},
  author={Erin Hengel and Eunyoung Moon},
  publisher = {Mimeo},
  year={2020}
}

@article{fortunato2018,
author = {Fortunato, Santo and Bergstrom, Carl and Borner, Katy and Evans, James and Helbing, Dirk and Milojevic, Stasa and Petersen, Alexander and Radicchi, Filippo and Sinatra, Roberta and Uzzi, Brian and Vespignani, Alessandro and Waltman, Ludo and Wang, Dashun and Barabasi, Albert-Laszlo},
year = {2018},
month = {03},
pages = {eaao0185},
title = {Science of science},
volume = {359},
journal = {Science},
doi = {10.1126/science.aao0185}
}

@article {lerchenmueller2019,
	author = {Lerchenmueller, Marc J and Sorenson, Olav and Jena, Anupam B},
	title = {Gender differences in how scientists present the importance of their research: observational study},
	volume = {367},
	elocation-id = {l6573},
	year = {2019},
	doi = {10.1136/bmj.l6573},
	publisher = {BMJ Publishing Group Ltd},
	URL = {https://www.bmj.com/content/367/bmj.l6573},
	eprint = {https://www.bmj.com/content/367/bmj.l6573.full.pdf},
	journal = {BMJ}
}

@article{kim2022,
title = {Gendered knowledge in fields and academic careers},
journal = {Research Policy},
volume = {51},
number = {1},
pages = {104411},
year = {2022},
issn = {0048-7333},
doi = {https://doi.org/10.1016/j.respol.2021.104411},
url = {https://www.sciencedirect.com/science/article/pii/S0048733321002079},
author = {Lanu Kim and Daniel Scott Smith and Bas Hofstra and Daniel A. McFarland}
}

@article{funk2017,
author = {Funk, Russell J. and Owen-Smith, Jason},
title = {A Dynamic Network Measure of Technological Change},
journal = {Management Science},
volume = {63},
number = {3},
pages = {791-817},
year = {2017},
doi = {10.1287/mnsc.2015.2366},

URL = { 
        https://doi.org/10.1287/mnsc.2015.2366
    
},
eprint = { 
        https://doi.org/10.1287/mnsc.2015.2366
    
}}

@article{birkle2020,
    author = {Birkle, Caroline and Pendlebury, David A. and Schnell, Joshua and Adams, Jonathan},
    title = "{Web of Science as a data source for research on scientific and scholarly activity}",
    journal = {Quantitative Science Studies},
    volume = {1},
    number = {1},
    pages = {363-376},
    year = {2020},
    month = {02},
    issn = {2641-3337},
    doi = {10.1162/qss_a_00018},
    url = {https://doi.org/10.1162/qss\_a\_00018},
    eprint = {https://direct.mit.edu/qss/article-pdf/1/1/363/1760864/qss\_a\_00018.pdf},
}

@book{jaffe2002,
  title={Patents, Citations, and Innovations: A Window on the Knowledge Economy},
  author={Adam B. Jaffe and Manuel Trajtenberg},
  publisher = {The MIT Press},
  year={2002}
}

@article{argamon2003,
author = {Argamon, Shlomo and Fine, Jonathan and Shimoni, Anat},
year = {2003},
month = {12},
pages = {},
title = {Gender, Genre, and Writing Style in Formal Written Texts},
volume = {23},
journal = {Text},
doi = {10.1515/text.2003.014}
}

@article{biber1986,
 ISSN = {00978507, 15350665},
 URL = {http://www.jstor.org/stable/414678},
 author = {Douglas Biber},
 journal = {Language},
 number = {2},
 pages = {384--414},
 publisher = {Linguistic Society of America},
 title = {Spoken and Written Textual Dimensions in English: Resolving the Contradictory Findings},
 volume = {62},
 year = {1986}
}

@article{marckworthbaker1974,
    title = "A Discriminant Function Analysis of Co-Variation of a Number of Syntactic Devices in Five Prose Genres",
    author = "Marckworth, Mary Lois  and
      Baker, Wm. J.",
    journal = "American Journal of Computational Linguistics",
    month = dec,
    year = "1974",
    note = "Microfiche 11",
    url = "https://aclanthology.org/J74-3006",
}

@book{biber1988, place={Cambridge}, title={Variation across Speech and Writing}, DOI={10.1017/CBO9780511621024}, publisher={Cambridge University Press}, author={Biber, Douglas}, year={1988}}

@book{biber_1998, place={Cambridge}, title={Corpus Linguistics Investigating Language Structure and Use}, DOI={10.1017/CBO9780511804489}, publisher={Cambridge University Press}, author={Biber, Douglas and Conrad, Susan and Reppen, Randi}, year={1998}}

@book{pennebaker2011,
  title={The Secret Life of Pronouns: What Our Words Say About Us},
  author={Pennebaker, J.W.},
  isbn={9781608194803},
  lccn={2011001289},
  url={https://books.google.com/books?id=Avz4rthHySEC},
  year={2011},
  publisher={Bloomsbury USA}
}

@inproceedings{garimella2016,
    title = "Zooming in on Gender Differences in Social Media",
    author = "Garimella, Aparna  and
      Mihalcea, Rada",
    booktitle = "Proceedings of the Workshop on Computational Modeling of People{'}s Opinions, Personality, and Emotions in Social Media ({PEOPLES})",
    month = dec,
    year = "2016",
    address = "Osaka, Japan",
    publisher = "The COLING 2016 Organizing Committee",
    url = "https://aclanthology.org/W16-4301",
    pages = "1--10"
}

@misc{honnibal2017,
    AUTHOR = {Honnibal, Matthew and Montani, Ines},
    TITLE  = {{spaCy 2}: Natural language understanding with {B}loom embeddings, convolutional neural networks and incremental parsing},
    year   = {2017}
}

@misc{weischedel2013,
  author = {Weischedel, Ralph and Palmer,Martha and Marcus, Mitchell and Hovy, Eduard and Pradhan, Sameer and Ramshaw, Lance and Xue, Nianwen and Taylor, Ann and Kaufman, Jeff and Franchini, Michelle and El-Bachouti, Mohammed and Belvin, Robert and Houston, Ann},
  title = {{OntoNotes Release 5.0 LDC2013T19. Web Download. Philadelphia: Linguistic Data Consortium}},
  year = {2013}
}

@article{thomson2001,
author = {Rob Thomson and Tamar Murachver and James Green},
title ={Where Is the Gender in Gendered Language?},
journal = {Psychological Science},
volume = {12},
number = {2},
pages = {171-175},
year = {2001},
doi = {10.1111/1467-9280.00329},
    note ={PMID: 11340928},

URL = { 
        https://doi.org/10.1111/1467-9280.00329
    
},
eprint = { 
        https://doi.org/10.1111/1467-9280.00329
}
}

@article{mulac2006,
    author = {Mulac, Anthony and Bradac, James J. and Gibbons, Pamela},
    title = "{Empirical Support for the Gender-as-Culture Hypothesis: An Intercultural Analysis of Male/Female Language Differences}",
    journal = {Human Communication Research},
    volume = {27},
    number = {1},
    pages = {121-152},
    year = {2006},
    month = {01},
    issn = {0360-3989},
    doi = {10.1111/j.1468-2958.2001.tb00778.x},
    url = {https://doi.org/10.1111/j.1468-2958.2001.tb00778.x},
    eprint = {https://academic.oup.com/hcr/article-pdf/27/1/121/22338892/jhumcom0121.pdf},
}

@article{ceci2014,
author = {Stephen J. Ceci and Donna K. Ginther and Shulamit Kahn and Wendy M. Williams},
title ={Women in Academic Science: A Changing Landscape},
journal = {Psychological Science in the Public Interest},
volume = {15},
number = {3},
pages = {75-141},
year = {2014},
doi = {10.1177/1529100614541236},
    note ={PMID: 26172066},

URL = { 
        https://doi.org/10.1177/1529100614541236
    
},
eprint = { 
        https://doi.org/10.1177/1529100614541236
    
}
}

@ARTICLE{su2015,
  
AUTHOR={Su, Rong and Rounds, James},   
	 
TITLE={All STEM fields are not created equal: People and things interests explain gender disparities across STEM fields},      
	
JOURNAL={Frontiers in Psychology},      
	
VOLUME={6},      

PAGES={189},     
	
YEAR={2015},      
	  
URL={https://www.frontiersin.org/article/10.3389/fpsyg.2015.00189},       
	
DOI={10.3389/fpsyg.2015.00189},      
	
ISSN={1664-1078}
}

@article{caplar2017,
   title={Quantitative evaluation of gender bias in astronomical publications from citation counts},
   volume={1},
   ISSN={2397-3366},
   url={http://dx.doi.org/10.1038/s41550-017-0141},
   DOI={10.1038/s41550-017-0141},
   number={6},
   journal={Nature Astronomy},
   publisher={Springer Science and Business Media LLC},
   author={Caplar, Neven and Tacchella, Sandro and Birrer, Simon},
   year={2017},
   month={May}
}

@article {dworkin2020,
	author = {Dworkin, Jordan D. and Linn, Kristin A. and Teich, Erin G. and Zurn, Perry and Shinohara, Russell T. and Bassett, Danielle S.},
	title = {The extent and drivers of gender imbalance in neuroscience reference lists},
	elocation-id = {2020.01.03.894378},
	year = {2020},
	doi = {10.1101/2020.01.03.894378},
	publisher = {Cold Spring Harbor Laboratory},
	URL = {https://www.biorxiv.org/content/early/2020/01/11/2020.01.03.894378},
	eprint = {https://www.biorxiv.org/content/early/2020/01/11/2020.01.03.894378.full.pdf},
	journal = {bioRxiv}
}

@article{zurn2020,
title = {The Citation Diversity Statement: A Practice of Transparency, A Way of Life},
journal = {Trends in Cognitive Sciences},
volume = {24},
number = {9},
pages = {669-672},
year = {2020},
issn = {1364-6613},
doi = {https://doi.org/10.1016/j.tics.2020.06.009},
url = {https://www.sciencedirect.com/science/article/pii/S1364661320301649},
author = {Perry Zurn and Danielle S. Bassett and Nicole C. Rust}
}

@book{clements2021,
  editor = {Clements, G. and Petray, M.J.},
  year = {2021},
  title = {Linguistic Discrimination in U.S Higher Education: Power, Prejudice, Impacts, and Remedies},
  edition = {1st},
  publisher = {Routledge},
  url = {https://doi.org/10.4324/9780367815103},
  doi = {10.4324/9780367815103},
  language = {en}
}

@article{vasarhelyi2021,
author = {Orsolya Vásárhelyi  and Igor Zakhlebin  and Staša Milojević  and Emőke-Ágnes Horvát },
title = {Gender inequities in the online dissemination of scholars work},
journal = {Proceedings of the National Academy of Sciences},
volume = {118},
number = {39},
pages = {e2102945118},
year = {2021},
doi = {10.1073/pnas.2102945118},
URL = {https://www.pnas.org/doi/abs/10.1073/pnas.2102945118},
eprint = {https://www.pnas.org/doi/pdf/10.1073/pnas.2102945118}}

@article{king2017,
author = {Molly M. King and Carl T. Bergstrom and Shelley J. Correll and Jennifer Jacquet and Jevin D. West},
title ={Men Set Their Own Cites High: Gender and Self-citation across Fields and over Time},
journal = {Socius},
volume = {3},
number = {},
pages = {2378023117738903},
year = {2017},
doi = {10.1177/2378023117738903},

URL = { 
        https://doi.org/10.1177/2378023117738903
    
},
eprint = { 
        https://doi.org/10.1177/2378023117738903
    
}}

@article{ni2021,
author = {Chaoqun Ni  and Elise Smith  and Haimiao Yuan  and Vincent Larivière  and Cassidy R. Sugimoto },
title = {The gendered nature of authorship},
journal = {Science Advances},
volume = {7},
number = {36},
pages = {eabe4639},
year = {2021},
doi = {10.1126/sciadv.abe4639},
URL = {https://www.science.org/doi/abs/10.1126/sciadv.abe4639},
eprint = {https://www.science.org/doi/pdf/10.1126/sciadv.abe4639}}

\clearpage
\pagestyle{empty}  
\clearpage
\begin{figure}[tp]
\centering
\includegraphics[width=0.78\textwidth]{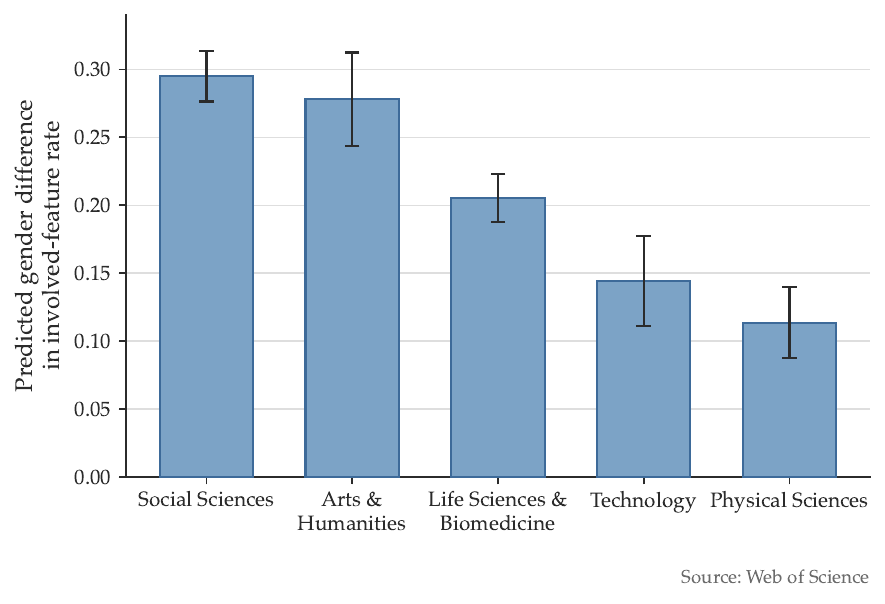}
\caption{\textbf{The gender difference in writing style is largest in the Social Sciences and Arts \& Humanities and smallest in the Physical Sciences.} Predicted difference between women and men in the involved-feature rate of single-authored paper abstracts, estimated separately by Web of Science research area (Supplementary Table \ref{detailed_papers}). Bars show the coefficient on the female indicator from ordinary least squares regressions with sub-field and year fixed effects; error bars are 95\% confidence intervals. Fields are ordered by effect size.}
\label{fig:Coefficients}
\end{figure}
\clearpage
\newgeometry{margin=0.5in}
\begin{landscape}
\begin{figure}[tp]
\centering
\includegraphics[width=\textwidth]{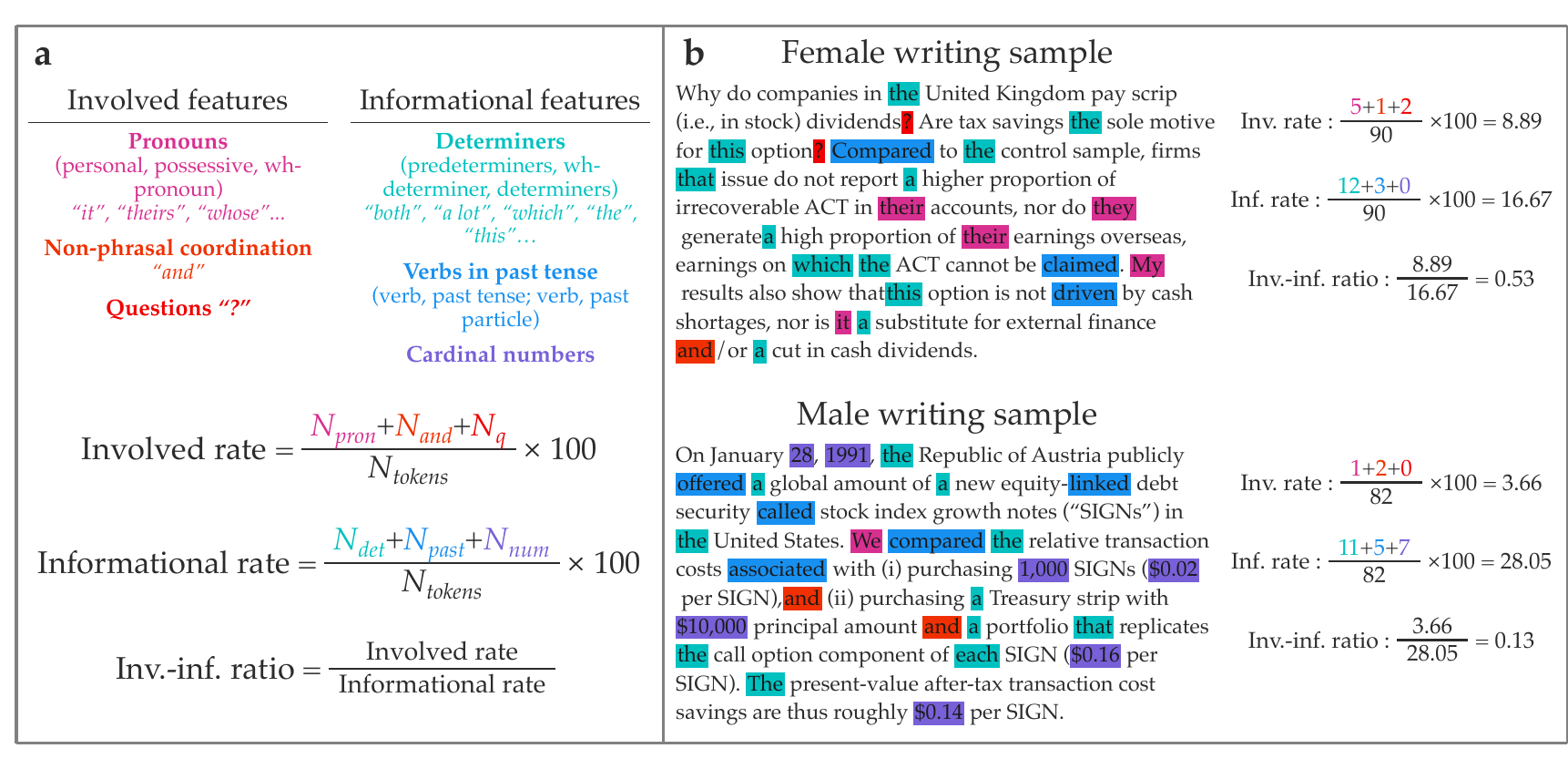}
\caption{\textbf{Involved and informational linguistic features and calculations.} \textbf{(a)} This panel specifies the involved and informational features used in our analysis and displays the formulas used to calculate the involved rate, the informational rate, and the Involved--Informational Ratio. \textbf{(b)} This panel presents two example texts from abstracts; one authored by a female and the other by a male. The features are highlighted, with colors corresponding to those denoting the features in panel a. On the right-hand side, the involved rate, informational rate, and Involved--Informational Ratio are calculated for the texts displayed on the left-hand side. Full definitions of each feature are given in Supplementary Table \ref{table_features}.}
\label{fig:gender_schematic}
\end{figure}
\end{landscape}
\newgeometry{margin=1in}
\clearpage
\newgeometry{margin=0.5in}
\begin{landscape}
\tikzstyle{mybox} = [align=left,draw=gray!55, thin, rectangle]

\begin{figure}[tp]
\centering
\begin{tikzpicture}
    \node [mybox, minimum height=6in, anchor=north] at (-1, 8.5){%
        \begin{minipage}{0.48\textwidth}
\fontsize{8.25pt}{8.6pt}\selectfont
\setlength{\lineskiplimit}{-4pt}
Why do companies in \hlblue{the} United Kingdom pay scrip (i.e., in stock) dividends\hlorange{?} Are tax savings \hlblue{the} sole motive for \hlblue{this} option\hlorange{?} If so, are \hlblue{all} tax-loss companies offering \hlblue{this} option\hlorange{?} Or is \hlblue{this} option \hlblue{driven} by other motives, such as cash savings, signaling, \hlorange{and} agency conflicts\hlorange{?} In \hlblue{this} study, \hlorange{I} provide \hlblue{some} insights into \hlblue{the} motivation for scrip-dividend payment by comparing \hlblue{the} operational performance \hlorange{and} other characteristics of \hlblue{all} companies \hlblue{that} \hlblue{distributed} scrip dividends with \hlblue{those} of \hlblue{a} control group of non-scrip-paying, but otherwise similar, firms.

\hspace{1em} Scrip dividends are \hlblue{offered} by \hlblue{an} increasing number of companies in \hlblue{the} United Kingdom as \hlblue{an} option whereby shareholders are able to choose between receiving dividends in cash or \hlblue{the} equivalent in \hlblue{the} form of shares (scrip). Companies stress tax savings when \hlorange{they} offer \hlblue{this} option because, unlike cash dividends, \hlblue{the} scrip option is not subject to \hlblue{a} payment of \hlblue{the} advanced corporation tax (ACT).

\hspace{1em} \hlorange{I} find that there are \hlblue{no} significant differences in \hlblue{the} scrip \hlorange{and} non-scrip-paying firms' tax exposures. \hlblue{Compared} to \hlblue{the} control sample, firms \hlblue{that} issue scrip dividends do not report \hlblue{a} higher proportion of irrecoverable ACT in \hlorange{their} accounts, nor do \hlorange{they} generate \hlblue{a} high proportion of \hlorange{their} earnings overseas, earnings on \hlblue{which} \hlblue{the} ACT cannot be \hlblue{claimed}. \hlorange{My} results also show that \hlblue{this} option is not \hlblue{driven} by cash shortages, nor is \hlorange{it} \hlblue{a} substitute for external finance and/or \hlblue{a} cut in cash dividends. However, \hlorange{I} find that firms \hlblue{that} pay scrip dividends are, on average, large, \hlorange{and} have high dividend yield, suggesting that \hlblue{the} cash \hlblue{saved} is substantial. However, \hlblue{these} firms already have high cash-flow balances \hlorange{and} low growth opportunities. \hlblue{The} overall results suggest that firms in \hlblue{the} UK do not appear to be making \hlblue{the} optimal financial choice for \hlorange{their} dividend policies.

\hspace{1em} Like stock dividends \hlorange{and} dividend reinvestment plans in \hlblue{the} US. scrip dividends provide \hlblue{both} firm \hlorange{and} shareholders with many benefits. Companies are able to retain cash without altering \hlorange{their} payout policies or raising new funds in \hlblue{the} capital markets; thus, companies \hlblue{that} use \hlblue{this} option can save on borrowing costs, underwriting fees, other issue costs, \hlorange{and} avoid negative signals of new equity. At \hlblue{the} same time, \hlorange{they} provide \hlorange{their} shareholders with \hlblue{the} opportunity to increase \hlorange{their} holdings without incurring \hlblue{any} transaction costs. In addition, \hlblue{the} UK institutional setting \hlorange{and} relevant tax code provisions allow companies to derive tax benefits from \hlblue{this} option. Unlike cash dividends, \hlblue{this} option is not subject to ACT. \hlblue{This} tax is first \hlblue{paid} when \hlblue{a} company declares cash dividends \hlorange{and} is \hlblue{deducted} from \hlorange{its} corporation-tax liability \hlblue{nine} months after \hlblue{the} accounting year-end if dividends are \hlblue{paid} from domestic earnings \hlorange{and} if taxable profit is higher than gross cash dividends. If \hlblue{these} \hlblue{two} conditions are not \hlblue{met}, surplus ACT is \hlblue{carried} in \hlblue{the} accounts \hlorange{and} \hlblue{set} off against preceding or immediately following periods. For companies with \hlblue{no} previous or foreseeable future taxable profit, ACT is \hlblue{written} off as \hlblue{a} loss against reserves. Thus, companies with prospective ACT loss are more likely to pay scrip, rather than cash, dividends.

\hspace{1em} However, scrip dividends are not costless. Firms incur administrative costs in Petting up \hlorange{and} running \hlblue{the} scheme, such as \hlblue{the} costs of advertising \hlblue{the} option, preparing \hlorange{and} mailing \hlblue{the} scrip-dividend prospectus, \hlorange{and} printing \hlorange{and} allocating \hlblue{the} new shares. Moreover, scrip dividends impose costs on \hlblue{the} firm's shareholders. \hlblue{This} option is not \hlblue{offered} to foreign investors. For \hlblue{some} domestic investors, \hlorange{it} is \hlblue{taxed} differently from cash dividends. While \hlblue{the} tax credit \hlblue{associated} with cash dividends can be \hlblue{claimed} by \hlblue{all} investors, \hlblue{the} tax credit on scrip dividends can only be \hlblue{claimed} by tax-paying individual investors. Individuals who have reliefs \hlorange{and} allowances in excess of \hlorange{their} income, as well as corporate investors, forgo \hlblue{the} tax credit when \hlorange{they} opt for scrip dividends. Thus, corporate \hlorange{and} nontaxable individual investors are likely to prefer cash dividends; individual shareholders \hlblue{taxed} at \hlblue{a} higher rate of income tax may opt for scrip dividends for \hlblue{which} \hlblue{the} firm issues additional shares. However, such additional shares could result in \hlblue{a} dilution of control for shareholders who opt for cash dividends, \hlorange{and} \hlblue{a} loss in future dividend increases if scrip dividends limit \hlblue{the} scope for such increases.

\hspace{1em} Why, then, do companies issue scrip dividends\hlorange{?} \hlblue{One} possibility is that individual tax-paying investors request \hlblue{this} option from \hlorange{their} companies, \hlblue{which} suggests that companies are subject to monitoring by atomistic shareholders while large shareholders, such as corporate investors, are passive owners.

\hspace{1em} \hlorange{I} suggest \hlblue{a} number of areas for further research \hlblue{that} will assess \hlblue{the} extent to \hlblue{which} \hlblue{the} granting of \hlblue{this} option is \hlblue{driven} by nonfinancial considerations.

        \end{minipage}
        };
    \node [mybox, minimum height=6in] at (12.5, 0.5){%
        \begin{minipage}{0.48\textwidth}
\fontsize{8.25pt}{8.6pt}\selectfont
\setlength{\lineskiplimit}{-4pt}

\hlblue{The} rich variety of securities innovations in recent years continues to intrigue academicians \hlorange{and} practitioners alike. \hlblue{The} array of innovative securities includes \hlblue{a} variety of equity-\hlblue{linked} debt securities. On January \hlblue{28}, \hlblue{1991}, \hlblue{the} Republic of Austria publicly \hlblue{offered} \$\hlblue{100} \hlblue{million} principal amount of \hlblue{a} new equity-\hlblue{linked} debt security \hlblue{called} stock index growth notes (''SIGNs'') in \hlblue{the} United States. \hlblue{The} SIGNs \hlblue{were} \hlblue{scheduled} to mature in approximately \hlblue{5.5} years. \hlorange{They} would make \hlblue{no} payments of interest prior to maturity. If \hlblue{the} value of \hlblue{the} S\&P \hlblue{500} is below \hlblue{336.69} on \hlblue{the} maturity date, \hlblue{the} holder will receive \$\hlblue{10}, \hlblue{the} initial offering price. If \hlblue{the} value exceeds \hlblue{336.69}, \hlblue{the} holder will get \$\hlblue{10} plus \$\hlblue{10} \hlblue{multiplied} by \hlblue{the} percentage appreciation in \hlblue{the} S\&P \hlblue{500} above \hlblue{336.69}. SIGNS may thus be \hlblue{characterized} as \hlblue{a} package consisting of (i) \hlblue{a} \hlblue{5.5}-year triple-A-\hlblue{rated} \hlblue{zero} coupon note plus (ii) \hlblue{a} \hlblue{5.5}-year European call option, or warrant, on \hlblue{the} S\&P \hlblue{500} with \hlblue{a} strike price of \hlblue{336.69}.

\hspace{1em} \hlblue{The} value of \hlblue{each} SIGN can be \hlblue{expressed} as \hlblue{the} sum of (i) \hlblue{the} value of \hlblue{the} \hlblue{zero} coupon bond component plus (ii) \hlblue{the} value of \hlblue{the} call option component plus (iii) \hlblue{the} value resulting from interest on \hlblue{the} SIGNs not being taxable until \hlblue{the} maturity date plus (iv) \hlblue{the} reduction, if \hlblue{any}, in transaction costs vis-\hlblue{a}-vis acquiring \hlblue{a} comparable \hlblue{zero} coupon bond \hlorange{and} purchasing call options on \hlblue{the} S\&P \hlblue{500} separately plus (v) \hlblue{the} value, if \hlblue{any}, attributable to creating \hlblue{an} investment alternative \hlblue{that} is not otherwise available to \hlblue{the} investors who \hlblue{purchased} \hlblue{the} SIGNs.

\hspace{1em} On January \hlblue{28}. \hlblue{1991}, \hlblue{the} Republic of Austria \hlblue{had} outstanding \hlblue{an} issue of \hlblue{zero} coupon bonds maturing July \hlblue{17}, \hlblue{1995}, \hlblue{that} \hlblue{was} yielding \hlblue{8.05}\% per annum semiannually \hlblue{compounded}. \hlblue{Based} on \hlblue{this} yield, \hlblue{the} \hlblue{zero} coupon bond component of \hlblue{each} SIGN \hlblue{was} worth approximately \$\hlblue{6.48}. \hlblue{The} call option component, \hlblue{valued} using \hlblue{the} Black-Scholes model \hlblue{modified} to account for dividend-paying stocks, \hlblue{was} \hlblue{estimated} to be \$\hlblue{2.30}.

\hspace{1em} \hlblue{The} value of \hlblue{the} tax arbitrage \hlblue{created} through \hlblue{the} introduction of SIGNs equals \hlblue{the} sum of (i) \hlblue{the} tax benefit resulting from \hlblue{the} investor's ability to defer income taxes on \hlblue{the} amortization of \hlblue{the} \hlblue{zero} coupon bond component (\$\hlblue{0.11} per SIGN) plus (ii) \hlblue{the} tax benefit resulting from \hlblue{the} investor's ability to defer income taxes \hlblue{that} would otherwise be \hlblue{incurred} on \hlblue{the} option component under \hlblue{the} mark-to-market rules (\$\hlblue{0.09} per SIGN). \hlblue{The} total value of \hlblue{the} tax arbitrage is \$\hlblue{0.20} per SIGN, just \hlblue{two} percent of \hlblue{the} public offering price per SIGN.

\hspace{1em} \hlorange{I} \hlblue{compared} \hlblue{the} relative transaction costs \hlblue{associated} with (i) purchasing \hlblue{1,000} SIGNs (\$\hlblue{0.02} per SIGN), \hlorange{and} (ii) purchasing \hlblue{a} Treasury strip with \$\hlblue{10,000} principal amount \hlorange{and} \hlblue{a} portfolio \hlblue{that} replicates \hlblue{the} call option component of \hlblue{each} SIGN (\$\hlblue{0.16} per SIGN). \hlblue{The} present-value after-tax transaction cost savings are thus roughly \$\hlblue{0.14} per SIGN.

\hspace{1em} \hlblue{The} creation of \hlblue{a} new security can benefit investors if \hlorange{it} creates new risk-return combinations. Call options on \hlblue{the} S\&P \hlblue{500} with \hlblue{a} time to expiration of \hlblue{5.5} years are not available on \hlblue{any} of \hlblue{the} options exchanges. \hlblue{An} investor who \hlblue{wished} to duplicate \hlblue{the} return stream of \hlblue{the} SIGNs could try to purchase \hlblue{a} \hlblue{customized} option in \hlblue{the} over-\hlblue{the}-counter market or employ \hlblue{a} dynamic replication strategy. \hlblue{The} SIGNs \hlblue{were} \hlblue{sold} in small denominations primarily to retail investors, who would not have \hlblue{had} access to \hlblue{the} over-\hlblue{the}-counter market on account of \hlblue{the} small transaction size. Dynamic replication would be prohibitively expensive for at least \hlblue{some} retail investors because of transaction costs.

\hspace{1em} \hlblue{The} SIGNs \hlblue{were} \hlblue{sold} to investors at \hlblue{a} price of \$\hlblue{10.00}. \hlblue{Based} on \hlblue{the} \hlblue{four} \hlblue{estimated} component values. \hlblue{the} \hlblue{implied} value attributable to creating \hlblue{a} new investment alternative is \$\hlblue{0.88} per SIGN. However, \hlblue{this} value also includes \hlblue{the} effect of \hlblue{any} mispricing \hlblue{that} might have \hlblue{occurred}. \hlblue{The} \hlblue{predicted} value of \hlblue{the} SIGNs \hlblue{was} \hlblue{8.80}\% below \hlblue{the} market price on January \hlblue{28}, \hlblue{1991}. At \hlblue{the} end of February \hlblue{1991}, \hlblue{the} \hlblue{predicted} value \hlblue{was} only \hlblue{1.60}\% below \hlblue{the} market price. At \hlblue{the} end of March \hlblue{1991}, \hlblue{the} \hlblue{predicted} value \hlblue{was} \$\hlblue{10.03}, \hlblue{which} slightly \hlblue{exceeded} \hlblue{the} \$\hlblue{10,00} market price. Thereafter, \hlblue{the} \hlblue{predicted} value \hlorange{and} \hlblue{the} market price approximate \hlblue{one} \hlblue{another}. \hlblue{This} price behavior suggests that investors may have initially \hlblue{overvalued} \hlblue{the} SIGNs but that \hlblue{the} mispricing \hlblue{was} \hlblue{eliminated} within roughly \hlblue{one} to \hlblue{two} months. \hlorange{It} also suggests that \hlblue{the} value of \hlblue{the} innovation should be \hlblue{attributed} primarily to \hlblue{the} value of \hlblue{the} tax arbitrage \hlorange{and} to \hlblue{the} reduction in after-tax transaction costs \hlorange{it} \hlblue{made} possible.

\hspace{1em} \hlblue{The} Republic of Austria \hlblue{used} approximately \hlblue{30}\% of \hlblue{the} \$\hlblue{9.50} net proceeds from \hlblue{the} sale of \hlblue{the} SIGNs issue to hedge \hlorange{its} contingent S\&P \hlblue{500} liability, leaving proceeds net of hedging costs amounting to \$\hlblue{6.65}. \hlblue{These} net proceeds exceed \hlblue{the} \$\hlblue{6.48} value of \hlblue{the} bond component of \hlblue{each} SIGN. \hlblue{The} profit of \$\hlblue{0.17} per SIGN effectively \hlblue{reduced} \hlblue{the} Republic of Austria's cost of issuing \hlblue{5.5}-year \hlblue{zero} coupon debt to \hlblue{7.56}\%, resulting in \hlblue{a} savings of \hlblue{49} basis points per annum.
        \end{minipage}
        };

\node [rectangle,minimum width=1cm,inner sep=0pt,label=0:{\large \textbf{Female}}] (female) at (-1, 8.9) {};
\node [rectangle,minimum width=1cm,inner sep=0pt,label=0:{\large \textbf{Male}}] (male) at (11, 8.9) {};

\node [rectangle,minimum width=0.7cm,fill=orange,draw=orange,thick,inner sep=0pt,minimum height=0.22cm,label={[label distance=-1pt]0:{\small Involved}}] at (-3.2,-7.42) {};
\node [rectangle,minimum width=0.7cm,fill=blue,draw=blue,thick,inner sep=0pt,minimum height=0.22cm,label={[label distance=-1pt]0:{\small Informational}}] at (-0.6,-7.42) {};

\end{tikzpicture}
\caption{\textbf{Illustrative writing samples by author gender.} Excerpts from two single-authored abstracts matched on journal (\emph{Financial Management Journal}), field (Business \& Economics), and period (1990s): one written by a female author (left) and one by a male author (right). Involved features (e.g., questions, first-person pronouns, the connective ``and'') are highlighted in orange; informational features (e.g., cardinal numbers, determiners, past-tense verbs) in blue.}
\label{fig:comparison}
\end{figure}
\end{landscape}
\newgeometry{margin=1in}
\clearpage
\begin{figure}[tp]
\centering
\includegraphics[width=\textwidth]{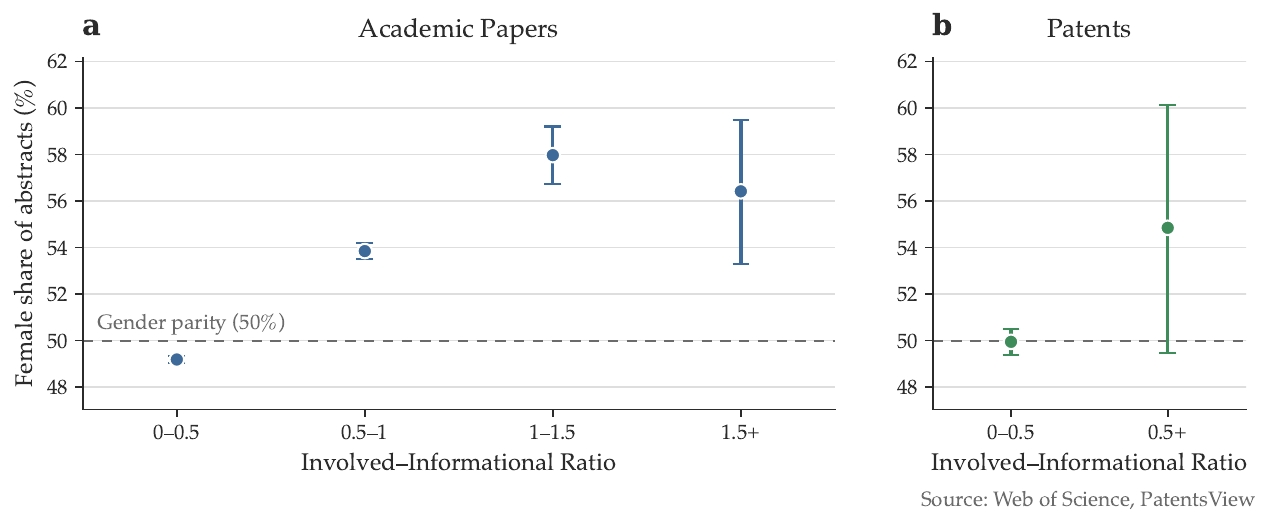}
\caption{\textbf{Abstracts with a higher Involved--Informational Ratio are increasingly written by women.} Female share of single-authored abstracts within bins of the Involved--Informational Ratio, for \textbf{(a)} academic papers and \textbf{(b)} patents. Points give the percentage of abstracts in each bin authored or invented by women; error bars are 95\% Wilson confidence intervals. The dashed line marks gender parity (50\%): points whose intervals lie above it indicate a female-majority share. For papers the female share rises from just below parity at low ratios to well above it at higher ratios; for patents the same shift is weaker and less precisely estimated.}
\label{fig:academic papers_patents_bars}
\end{figure}
\clearpage
\newgeometry{margin=0.5in}
\begin{landscape}
\begin{figure}[p]
\centering
\includegraphics[width=\textwidth]{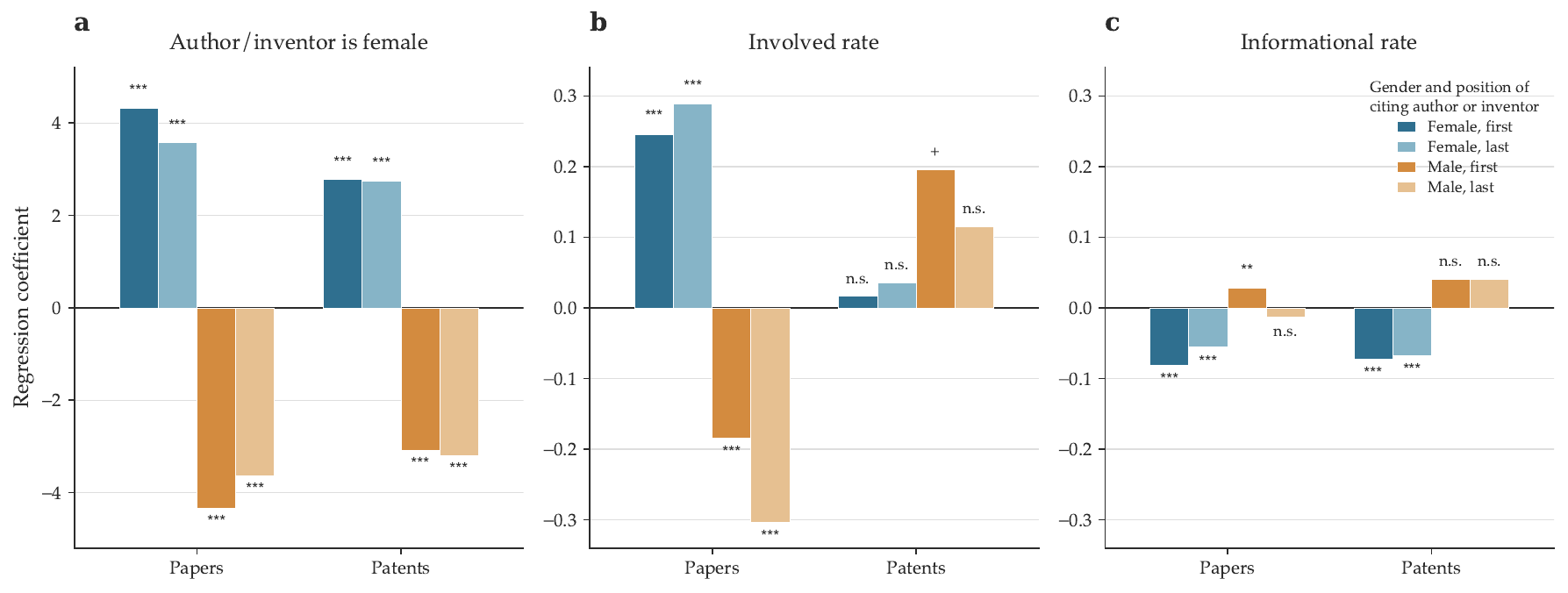}
\caption{\textbf{Regression coefficients for citation patterns.} The underlying estimates are reported in Table \ref{table:RegressionsCitationsByGenderOfCitingAuthorPapersSeparateMeasuresV4Unstandardized} (papers) and Table \ref{table:RegressionsCitationsByGenderOfCitingInventorPatentsSeparateMeasuresV4Unstandardized} (patents). This panel displays plots that illustrate the magnitude and direction of the regression coefficients when the outcome variable is the gender of the first or last citing author (for papers) or inventor (for patents). \textbf{(a)} This plot reveals that papers written by
female authors receive a greater share of citations from papers with a female first author and papers with a female last author; the opposite
pattern holds for papers written by men. \textbf{(b)} This plot indicates that papers with higher rates of involved features tend to be cited at a higher rate by papers with a female first author and papers with a female last author; whereas papers with higher rates of involved features are cited at a lower rate by papers with a male first author. The involved rate was an insignificant predictor in the patent models, except for male first inventors, where it was marginally significant. \textbf{(c)} This plot indicates that papers with higher rates of informational features tend to be cited at a lower rate by papers with a female first author and a female last author, and at a higher rate by papers with a male first author (the association for male last authors is not significant). For patents, higher informational rates are likewise associated with fewer citations from female first and last inventors. \emph{Note:} The regression coefficients on the y-axis display a wider range for (a) compared to (b) and (c). The stars above each bar denote levels of significance (\({+}p<0.10; {*}p<0.05; {**}p<0.01; {***}p<0.001\)), and `n.s.' indicates `not significant'.}
\label{fig:citation_analysis}
\end{figure}
\end{landscape}
\newgeometry{margin=1in}
\clearpage

\clearpage
\clearpage
\begin{table}[tp]
\centering
\caption{Distribution of individual involved and informational features by author and inventor gender}\label{table:2}
\begin{adjustbox}{max width=\textwidth,center}
\begin{threeparttable}
\begin{tabular}{lccccc}
\toprule
 & \multicolumn{2}{c}{Female} & \multicolumn{2}{c}{Male} & \\
\cmidrule(lr){2-3}\cmidrule(lr){4-5}
 & Mean & SD & Mean & SD & $p$-value \\
\midrule
\multicolumn{6}{l}{\textbf{Academic papers}} \\
\multicolumn{6}{l}{\textit{Informational features}} \\
Numbers & 1.52 & 1.81 & 1.54 & 1.85 & $<$0.001 \\
Determiners & 9.93 & 3.07 & 10.14 & 3.08 & $<$0.001 \\
Past tense & 4.50 & 2.47 & 4.45 & 2.41 & $<$0.001 \\
\multicolumn{6}{l}{\textit{Involved features}} \\
Questions & 0.03 & 0.19 & 0.03 & 0.19 & 0.129 \\
``and'' & 3.16 & 1.49 & 2.94 & 1.47 & $<$0.001 \\
Pronoun & 1.60 & 1.53 & 1.58 & 1.51 & $<$0.001 \\
\midrule
\multicolumn{6}{l}{\textbf{Patents}} \\
\multicolumn{6}{l}{\textit{Informational features}} \\
Numbers & 1.30 & 2.23 & 1.29 & 2.15 & 0.766 \\
Determiners & 15.91 & 3.71 & 15.44 & 3.76 & $<$0.001 \\
Past tense & 4.39 & 2.09 & 4.47 & 2.13 & $<$0.001 \\
\multicolumn{6}{l}{\textit{Involved features}} \\
Questions & NA & NA & NA & NA & NA \\
``and'' & 2.75 & 1.50 & 2.60 & 1.43 & $<$0.001 \\
Pronoun & 0.33 & 0.62 & 0.34 & 0.60 & 0.516 \\
\bottomrule
\end{tabular}
\begin{tablenotes}\footnotesize
\item Means and standard deviations (SD) of each writing-style feature for female- and male-authored papers and female- and male-invented patents. p-values are from two-sample $t$-tests comparing female and male means. NA denotes a feature not present in patent abstracts.
\end{tablenotes}
\end{threeparttable}
\end{adjustbox}
\end{table}
\clearpage
\begin{table}[tp]
\centering
\caption{Distribution of aggregate involved and informational measures by author and inventor gender}\label{table:Features_academic papers_and_patents}
\begin{adjustbox}{max width=\textwidth,center}
\begin{threeparttable}
\begin{tabular}{lccccc}
\toprule
 & \multicolumn{2}{c}{Female} & \multicolumn{2}{c}{Male} & \\
\cmidrule(lr){2-3}\cmidrule(lr){4-5}
 & Mean & SD & Mean & SD & $p$-value \\
\midrule
\multicolumn{6}{l}{\textbf{Academic papers}} \\
Involved (normalized) & 4.78 & 2.14 & 4.55 & 2.09 & $<$0.001 \\
Informational (normalized) & 15.96 & 4.04 & 16.13 & 4.02 & $<$0.001 \\
Ratio (normalized) & 0.33 & 0.21 & 0.31 & 0.20 & $<$0.001 \\
\midrule
\multicolumn{6}{l}{\textbf{Patents}} \\
Involved (normalized) & 3.08 & 1.60 & 2.94 & 1.53 & $<$0.001 \\
Informational (normalized) & 21.60 & 4.07 & 21.20 & 4.09 & $<$0.001 \\
Ratio (normalized) & 0.15 & 0.10 & 0.15 & 0.13 & 0.006 \\
\bottomrule
\end{tabular}
\begin{tablenotes}\footnotesize
\item Means and standard deviations (SD) of the aggregate involved and informational measures and their ratio (all normalized) for female- and male-authored papers and female- and male-invented patents. p-values are from two-sample $t$-tests comparing female and male means.
\end{tablenotes}
\end{threeparttable}
\end{adjustbox}
\end{table}
\clearpage
\begin{table}[tp]
\centering
\caption{Writing style by author gender, single-authored papers}\label{table:coefficients_academic papers_rate}
\begin{adjustbox}{max width=\textwidth,center}
\begin{threeparttable}
\begin{tabular}{lcccccc}
\toprule
 & \multicolumn{2}{c}{Involved--Informational Ratio} & \multicolumn{2}{c}{Involved rate} & \multicolumn{2}{c}{Informational rate} \\
\cmidrule(lr){2-3}\cmidrule(lr){4-5}\cmidrule(lr){6-7}
 & (1) & (2) & (3) & (4) & (5) & (6) \\
\midrule
Author is female (1 = Yes)                                                                                                                            & 0.0209*** & 0.0209*** & 0.2320*** & 0.2321*** & -0.1765*** & -0.1771*** \\
                                                                                                                                                                     & (0.0006) & (0.0005) & (0.0059) & (0.0054) & (0.0113) & (0.0107) \\
Constant                                                                                                                                               & 0.3131*** & 0.1408*** & 4.5491*** & 2.7018*** & 16.1328*** & 19.8950*** \\
                                                                                                                          & (0.0004) & (0.0098) & (0.0041) & (0.1171) & (0.0079) & (0.3192) \\
\midrule
Field fixed effects &  & \checkmark &  & \checkmark &  & \checkmark \\
Year fixed effects &  & \checkmark &  & \checkmark &  & \checkmark \\
\midrule
$R^2$ & 0.0025       & 0.1271          & 0.0030       & 0.1692       & 0.0005        & 0.0991             \\
Observations & 512,940       & 512,520       & 512,940        & 512,520        & 512,940       & 512,520       \\
\bottomrule
\end{tabular}
\begin{tablenotes}\footnotesize
\item Ordinary least squares regressions of each writing-style measure on a female indicator for single-authored academic papers. Columns (1), (3), and (5) include no fixed effects; columns (2), (4), and (6) add field and year fixed effects (\checkmark). The fixed-effects columns drop 420 papers (210 per gender) that lack a field classification, leaving 512,520 observations; the no-fixed-effects columns retain the full matched sample of 512,940. Detailed results by field appear in Table~\ref{detailed_papers}, and the involved-rate differences are visualized in Figure~\ref{fig:Coefficients}. Robust standard errors in parentheses. +$p<0.10$; *$p<0.05$; **$p<0.01$; ***$p<0.001$.
\end{tablenotes}
\end{threeparttable}
\end{adjustbox}
\end{table}
\clearpage
\begin{table}[tp]
\centering
\caption{Writing style by inventor and lawyer gender, single-inventor patents}\label{table:coefficients_patents_rate}
\begin{adjustbox}{max width=\textwidth,center}
\begin{threeparttable}
\begin{tabular}{lcccccc}
\toprule
 & \multicolumn{2}{c}{Involved--Informational Ratio} & \multicolumn{2}{c}{Involved rate} & \multicolumn{2}{c}{Informational rate} \\
\cmidrule(lr){2-3}\cmidrule(lr){4-5}\cmidrule(lr){6-7}
 & (1) & (2) & (3) & (4) & (5) & (6) \\
\midrule
Inventor is female (1 = Yes) & 0.0033** & 0.0034** & 0.1436*** & 0.1436*** & 0.4163*** & 0.4086*** \\
                                            & (0.0013) & (0.0013) & (0.0175) & (0.0173) & (0.0454) & (0.0434) \\
Lawyer is female (1 = Yes)   & 0.0150*** & 0.0076** & 0.0234 & 0.0240 & -1.2126*** & -0.6160*** \\
                                            & (0.0028) & (0.0028) & (0.0357) & (0.0356) & (0.0970) & (0.0922) \\
Constant                                 & 0.1478*** & 0.2094*** & 2.9386*** & 3.1519*** & 21.2738*** & 17.5371*** \\
                                     & (0.0010) & (0.0202) & (0.0123) & (0.2253) & (0.0326) & (0.6345) \\
\midrule
Field fixed effects &  & \checkmark &  & \checkmark &  & \checkmark \\
Year fixed effects &  & \checkmark &  & \checkmark &  & \checkmark \\
\midrule
$R^2$ & 0.0013       & 0.0297                  & 0.0021       & 0.0219       & 0.0080        & 0.0989        \\
Observations & 32,106        & 32,106        & 32,106         & 32,106         & 32,106        & 32,106         \\
\bottomrule
\end{tabular}
\begin{tablenotes}\footnotesize
\item Ordinary least squares regressions of each writing-style measure on inventor- and lawyer-female indicators for single-inventor patents. Columns (1), (3), and (5) include no fixed effects; columns (2), (4), and (6) add field and year fixed effects (\checkmark). Detailed results by patent category appear in Table~\ref{detailed_patents}. Robust standard errors in parentheses. +$p<0.10$; *$p<0.05$; **$p<0.01$; ***$p<0.001$.
\end{tablenotes}
\end{threeparttable}
\end{adjustbox}
\end{table}
\clearpage
 
\begin{landscape}
 
\thispagestyle{empty}

{
 
\scriptsize
 
\setlength{\tabcolsep}{2pt}
 
\renewcommand{\arraystretch}{1}
 
\begin{table}[tp]\centering
 
\scalebox{0.8}{%
 
\begin{threeparttable}
\def\sym#1{\ifmmode^{#1}\else\(^{#1}\)\fi}
\caption{Citations to sample papers as a function of author gender and writing style. These coefficients are visualized in Figure \ref{fig:citation_analysis}; standardized coefficients are reported in Supplementary Table \ref{table:RegressionsCitationsByGenderOfCitingAuthorPapersSeparateMeasuresV4Standardized}}
\label{table:RegressionsCitationsByGenderOfCitingAuthorPapersSeparateMeasuresV4Unstandardized}
\begin{tabular}{l*{8}{c}}
\toprule
 
 
                                             &\multicolumn{2}{c}{\shortstack{Citations from female first author}}&\multicolumn{2}{c}{\shortstack{Citations from male first author}}&\multicolumn{2}{c}{\shortstack{Citations from female last author}}&\multicolumn{2}{c}{\shortstack{Citations from male last author}}\\\cmidrule(lr){2-3}\cmidrule(lr){4-5}\cmidrule(lr){6-7}\cmidrule(lr){8-9}
                                             &\multicolumn{1}{c}{(1)}   &\multicolumn{1}{c}{(2)}   &\multicolumn{1}{c}{(3)}   &\multicolumn{1}{c}{(4)}   &\multicolumn{1}{c}{(5)}   &\multicolumn{1}{c}{(6)}   &\multicolumn{1}{c}{(7)}   &\multicolumn{1}{c}{(8)}   \\
\midrule
Author is female (1 = Yes)                   & 4.4000*** & 4.3287*** & -4.3791*** & -4.3315*** & 3.6482*** & 3.5716*** & -3.7085*** & -3.6403*** \\
                                             &       (0.0687)   &       (0.0688)   &       (0.0759)   &       (0.0760)   &       (0.0658)   &       (0.0658)   &       (0.0750)   &       (0.0752)   \\
Paper is cited 0 times (1 = Yes)             & -29.1462*** & -29.1109*** & -51.1809*** & -51.1884*** & -25.8153*** & -25.7966*** & -55.3740*** & -55.3530*** \\
                                             &       (0.0768)   &       (0.0771)   &       (0.0827)   &       (0.0832)   &       (0.0747)   &       (0.0751)   &       (0.0822)   &       (0.0827)   \\
Involved rate                                &  & 0.2454*** &  & -0.1841*** &  & 0.2882*** &  & -0.3035*** \\
                                             &                  &       (0.0186)   &                  &       (0.0202)   &                  &       (0.0180)   &                  &       (0.0200)   \\
Informational rate                           &  & -0.0810*** &  & 0.0275** &  & -0.0553*** &  & -0.0129 \\
                                             &                  &       (0.0093)   &                  &       (0.0104)   &                  &       (0.0089)   &                  &       (0.0102)   \\
Constant                                     & 26.9824*** & 27.1585*** & 52.9872*** & 53.3842*** & 23.5475*** & 23.1212*** & 57.4164*** & 58.9967*** \\
                                             &       (0.0684)   &       (0.1989)   &       (0.0783)   &       (0.2197)   &       (0.0658)   &       (0.1897)   &       (0.0772)   &       (0.2170)   \\\midrule
Field fixed effects                          &            \checkmark   &            \checkmark   &            \checkmark   &            \checkmark   &            \checkmark   &            \checkmark   &            \checkmark   &            \checkmark   \\
Year fixed effects                           &            \checkmark   &            \checkmark   &            \checkmark   &            \checkmark   &            \checkmark   &            \checkmark   &            \checkmark   &            \checkmark   \\
\midrule
$R^2$                                    &         0.3080   &         0.3084   &         0.4718   &         0.4720   &         0.2743   &         0.2748   &         0.5242   &         0.5244   \\
Observations                                            &         512,520   &         512,520   &         512,520   &         512,520   &         512,520   &         512,520   &         512,520   &         512,520   \\
\midrule Wald tests for linguistic predictors&                  &                  &                  &                  &                  &                  &                  &                  \\
F                                            &                  &       166.5310   &                  &        55.5259   &                  &       186.4656   &                  &       119.5908   \\
d.f.                                         &                  &         2.0000   &                  &         2.0000   &                  &         2.0000   &                  &         2.0000   \\
p-value                                      &                  &         0.0000   &                  &         0.0000   &                  &         0.0000   &                  &         0.0000   \\
 
\\
 
\bottomrule
 
\end{tabular}
 
\begin{tablenotes}
 
\item \emph{Notes:} Regression results for predicted outcomes if author is female, with the following dependent variables: rate of citations from papers with a female, male first author and with a female, male last author. These results represent data across all fields. Robust standard errors in parentheses; p-values are two-tailed. \checkmark\ indicates the fixed effect is included. 
\item +$p<0.10$; *$p<0.05$; **$p<0.01$; ***$p<0.001$
 
\end{tablenotes}
 
\end{threeparttable}
}
\end{table}
 
}%

\end{landscape}
 


\begin{landscape}

\thispagestyle{empty}

{

\scriptsize

\setlength{\tabcolsep}{2pt}

\renewcommand{\arraystretch}{1}

\clearpage

\begin{table}[tp]\centering

\scalebox{0.8}{%

\begin{threeparttable}
\def\sym#1{\ifmmode^{#1}\else\(^{#1}\)\fi}
\caption{Citations to sample patents as a function of inventor gender and writing style. These coefficients are visualized in Figure \ref{fig:citation_analysis}; standardized coefficients are reported in Supplementary Table \ref{table:RegressionsCitationsByGenderOfCitingInventorPatentsSeparateMeasuresV4Standardized}}
\label{table:RegressionsCitationsByGenderOfCitingInventorPatentsSeparateMeasuresV4Unstandardized}
\begin{tabular}{l*{8}{c}}
\toprule

 
                                             &\multicolumn{2}{c}{\shortstack{Citations from female first inventor}}&\multicolumn{2}{c}{\shortstack{Citations from male first inventor}}&\multicolumn{2}{c}{\shortstack{Citations from female last inventor}}&\multicolumn{2}{c}{\shortstack{Citations from male last inventor}}\\\cmidrule(lr){2-3}\cmidrule(lr){4-5}\cmidrule(lr){6-7}\cmidrule(lr){8-9}
                                             &\multicolumn{1}{c}{(1)}   &\multicolumn{1}{c}{(2)}   &\multicolumn{1}{c}{(3)}   &\multicolumn{1}{c}{(4)}   &\multicolumn{1}{c}{(5)}   &\multicolumn{1}{c}{(6)}   &\multicolumn{1}{c}{(7)}   &\multicolumn{1}{c}{(8)}   \\
\midrule
Inventor is female (1 = Yes) &         2.7496***&         2.7766***&        -3.0418***&        -3.0866***&         2.7250***&         2.7473***&        -3.1663***&        -3.1995***\\
                                             &       (0.1558)   &       (0.1571)   &       (0.3436)   &       (0.3450)   &       (0.1604)   &       (0.1613)   &       (0.3410)   &       (0.3420)   \\
Lawyer is female (1 = Yes)                   & 0.3591 & 0.3145 & 0.1925 & 0.2128 & 0.0685 & 0.0263 & 0.4741 & 0.4965 \\
                                             &       (0.3311)   &       (0.3326)   &       (0.6869)   &       (0.6873)   &       (0.3322)   &       (0.3333)   &       (0.6819)   &       (0.6826)   \\
Patent is cited 0 times (1 = Yes)            & -5.0332*** & -5.0214*** & -41.9853*** & -41.9863*** & -5.5293*** & -5.5177*** & -41.4186*** & -41.4218*** \\
                                             &       (0.1303)   &       (0.1301)   &       (0.2893)   &       (0.2896)   &       (0.1378)   &       (0.1377)   &       (0.2862)   &       (0.2864)   \\
Involved rate                                &  & 0.0163 &  & 0.1959+ &  & 0.0355 &  & 0.1151 \\
                                             &                  &       (0.0525)   &                  &       (0.1120)   &                  &       (0.0549)   &                  &       (0.1104)   \\
Informational rate                           &  & -0.0722*** &  & 0.0408 &  & -0.0675*** &  & 0.0409 \\
                                             &                  &       (0.0199)   &                  &       (0.0447)   &                  &       (0.0201)   &                  &       (0.0441)   \\
Constant                                     & 3.6513*** & 5.1323*** & 43.8262*** & 42.3851*** & 4.0874*** & 5.4122*** & 43.4562*** & 42.2496*** \\
                                             &       (0.1113)   &       (0.4785)   &       (0.3117)   &       (1.0902)   &       (0.1179)   &       (0.4981)   &       (0.3081)   &       (1.0783)   \\\midrule
Field fixed effects                          &            \checkmark   &            \checkmark   &            \checkmark   &            \checkmark   &            \checkmark   &            \checkmark   &            \checkmark   &            \checkmark   \\
Year fixed effects                           &            \checkmark   &            \checkmark   &            \checkmark   &            \checkmark   &            \checkmark   &            \checkmark   &            \checkmark   &            \checkmark   \\
\midrule
$R^2$                                    &         0.0585   &         0.0589   &         0.2962   &         0.2963   &         0.0572   &         0.0575   &         0.2964   &         0.2964   \\
Observations                                            &          32,106   &          32,106   &          32,106   &          32,106   &          32,106   &          32,106   &          32,106   &          32,106   \\
\midrule Wald tests for linguistic predictors&                  &                  &                  &                  &                  &                  &                  &                  \\
F                                            &                  &         6.7987   &                  &         1.7288   &                  &         6.3657   &                  &         0.8430   \\
d.f.                                         &                  &         2.0000   &                  &         2.0000   &                  &         2.0000   &                  &         2.0000   \\
p-value                                      &                  &         0.0011   &                  &         0.1775   &                  &         0.0017   &                  &         0.4304   \\
 
\\

\bottomrule

\end{tabular}

\begin{tablenotes}

\item \emph{Notes:} Regression results for predicted outcomes if inventor or patent lawyer is female, with the following dependent variables: rate of citations from patents with a female or male first inventor and with a female or male last inventor. Robust standard errors in parentheses; p-values are two-tailed. \checkmark\ indicates the fixed effect is included. 
\item +$p<0.10$; *$p<0.05$; **$p<0.01$; ***$p<0.001$

\end{tablenotes}

\end{threeparttable}
}
\end{table}

}%

\end{landscape}


\pagebreak
\renewcommand\thefigure{S\arabic{figure}}  
\renewcommand{\figurename}{Fig.}
\renewcommand{\thetable}{S\arabic{table}}
\setcounter{figure}{0}
\setcounter{table}{0}
\clearpage
\appendix

\setcounter{section}{0}
\section*{Supplementary Materials}

\subsection*{Index of tables and figures}
\begin{figure}[tp]
\centering
\includegraphics[width=0.85\textwidth]{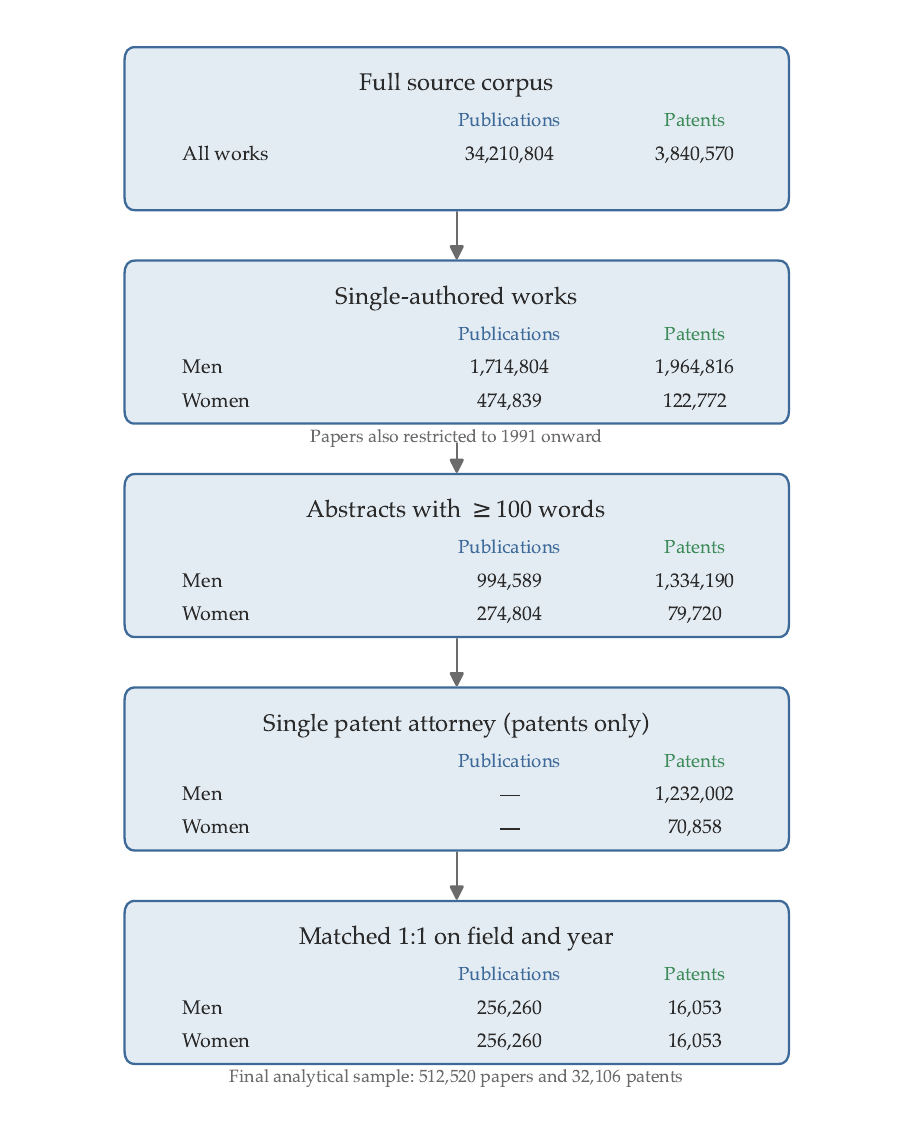}
\caption{\textbf{Construction of the matched analytical sample.} Inclusion criteria applied to the full Web of Science and PatentsView corpora, with counts shown separately for single-authored works by men and women at each stage. Papers are restricted to 1991 onward; patents are additionally restricted to those with a single named attorney. The final stage matches men's and women's works one-to-one on field and publication or grant year, yielding 256,260 papers and 16,053 patents per gender (512,520 papers and 32,106 patents in total). These paper counts reflect works with a complete field classification, as used in the field-fixed-effects models; the descriptive tables and no-fixed-effects models additionally retain 210 papers per gender that lack a field code, for a full matched paper sample of 256,470 per gender (512,940 total).}
\label{fig:Corpus_Size}
\end{figure}

\begin{landscape}
\begin{figure}[p]
\centering
\includegraphics[width=\textwidth]{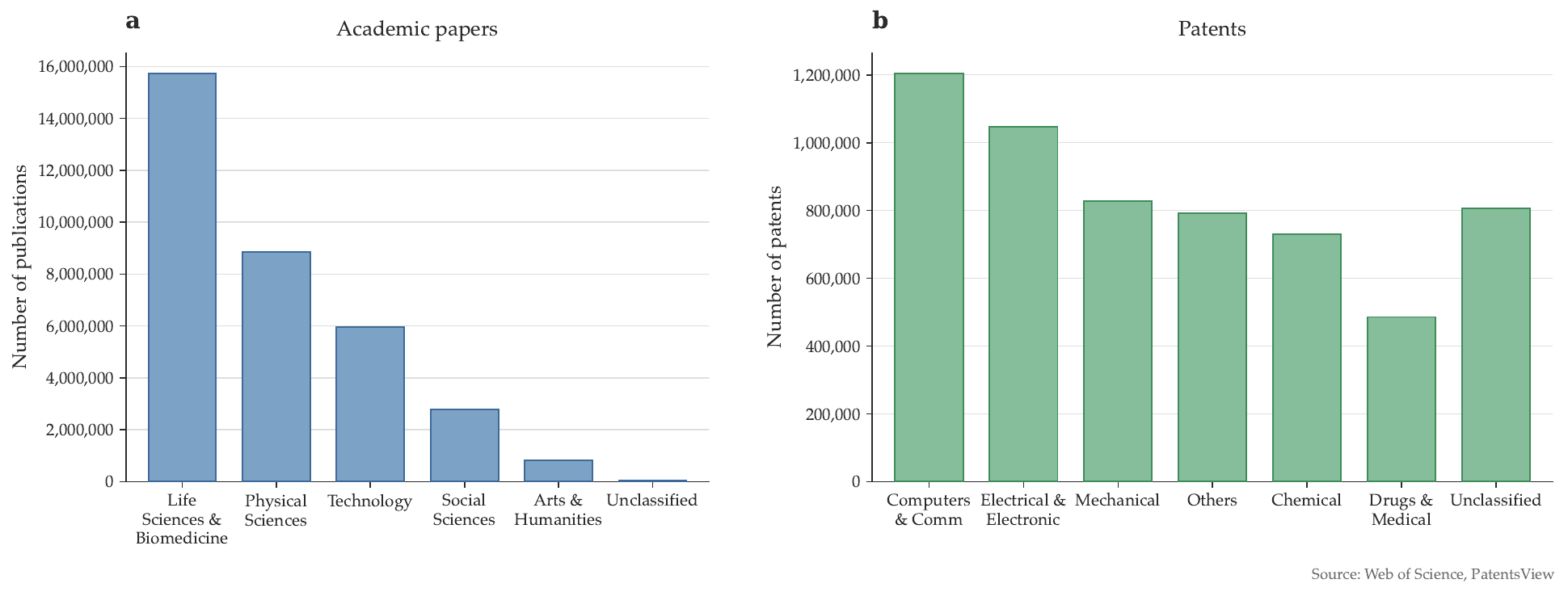}
\caption{\textbf{Field composition of the full Web of Science and PatentsView corpora.} Number of works in each field across the complete source corpora, prior to sampling. \textbf{(a)} Academic papers by Web of Science research area; \textbf{(b)} patents by NBER technology category. Categories are ordered by frequency. ``Unclassified'' denotes works with no assigned field (39,095 papers, 0.1\%; 806,053 patents, the latter including 51 records in the residual ``Not Classified'' code) and is shown last for completeness. For patents, ``Others'' is the standard NBER miscellaneous category and is distinct from ``Unclassified.''}
\label{fig:academic papers_all_field}
\end{figure}
\end{landscape}

\begingroup
\setlength{\parindent}{0pt}
\setlength{\parskip}{1.5pt}

{\bfseries Sample construction}\par\nobreak
\begingroup\leftskip=1.7em
\hangindent=1.8em Figure~\ref{fig:Corpus_Size}. Identification of the analytical sample.\par
\hangindent=1.8em Figure~\ref{fig:academic papers_all_field}. Field composition of the full Web of Science and PatentsView corpora.\par
\hangindent=1.8em Table~\ref{table_features}. Construction of the involved and informational writing-style features.\par
\endgroup
\addvspace{7pt}
{\bfseries Collaborative authorship}\par\nobreak
\begingroup\leftskip=1.7em
\hangindent=1.8em Table~\ref{tab:c_pap_m}. Writing style by author gender in collaboratively authored papers, matched sample.\par
\hangindent=1.8em Table~\ref{tab:c_pat_m}. Writing style by inventor gender in collaboratively invented patents, matched sample.\par
\endgroup
\addvspace{7pt}
{\bfseries Disciplinary differences}\par\nobreak
\begingroup\leftskip=1.7em
\hangindent=1.8em Table~\ref{tab:disc_pap}. Writing style by author gender and field in collaboratively authored papers.\par
\hangindent=1.8em Table~\ref{tab:disc_pat}. Writing style by inventor gender and field in collaboratively invented patents.\par
\endgroup
\addvspace{7pt}
{\bfseries Robustness to the matched-sample design}\par\nobreak
\begingroup\leftskip=1.7em
\hangindent=1.8em Table~\ref{tab:c_pap_f}. Writing style by author gender, full unmatched sample of papers.\par
\hangindent=1.8em Table~\ref{tab:c_pat_f}. Writing style by inventor gender, full unmatched sample of patents.\par
\endgroup
\addvspace{7pt}
{\bfseries Patent claims}\par\nobreak
\begingroup\leftskip=1.7em
\hangindent=1.8em Table~\ref{tab:claims_compare}. Abstract versus claims writing-style rates.\par
\hangindent=1.8em Table~\ref{tab:claims_reg}. Writing style in patent claims, by inventor and lawyer gender.\par
\endgroup
\addvspace{7pt}
{\bfseries Full-text analysis}\par\nobreak
\begingroup\leftskip=1.7em
\hangindent=1.8em Table~\ref{tab:ft_solo}. Writing style by author gender across the full text of single-authored papers.\par
\hangindent=1.8em Table~\ref{tab:ft_collab}. Writing style by first-author gender across the full text of collaboratively authored papers.\par
\hangindent=1.8em Table~\ref{tab:ft_collab_last}. Writing style by last-author gender across the full text of collaboratively authored papers.\par
\endgroup
\addvspace{7pt}
{\bfseries Gender differences by field, full regression results}\par\nobreak
\begingroup\leftskip=1.7em
\hangindent=1.8em Table~\ref{detailed_papers}. Gender differences in writing style by field, academic papers.\par
\hangindent=1.8em Table~\ref{detailed_patents}. Gender differences in writing style by field, patents.\par
\endgroup
\addvspace{7pt}
{\bfseries Citation regressions, standardized coefficients}\par\nobreak
\begingroup\leftskip=1.7em
\hangindent=1.8em Table~\ref{table:RegressionsCitationsByGenderOfCitingAuthorPapersSeparateMeasuresV4Standardized}. Citations to sample papers as a function of author gender and writing style, standardized coefficients.\par
\hangindent=1.8em Table~\ref{table:RegressionsCitationsByGenderOfCitingInventorPatentsSeparateMeasuresV4Standardized}. Citations to sample patents as a function of inventor gender and writing style, standardized coefficients.\par
\endgroup
\addvspace{7pt}
{\bfseries Citation patterns controlling for content similarity}\par\nobreak
\begingroup\leftskip=1.7em
\hangindent=1.8em Table~\ref{table:risk_papers_citations}. Involved and informational language predict the gender of citing authors, net of content similarity.\par
\hangindent=1.8em Table~\ref{table:risk_patents_citations}. Involved and informational language predict the gender of citing inventors, net of content similarity.\par
\hangindent=1.8em Table~\ref{table:risk_papers_citations_k50}. Involved and informational language predict the gender of citing authors, net of content similarity, using 50 nearest neighbors.\par
\hangindent=1.8em Table~\ref{table:risk_patents_citations_k50}. Involved and informational language predict the gender of citing inventors, net of content similarity, using 50 nearest neighbors.\par
\hangindent=1.8em Table~\ref{table:risk_papers_citations_k150}. Involved and informational language predict the gender of citing authors, net of content similarity, using 150 nearest neighbors.\par
\hangindent=1.8em Table~\ref{table:risk_patents_citations_k150}. Involved and informational language predict the gender of citing inventors, net of content similarity, using 150 nearest neighbors.\par
\endgroup
\endgroup

\clearpage

\newgeometry{margin=0.5in}
\newgeometry{margin=1in}

\clearpage
\begin{table}[tp]
\centering
\caption{Construction of the involved and informational writing-style features}\label{table_features}
\begin{adjustbox}{max width=\textwidth,center}
\begin{threeparttable}
\begin{tabular}{lllll}
\toprule
Tag & Source & Description & Feature & Dimension \\
\midrule
CD & spaCy & Cardinal number & Number & Informational \\
\midrule
PDT & spaCy & Predeterminer (e.g.\ both) & Determiner & Informational \\
WDT &  & Wh-determiner (e.g.\ which) &  &  \\
DT &  & Determiner &  &  \\
\midrule
VBD & spaCy & Verb, past tense & Past tense & Informational \\
VBN &  & Verb, past participle &  &  \\
\midrule
PRP & spaCy & Personal pronoun & Pronoun & Involved \\
PRP\$ &  & Possessive pronoun &  &  \\
WP &  & Wh-pronoun &  &  \\
WP\$ &  & Possessive wh-pronoun &  &  \\
\midrule
AND & string match & Connector ``and'' & Connector & Involved \\
\midrule
? & string match & Question mark & Question & Involved \\
\bottomrule
\end{tabular}
\begin{tablenotes}\footnotesize
\item Features are computed with the spaCy library in Python \citep{honnibal2017}. Part-of-speech tags follow the OntoNotes~5 version of the Penn Treebank tag set \citep{weischedel2013}.
\end{tablenotes}
\end{threeparttable}
\end{adjustbox}
\end{table}
\clearpage

\section{Gender coding}

In this appendix, we describe our procedure for coding the gender of paper authors and patent inventors and lawyers. For papers, we coded author gender based on author first and (when available) middle names, as reported in Web of Science, using the genderize.io application programming interface (API). While similar approaches have been widely used in prior work, we recognize that author names offer only an imperfect proxy for gender, and this limitation should be kept in mind when interpreting the results of our study. Given a name, the genderize.io API returns the most likely gender and an associated probability score. The genderize.io API makes gender suggestions based on worldwide data from hundreds of millions of social media profiles; as such, the service makes a significant improvement over prior approaches, which are typically based on records from the United States Social Security Administration or similar government entities, and that therefore perform poorly on international samples. For our study, we used a probability cutoff of 0.9 for assigning gender; gender suggestions below that level were set to ``unknown'' and excluded from the study. When possible, we assigned gender by looking across the academic papers of an author. For example, if the same author had published one paper under the name ``J. Smith'' and the other under the name ``John Smith'', we would assign the gender of the author as ``male'' in both cases, using the name from the latter paper to propagate a gender to the second.

For patents, we assign inventor gender by using the coding provided by the United States Patent and Trademark Office, via its Patents View database. Similar to our approach for papers, the Patents View database assigns gender based on inventor names. Details on the algorithm are given in \citep{toole2021}. Patents View does not assign gender to patent attorneys. Therefore, we assigned gender ourselves, using an approach analogous to the one described above for papers.

\section{Collaborative authorship}\label{sec:collab}
\subsection*{Data and methods}
We extend the writing-style analysis to collaboratively authored work. For academic papers we
draw on the same Web of Science corpus, retaining English-language abstracts of at least 100
tokens written by more than one author; for patents we use multi-inventor utility patents from
PatentsView under the same criteria. Author and inventor gender are coded from first names as
in the main analysis. Because a collaborative work has more than one contributor, we measure
gender at two positions---the first and the last author (inventor)---and analyze each
separately. We estimate the same models as in the main text, regressing the
Involved--Informational Ratio, the involved rate, and the informational rate on a first- (or
last-) author (inventor) female indicator with field and year fixed effects, on the
field-by-year matched, gender-balanced sample used in the main analysis
(Tables~\ref{tab:c_pap_m}--\ref{tab:c_pat_m}). The same models estimated on the full sample are
reported in Section~\ref{sec:full_samples} in the Supplementary Materials.

Because our interest here is the gender of the inventors, and because most collaboratively
invented patents are prosecuted by organizational attorneys (law firms) rather than a single
named individual---so that attorney gender cannot be assigned for the large majority of
multi-inventor patents---the collaborative patent models do not include lawyer gender. The
single-inventor analysis in the main text, where a single attorney can be identified, retains
the lawyer-gender control.

\subsection*{Results}
For academic papers, the gender differences documented in the main analysis hold among collaboratively authored work, at both author positions---women use more involved features, fewer informational features, and a higher Involved--Informational Ratio, all significant at $p<0.001$. The magnitudes are roughly half the single-authored estimates, consistent with an individual's writing style being partly obscured when several authors contribute to a text. For patents, the same pattern holds---female inventors, whether first or last, write with more involved features and a higher Involved--Informational Ratio. The one way in which the single-inventor analysis differs from the collaborative analyses (for both female first inventors and female last inventors) concerns the informational rate measure. In the single-inventor analysis, informational rate is positively associated with female inventors, whereas in the collaborative models it is negatively associated with female inventors. Notably, informational rate is also negatively associated with female lawyers in the single-inventor analysis.

\clearpage
\begin{table}[tp]
\centering
\caption{Writing style by author gender in collaboratively authored papers, matched sample}\label{tab:c_pap_m}
\begin{adjustbox}{max width=\textwidth,center}
\begin{threeparttable}
\begin{tabular}{lccc}
\toprule
 & Involved--Informational Ratio & Involved rate & Informational rate \\
\cmidrule(lr){2-2}\cmidrule(lr){3-3}\cmidrule(lr){4-4}
 & (1) & (2) & (3) \\
\midrule
\multicolumn{4}{l}{\textbf{Single-authored}} \\
Author is female (1 = Yes) & 0.0209*** & 0.2321*** & -0.1771*** \\
 & (0.0005) & (0.0054) & (0.0107) \\
$R^2$ & 0.1271 & 0.1692 & 0.0991 \\
Observations & 512,520 & 512,520 & 512,520 \\
\midrule
\multicolumn{4}{l}{\textbf{Collaborative, first author}} \\
First author is female (1 = Yes) & 0.0082*** & 0.1087*** & -0.1262*** \\
 & (0.0001) & (0.0015) & (0.0036) \\
$R^2$ & 0.0803 & 0.0601 & 0.0791 \\
Observations & 4,263,357 & 4,263,358 & 4,263,358 \\
\midrule
\multicolumn{4}{l}{\textbf{Collaborative, last author}} \\
Last author is female (1 = Yes) & 0.0108*** & 0.1386*** & -0.1091*** \\
 & (0.0002) & (0.0018) & (0.0043) \\
$R^2$ & 0.0851 & 0.0662 & 0.0802 \\
Observations & 2,900,705 & 2,900,706 & 2,900,706 \\
\midrule
Field fixed effects & \checkmark & \checkmark & \checkmark \\
Year fixed effects & \checkmark & \checkmark & \checkmark \\
\bottomrule
\end{tabular}
\begin{tablenotes}\footnotesize
\item The single-authored panel reproduces the corresponding estimates from the main analysis. Collaborative models use all multi-authored papers. Robust standard errors (in parentheses); columns numbered (1)--(3). \checkmark\ indicates the fixed effect is included. +$p<0.10$; *$p<0.05$; **$p<0.01$; ***$p<0.001$.
\end{tablenotes}
\end{threeparttable}
\end{adjustbox}
\end{table}
\clearpage
\begin{table}[tp]
\centering
\caption{Writing style by inventor gender in collaboratively invented patents, matched sample}\label{tab:c_pat_m}
\begin{adjustbox}{max width=\textwidth,center}
\begin{threeparttable}
\begin{tabular}{lccc}
\toprule
 & Involved--Informational Ratio & Involved rate & Informational rate \\
\cmidrule(lr){2-2}\cmidrule(lr){3-3}\cmidrule(lr){4-4}
 & (1) & (2) & (3) \\
\midrule
\multicolumn{4}{l}{\textbf{Single-inventor}} \\
Inventor is female (1 = Yes) & 0.0034** & 0.1436*** & 0.4086*** \\
 & (0.0013) & (0.0173) & (0.0434) \\
Lawyer is female (1 = Yes) & 0.0076** & 0.0240 & -0.6160*** \\
 & (0.0028) & (0.0356) & (0.0922) \\
$R^2$ & 0.0297 & 0.0219 & 0.0989 \\
Observations & 32,106 & 32,106 & 32,106 \\
\midrule
\multicolumn{4}{l}{\textbf{Collaborative, first inventor}} \\
First inventor is female (1 = Yes) & 0.0037*** & 0.0433*** & -0.1202*** \\
 & (0.0004) & (0.0061) & (0.0161) \\
$R^2$ & 0.0679 & 0.0175 & 0.1583 \\
Observations & 249,876 & 249,876 & 249,876 \\
\midrule
\multicolumn{4}{l}{\textbf{Collaborative, last inventor}} \\
Last inventor is female (1 = Yes) & 0.0045*** & 0.0413*** & -0.1778*** \\
 & (0.0004) & (0.0056) & (0.0149) \\
$R^2$ & 0.0758 & 0.0184 & 0.1703 \\
Observations & 293,540 & 293,540 & 293,540 \\
\midrule
Field fixed effects & \checkmark & \checkmark & \checkmark \\
Year fixed effects & \checkmark & \checkmark & \checkmark \\
\bottomrule
\end{tabular}
\begin{tablenotes}\footnotesize
\item The single-authored (single-inventor) panel reproduces the corresponding estimates from the main analysis. Collaborative patent models use all multi-inventor patents and do not include lawyer gender (see text). Robust standard errors (in parentheses); columns numbered (1)--(3). \checkmark\ indicates the fixed effect is included. +$p<0.10$; *$p<0.05$; **$p<0.01$; ***$p<0.001$.
\end{tablenotes}
\end{threeparttable}
\end{adjustbox}
\end{table}
\clearpage

\section{Disciplinary differences}
\subsection*{Data and methods}
The main analysis reports that the gender difference in writing style varies by field---larger
in the Social Sciences and Arts \& Humanities and smaller in the Physical Sciences (Figure~\ref{fig:Coefficients}; Tables~\ref{detailed_papers}). To test whether this disciplinary pattern is specific
to single-authored work, we re-estimate the writing-style models separately by field on the
collaborative matched sample, using the gender of the first and the last author (inventor) in
separate panels---by WoS research area for papers (Table~\ref{tab:disc_pap}) and by NBER category
for patents (Table~\ref{tab:disc_pat}), each with sub-field and year fixed effects. Fields are
listed in the same order as the single-authored by-field tables in the main analysis.

\subsection*{Results}
For academic papers, the disciplinary pattern is reproduced in collaboratively authored work.
When first author was female, the gender difference in the Involved--Informational Ratio and the involved rate is largest in the Social Sciences and Arts \& Humanities and smallest in the Physical Sciences, with Technology and the Life Sciences in between---the same ordering reported for single-authored work in the main analysis. For female last authors, the results are generally similar---the Involved--Informational Ratio and involved rate are larger in the Social Sciences and smaller in the Physical Sciences. However, the pattern for Arts \& Humanities is less pronounced, where the coefficient is not consistently larger than those observed in the Life Sciences \& Biomedicine or Technology. For patents, the Involved--Informational Ratio and involved rate are higher for female inventors across most NBER categories, at both the first- and last-inventor position, consistent with the weaker and more uniform pattern for patents noted in the main text.

\clearpage
\begin{table}[tp]
\centering
\caption{Writing style by author gender and field in collaboratively authored papers}\label{tab:disc_pap}
\begin{adjustbox}{max width=\textwidth,center}
\begin{threeparttable}
\begin{tabular}{lccc}
\toprule
 & Involved--Informational Ratio & Involved rate & Informational rate \\
\cmidrule(lr){2-2}\cmidrule(lr){3-3}\cmidrule(lr){4-4}
 & (1) & (2) & (3) \\
\midrule
\multicolumn{4}{l}{\textbf{First author}} \\
Physical Sciences (N~=~614,472) & 0.0028*** & 0.0811*** & 0.0912*** \\
 & (0.0003) & (0.0038) & (0.0094) \\
Technology (N~=~561,057) & 0.0056*** & 0.0936*** & -0.0266** \\
 & (0.0004) & (0.0040) & (0.0099) \\
Life Sciences \& Biomedicine (N~=~2,701,626) & 0.0087*** & 0.1045*** & -0.1808*** \\
 & (0.0002) & (0.0017) & (0.0045) \\
Arts \& Humanities (N~=~11,057) & 0.0140*** & 0.1554*** & -0.1787* \\
 & (0.0039) & (0.0399) & (0.0730) \\
Social Sciences (N~=~375,131) & 0.0143*** & 0.1735*** & -0.1662*** \\
 & (0.0006) & (0.0061) & (0.0122) \\
\midrule
\multicolumn{4}{l}{\textbf{Last author}} \\
Physical Sciences (N~=~396,510) & 0.0063*** & 0.1227*** & 0.0649*** \\
 & (0.0004) & (0.0048) & (0.0117) \\
Technology (N~=~379,670) & 0.0086*** & 0.1359*** & -0.0279* \\
 & (0.0004) & (0.0050) & (0.0121) \\
Life Sciences \& Biomedicine (N~=~1,775,191) & 0.0104*** & 0.1262*** & -0.1355*** \\
 & (0.0002) & (0.0022) & (0.0055) \\
Arts \& Humanities (N~=~10,063) & 0.0097* & 0.0899* & -0.0707 \\
 & (0.0042) & (0.0419) & (0.0762) \\
Social Sciences (N~=~339,262) & 0.0179*** & 0.1942*** & -0.2146*** \\
 & (0.0007) & (0.0064) & (0.0128) \\
\midrule
Sub-field fixed effects & \checkmark & \checkmark & \checkmark \\
Year fixed effects & \checkmark & \checkmark & \checkmark \\
\bottomrule
\end{tabular}
\begin{tablenotes}\footnotesize
\item Female coefficient for the first or last author, estimated separately by WoS research area on the collaborative matched sample with sub-field and year fixed effects. Robust standard errors (in parentheses); columns numbered (1)--(3). \checkmark\ indicates the fixed effect is included. +$p<0.10$; *$p<0.05$; **$p<0.01$; ***$p<0.001$.
\end{tablenotes}
\end{threeparttable}
\end{adjustbox}
\end{table}
\clearpage
\begin{table}[tp]
\centering
\caption{Writing style by inventor gender and field in collaboratively invented patents}\label{tab:disc_pat}
\begin{adjustbox}{max width=\textwidth,center}
\begin{threeparttable}
\begin{tabular}{lccc}
\toprule
 & Involved--Informational Ratio & Involved rate & Informational rate \\
\cmidrule(lr){2-2}\cmidrule(lr){3-3}\cmidrule(lr){4-4}
 & (1) & (2) & (3) \\
\midrule
\multicolumn{4}{l}{\textbf{First inventor}} \\
Computers \& Comm (N~=~81,008) & 0.0022** & 0.0417*** & 0.0162 \\
 & (0.0007) & (0.0105) & (0.0275) \\
Electrical \& Electronic (N~=~48,110) & 0.0036*** & 0.0766*** & 0.0310 \\
 & (0.0008) & (0.0135) & (0.0351) \\
Mechanical (N~=~24,950) & 0.0014 & 0.0258 & 0.0008 \\
 & (0.0010) & (0.0187) & (0.0482) \\
Drugs \& Medical (N~=~31,752) & 0.0061** & 0.0578** & -0.1842*** \\
 & (0.0020) & (0.0189) & (0.0499) \\
Chemical (N~=~36,588) & 0.0047*** & -0.0029 & -0.4243*** \\
 & (0.0013) & (0.0156) & (0.0448) \\
Others (N~=~27,466) & 0.0064*** & 0.0510** & -0.4178*** \\
 & (0.0011) & (0.0185) & (0.0463) \\
\midrule
\multicolumn{4}{l}{\textbf{Last inventor}} \\
Computers \& Comm (N~=~87,212) & 0.0021*** & 0.0386*** & 0.0261 \\
 & (0.0006) & (0.0101) & (0.0266) \\
Electrical \& Electronic (N~=~55,374) & 0.0034*** & 0.0602*** & -0.0401 \\
 & (0.0007) & (0.0125) & (0.0326) \\
Mechanical (N~=~29,420) & 0.0026** & 0.0271 & -0.1432** \\
 & (0.0009) & (0.0174) & (0.0445) \\
Drugs \& Medical (N~=~41,246) & 0.0073*** & 0.0326* & -0.2924*** \\
 & (0.0017) & (0.0165) & (0.0447) \\
Chemical (N~=~46,148) & 0.0077*** & 0.0301* & -0.5136*** \\
 & (0.0012) & (0.0140) & (0.0399) \\
Others (N~=~34,140) & 0.0061*** & 0.0559*** & -0.3594*** \\
 & (0.0010) & (0.0165) & (0.0416) \\
\midrule
Sub-field fixed effects & \checkmark & \checkmark & \checkmark \\
Year fixed effects & \checkmark & \checkmark & \checkmark \\
\bottomrule
\end{tabular}
\begin{tablenotes}\footnotesize
\item Female coefficient for the first or last inventor, estimated separately by NBER category on the collaborative matched sample with sub-field and year fixed effects. Robust standard errors (in parentheses); columns numbered (1)--(3). \checkmark\ indicates the fixed effect is included. +$p<0.10$; *$p<0.05$; **$p<0.01$; ***$p<0.001$.
\end{tablenotes}
\end{threeparttable}
\end{adjustbox}
\end{table}
\clearpage

\section{Robustness to the matched-sample design---the full samples}\label{sec:full_samples}
\subsection*{Data and methods}
The main analysis matches men's and women's work on field and year to balance composition. To
verify that the results do not depend on this matched, gender-balanced sample, we re-estimate
the same models on the \emph{full} samples of qualifying single- and multi-authored abstracts,
with field and year fixed effects (Tables~\ref{tab:c_pap_f}--\ref{tab:c_pat_f}). The full
samples are one to two orders of magnitude larger than the matched samples; the patent models
again use all multi-inventor patents, as described in Section~\ref{sec:collab}.

\subsection*{Results}
The results are robust to abandoning the matched design. For academic papers, every coefficient
retains its sign, significance, and approximate magnitude---among single- and collaboratively
authored work alike---so the result cannot be attributed to the gender-balanced matching. For
patents, the Involved--Informational Ratio and the involved rate remain positive and
significant for single- and multi-inventor work. The increase in the Involved--Informational Ratio associated with a lawyer being female in the matched sample is no longer statistically significant in the full-sample analysis of single-inventor patents. Another quantity that varies with the sample is the informational rate for single-inventor patents---positive in the matched sample, it is statistically indistinguishable from zero in the full sample. However, informational rate is negatively associated with female lawyers in the single-author analysis (as in our main analysis), and with female inventors in the collaborative first-author and collaborative last-author analyses. These negative associations generally indicate lower informational rates when a woman is involved in the writing process. This pattern is consistent with our hypotheses and aligns with the findings from the main analysis.

\clearpage
\begin{table}[tp]
\centering
\caption{Writing style by author gender, full unmatched sample of papers}\label{tab:c_pap_f}
\begin{adjustbox}{max width=\textwidth,center}
\begin{threeparttable}
\begin{tabular}{lccc}
\toprule
 & Involved--Informational Ratio & Involved rate & Informational rate \\
\cmidrule(lr){2-2}\cmidrule(lr){3-3}\cmidrule(lr){4-4}
 & (1) & (2) & (3) \\
\midrule
\multicolumn{4}{l}{\textbf{Single-authored}} \\
Author is female (1 = Yes) & 0.0262*** & 0.2762*** & -0.2992*** \\
 & (0.0006) & (0.0057) & (0.0112) \\
$R^2$ & 0.1033 & 0.1313 & 0.0785 \\
Observations & 672,319 & 672,319 & 672,319 \\
\midrule
\multicolumn{4}{l}{\textbf{Collaborative, first author}} \\
First author is female (1 = Yes) & 0.0080*** & 0.1070*** & -0.1214*** \\
 & (0.0001) & (0.0013) & (0.0031) \\
$R^2$ & 0.0870 & 0.0640 & 0.0826 \\
Observations & 6,659,006 & 6,659,009 & 6,659,009 \\
\midrule
\multicolumn{4}{l}{\textbf{Collaborative, last author}} \\
Last author is female (1 = Yes) & 0.0103*** & 0.1351*** & -0.0955*** \\
 & (0.0001) & (0.0015) & (0.0035) \\
$R^2$ & 0.0860 & 0.0636 & 0.0820 \\
Observations & 6,905,158 & 6,905,161 & 6,905,161 \\
\midrule
Field fixed effects & \checkmark & \checkmark & \checkmark \\
Year fixed effects & \checkmark & \checkmark & \checkmark \\
\bottomrule
\end{tabular}
\begin{tablenotes}\footnotesize
\item The single-authored panel re-estimates the analysis on the full, unmatched sample. Collaborative models use all multi-authored papers. Robust standard errors (in parentheses); columns numbered (1)--(3). \checkmark\ indicates the fixed effect is included. +$p<0.10$; *$p<0.05$; **$p<0.01$; ***$p<0.001$.
\end{tablenotes}
\end{threeparttable}
\end{adjustbox}
\end{table}
\clearpage
\begin{table}[tp]
\centering
\caption{Writing style by inventor gender, full unmatched sample of patents}\label{tab:c_pat_f}
\begin{adjustbox}{max width=\textwidth,center}
\begin{threeparttable}
\begin{tabular}{lccc}
\toprule
 & Involved--Informational Ratio & Involved rate & Informational rate \\
\cmidrule(lr){2-2}\cmidrule(lr){3-3}\cmidrule(lr){4-4}
 & (1) & (2) & (3) \\
\midrule
\multicolumn{4}{l}{\textbf{Single-inventor}} \\
Inventor is female (1 = Yes) & 0.0035*** & 0.0625*** & -0.0463 \\
 & (0.0009) & (0.0157) & (0.0377) \\
Lawyer is female (1 = Yes) & 0.0017 & -0.0327+ & -0.2965*** \\
 & (0.0012) & (0.0174) & (0.0458) \\
$R^2$ & 0.0290 & 0.0189 & 0.0840 \\
Observations & 206,420 & 206,420 & 206,420 \\
\midrule
\multicolumn{4}{l}{\textbf{Collaborative, first inventor}} \\
First inventor is female (1 = Yes) & 0.0031*** & 0.0397*** & -0.0974*** \\
 & (0.0003) & (0.0045) & (0.0118) \\
$R^2$ & 0.0563 & 0.0161 & 0.1248 \\
Observations & 1,812,298 & 1,812,299 & 1,812,299 \\
\midrule
\multicolumn{4}{l}{\textbf{Collaborative, last inventor}} \\
Last inventor is female (1 = Yes) & 0.0047*** & 0.0472*** & -0.1818*** \\
 & (0.0003) & (0.0042) & (0.0111) \\
$R^2$ & 0.0570 & 0.0162 & 0.1245 \\
Observations & 1,813,212 & 1,813,213 & 1,813,213 \\
\midrule
Field fixed effects & \checkmark & \checkmark & \checkmark \\
Year fixed effects & \checkmark & \checkmark & \checkmark \\
\bottomrule
\end{tabular}
\begin{tablenotes}\footnotesize
\item The single-authored (single-inventor) panel re-estimates the analysis on the full, unmatched sample. Collaborative patent models use all multi-inventor patents and do not include lawyer gender (see text). Robust standard errors (in parentheses); columns numbered (1)--(3). \checkmark\ indicates the fixed effect is included. +$p<0.10$; *$p<0.05$; **$p<0.01$; ***$p<0.001$.
\end{tablenotes}
\end{threeparttable}
\end{adjustbox}
\end{table}
\clearpage

\section{Patent claims}\label{sec:claims_supp}
\subsection*{Data and methods}
To test whether the writing-style differences extend to the most rigidly formatted patent text, we compared abstracts and claims directly. Using PatentsView, we assembled a matched sample of 7,616 patents for which both the abstract and the claims were available, and computed the involved and informational features on the \emph{independent} claim using the same procedure as in the main analysis. We then estimated the same models as for abstracts---the Involved--Informational Ratio, the involved rate, and the informational rate on inventor- and lawyer-female indicators, with grant-year and NBER-subcategory fixed effects and robust standard errors (Table~\ref{tab:claims_reg})---and paired comparisons of the abstract and claims rates within patent (Table~\ref{tab:claims_compare}).

\subsection*{Results}
Claims carry slightly more involved and informational language than abstracts (Table~\ref{tab:claims_compare}), but both rates rise by similar proportions, so the Involved--Informational Ratio is statistically indistinguishable between the two components. The gender pattern observed for abstracts is preserved in claims (Table~\ref{tab:claims_reg})---the predicted ratio remains greater for female than male inventors, and female inventors write at higher involved and informational rates. For patent lawyers, none of the three associations reach significance in the claims models, in contrast to the significant effects in abstracts---suggesting that the stylistic imprint of lawyers is more constrained by the rigid legal structure of claims.

\begin{table}[tp]\centering
\caption{Abstract versus claims writing-style rates}\label{tab:claims_compare}
\begin{threeparttable}\begin{tabular}{lccc}\toprule
 & Abstract & Claims & $t$ \\\midrule
Involved rate & 3.01 & 3.18 & -8.52*** \\
Informational rate & 21.48 & 21.84 & -8.09*** \\
Involved--Informational Ratio & 0.15 & 0.15 & 0.38 \\
\bottomrule\end{tabular}
\begin{tablenotes}\footnotesize\item Paired comparison over the matched sample of 7,616 patents; rates are per 100 tokens, and claims features are computed on the independent claim with the same parser used throughout the paper. Paired $t$-tests, abstract vs.\ claims, over the pooled sample. The Involved--Informational Ratio comparison is also statistically indistinguishable within inventor gender ($t=-0.26$, $p=.797$ for female; $t=0.81$, $p=.419$ for male inventors). +$p<0.10$; *$p<0.05$; **$p<0.01$; ***$p<0.001$.\end{tablenotes}
\end{threeparttable}\end{table}
\clearpage
\begin{table}[tp]\centering
\caption{Writing style in patent claims, by inventor and lawyer gender}\label{tab:claims_reg}
\begin{adjustbox}{max width=\textwidth,center}\begin{threeparttable}
\begin{tabular}{lccc}\toprule
 & Involved--Informational Ratio & Involved rate & Informational rate \\\midrule
Inventor is female (1 = Yes) & 0.0044** & 0.1153*** & 0.1777** \\
 & (0.0016) & (0.0303) & (0.0679) \\
Lawyer is female (1 = Yes) & -0.0003 & -0.0464 & -0.1857 \\
 & (0.0033) & (0.0619) & (0.1493) \\
\midrule Observations & 7,616 & 7,616 & 7,616 \\\bottomrule
\end{tabular}
\begin{tablenotes}\footnotesize\item Ordinary least squares regressions of each writing-style measure, computed on the independent claim, on inventor- and lawyer-female indicators, with grant-year and NBER-subcategory fixed effects, over the matched sample of 7,616 patents for which both abstract and claims text are available; a direct parallel to Table~\ref{table:coefficients_patents_rate}. Robust standard errors in parentheses. +$p<0.10$; *$p<0.05$; **$p<0.01$; ***$p<0.001$.\end{tablenotes}
\end{threeparttable}\end{adjustbox}\end{table}
\clearpage

\section{Full text}
\subsection*{Data and methods}
To examine whether these patterns extend beyond abstracts to the full text of papers, we matched our sample papers by DOI to the PubMed Central Open Access Subset, the largest corpus of full-text scientific articles available for systematic text analysis. For each matched article we retrieved the full text and parsed it into its constituent sections (introduction, methods, results, discussion). We computed the involved and informational features for the abstract, each section, and the full body using the same procedure as in the main analysis, and estimated the same models---the Involved--Informational Ratio, the involved rate, and the informational rate on the author-female indicator, with field and year fixed effects---separately for each section. Because open-access coverage is greatest for recent and biomedical work, the full-text corpus is drawn from the Life Sciences \& Biomedicine. Because this corpus is concentrated in a single broad field, there is little cross-field variation for which the matched, gender-balanced design of the main analysis would adjust; we therefore control for field and year with fixed effects rather than re-matching within this smaller corpus. We report single-authored papers (the population of the main analysis; Table~\ref{tab:ft_solo}) and collaboratively authored papers by first and last author (Tables~\ref{tab:ft_collab}--\ref{tab:ft_collab_last}).

\subsection*{Results}
The gender difference in writing style is not confined to abstracts. For single-authored
papers---the population of our main analysis---women use significantly more involved features
and a higher Involved--Informational Ratio in the full body of the paper, and in every section---the introduction, methods, results, and discussion (Table~\ref{tab:ft_solo}). The effect is
largest for single-authored work, as expected when the writing reflects a single individual,
and is present, if smaller, for collaboratively authored papers measured by either the first or
last author (Tables~\ref{tab:ft_collab}--\ref{tab:ft_collab_last}).

\section{Controlling for content similarity}
Papers and patents are often cited because their content relates to the work that cites them. While we agree that content similarity likely plays a significant role in citation decisions, we argue that linguistic features---specifically involved and informational rates---also influence citation patterns. Prior work in bibliometrics has similarly used content-based comparison sets and risk pools to account for topic similarity when examining citation dynamics and gendered citation behavior. For example, \citet{tekles2022} constructed a topic-based risk pool to examine gender homophily in citations, while \citet{ghiasi2018} used title and abstract similarity to match papers within journal issues and control for content similarity. To test whether writing style influences citations when controlling for content similarity, we conducted analyses incorporating a content-based risk set of potential citers. Specifically, we constructed a dataset that included both observed citations and up to ten content-matched non-citing papers and patents for each single-authored paper and patent in our analysis. For simplicity, we will refer to these single-authored papers and patents as focal works. The citing papers and patents that formed these counterfactual citations were “at risk” of citing the focal work because they were highly similar in content to it, but ultimately did not cite it. This design enabled us to test whether involved and informational rates predict citation outcomes beyond content similarity.

To identify the "risk set" of potential citers, we used the SBERT model (all-MiniLM-L6-v2) to generate 384-dimensional embeddings of each paper and patent abstract, capturing their semantic content \citep{reimers2019sbert}. To manage computational complexity, we limited the risk set candidates to papers and patents published in the same year and field as those that cited the focal works. We then used FAISS (Facebook AI Similarity Search), a library for efficient similarity search over high-dimensional vectors, to retrieve the top 100 nearest neighbors for each focal work based on cosine similarity between SBERT embeddings \citep{douze2024faiss}. FAISS ranks neighbors in descending cosine-similarity order. In the rare event of an exact similarity tie, ties are broken according to document ID order. Because SBERT embeddings are continuous 384-dimensional vectors, exact ties are exceedingly uncommon in practice and therefore unlikely to meaningfully affect the results.

From this set, we excluded papers published in the same year or earlier to ensure that all counterfactual citations could have plausibly cited the focal work. We also excluded papers and patents that actually cited the focal work. We then constructed counterfactual citations using the ten most semantically similar papers and patents. For focal works that received no observed citations, we also included ten counterfactual citations whenever possible. Not all of the focal works from the citation analyses in the main manuscript were included in these analyses, as the top 100 nearest neighbors did not always contain a risk set of citers published after the focal work’s publication year. Each focal work contributes a variable number of rows to the regression model, including up to ten counterfactual rows in addition to however many observed citations the work received. All rows enter the regression with equal weight; no explicit re-weighting procedure is applied. Dependence among rows associated with the same focal work is accounted for through clustered standard errors, where clustering is performed at the focal-work level. Although focal works with more observed citations contribute more objects to the analysis, this is intentional, as the regression is designed to reflect the empirical distribution of citations. As a result, 39,247 papers and 3 patents were excluded. Additionally, some focal works did not have ten associated counterfactual citations, as not all had ten risk-set citers among their 100 nearest neighbors that met our exclusion criteria (i.e., 23,398 papers and 790 patents). To test whether our findings were sensitive to the number of nearest neighbors selected, we reran the same analysis using 50 and 150 nearest neighbors separately and found similar results. See Tables \ref{table:risk_papers_citations_k50} and \ref{table:risk_patents_citations_k50} for analyses with 50 nearest neighbors and Tables \ref{table:risk_papers_citations_k150} and \ref{table:risk_patents_citations_k150} for analyses with 150 nearest neighbors.

To assess whether excluding focal papers with no valid counterfactuals biased our results, we compared excluded works to the works included in our analyses. We characterized the included and excluded works along three dimensions---author gender, linguistic feature means, and publication year. For publications, the gender distribution of excluded works was virtually identical to that of the included sample: 49.98\% female among excluded works versus 50.0\% among included works ($\chi^2 = 0.009$, $p = .93$, Cohen's $h = 0.001$). For patents, the difference was similarly negligible: 52.1\% of excluded works had female inventors versus 50.0\% among included works ($\chi^2 = 3.51$, $p = .06$, $h = 0.043$). When considering linguistic features, excluded publications had a modestly higher mean involved rate (5.08 vs.\ 4.67 among included works, $p < .001$, $g = 0.20$) and a slightly lower mean informational rate (15.63 vs.\ 16.04, $p < .001$, $g = 0.10$). For patents, involved rate means were 3.000 versus 3.012 ($p = .74$, $g = 0.008$), while informational rate means were 21.09 versus 21.40 ($p = .001$, $g = 0.077$). The dominant driver of exclusion was publication/grant year. Excluded publications were primarily from 2016 ($M = 2016.0$, $SD = 0.07$), compared to a mean publication year of 2008.0 ($SD = 6.76$) among included works ($p < .001$, $g = 1.28$). Excluded patents had a mean grant year of 2010.6, compared to 2000.0 among included works ($p < .001$, $g = 1.06$). The effect sizes of these year differences were large ($g > 1.0$ in both papers and patents). Although year, linguistic features, and gender measure are distinct constructs and are therefore not directly comparable, differences in gender composition and linguistic features were consistently small to negligible, whereas differences in publication year were substantial. Very recent focal works have fewer post-publication papers in the FAISS index that are eligible to serve as counterfactual citers (i.e., papers published after the focal work), making them less likely to obtain even a single valid counterfactual. Taken together, these results indicate that exclusion from the risk set is driven primarily by publication year rather than by gender or linguistic features.

We then analyzed gender differences in citation patterns. Using PanelOLS regressions, we modeled the likelihood that a citation came from a female or male first or last author or inventor. The dependent variable was coded as 1 if the citing work had a female (or male) first or last author or inventor and the citation was observed, and 0 otherwise. We included fixed effects for field and publication year. To account for variation in the number of observed and counterfactual citations---where each focal work is associated with a varying number of observations---we clustered standard errors by focal paper or patent. As shown in Table \ref{table:risk_papers_citations} (academic papers) and Table \ref{table:risk_patents_citations} (patents), many of the linguistic features remain significant predictors of citation behavior, even after controlling for content. For academic papers, the rate of involved language in abstracts was positively associated with citations from female first author (Table \ref{table:risk_papers_citations}; $\beta = 0.0015$, $p < 0.001$) and last author ($\beta = 0.0019$, $p < 0.001$). Conversely, the involved rate was negatively associated with citations from male first author ($\beta = -0.0045$, $p < 0.001$) and last author ($\beta = -0.0053$, $p < 0.001$). In contrast, the informational rate was negatively associated with citations from female first author ($\beta = -0.0031$, $p < 0.001$) and last author ($\beta = -0.0024$, $p < 0.001$). Unlike in the rate-based analysis reported in the main manuscript, this pattern did not reverse for papers with male first author. Instead, we find that informational rate is negatively associated with both male first author ($\beta = -0.0039$, $p < 0.001$) and last author ($\beta = -0.0050$, $p < 0.001$).

Turning to patents, the results generally mirrored those of academic papers but were weaker in magnitude (Table \ref{table:risk_patents_citations}). Informational rate was a significant predictor in all models and aligned with our hypothesized relationship between writing style and citation outcomes. Patents with a higher informational rate were less likely to be cited by patents with female first inventor ($\beta = -0.0005$, $p < 0.01$) and last inventor ($\beta = -0.0006$, $p < 0.001$). Conversely, patents with greater informational content were cited more frequently by patents with male first inventor ($\beta = 0.0012$, $p < 0.10$, marginally significant) and last inventor ($\beta = 0.0014$, $p < 0.05$). Involved language was generally not a significant predictor for patents.

Overall, these results closely align with the rate-based citation analyses presented in the main manuscript. One key difference is that the informational rate was negatively associated with citations from papers with male first and last authors, whereas it was positively associated with such citations in the main manuscript. This suggests that when content similarity is accounted for, high rates of informational language may make papers less appealing to both male and female citers. However, this was not the case for patents, where patents with male first or last inventor were more likely to cite patents with higher informational rates. Together, these findings underscore the nuanced role of linguistic style in shaping citation patterns and suggest that these effects may vary across document types. We advise against drawing strong conclusions from these findings without further consideration. This analysis does not include all focal papers from the original dataset, as some uncited papers did not yield counterfactual citations that met our criteria (i.e., none of the 100 most semantically similar papers were published after the focal paper). Additionally, focal papers were not equally represented, as some had more observed citations than others, and not all received the full set of ten counterfactuals. 

In general, the results indicate that even after controlling for content similarity, linguistic features remain meaningful predictors of gender-based citation dynamics. While largely consistent with our earlier findings, this analysis is best viewed as complementary to the rate-based models reported in the main manuscript.

\clearpage
\begin{table}[tp]
\centering
\caption{Writing style by author gender across the full text of single-authored papers}\label{tab:ft_solo}
\begin{adjustbox}{max width=\textwidth,center}
\begin{threeparttable}
\begin{tabular}{lccc}
\toprule
 & Involved--Informational Ratio & Involved rate & Informational rate \\
\cmidrule(lr){2-2}\cmidrule(lr){3-3}\cmidrule(lr){4-4}
 & (1) & (2) & (3) \\
\midrule
\multicolumn{4}{l}{\textbf{Single-authored}} \\
Abstract & 0.0300*** & 0.2669*** & -0.4250*** \\
 & (0.0045) & (0.0397) & (0.0821) \\
$R^2$ & 0.0720 & 0.0917 & 0.1183 \\
Observations & 12,164 & 12,164 & 12,164 \\
Introduction & 0.0221*** & 0.1976*** & -0.4951*** \\
 & (0.0032) & (0.0375) & (0.0775) \\
$R^2$ & 0.1769 & 0.2068 & 0.1180 \\
Observations & 7,418 & 7,418 & 7,418 \\
Methods & 0.0109*** & 0.2725*** & 0.3122* \\
 & (0.0031) & (0.0448) & (0.1307) \\
$R^2$ & 0.3138 & 0.3766 & 0.1409 \\
Observations & 3,719 & 3,719 & 3,719 \\
Results & 0.0233*** & 0.3615*** & -0.0691 \\
 & (0.0050) & (0.0712) & (0.1423) \\
$R^2$ & 0.1806 & 0.1635 & 0.1697 \\
Observations & 3,246 & 3,246 & 3,246 \\
Discussion & 0.0306*** & 0.2911*** & -0.4925*** \\
 & (0.0054) & (0.0478) & (0.0887) \\
$R^2$ & 0.1001 & 0.1314 & 0.1564 \\
Observations & 6,233 & 6,233 & 6,233 \\
Full body & 0.0195*** & 0.2629*** & -0.2403*** \\
 & (0.0023) & (0.0304) & (0.0645) \\
$R^2$ & 0.3217 & 0.3437 & 0.1793 \\
Observations & 8,257 & 8,257 & 8,257 \\
\bottomrule
\end{tabular}
\begin{tablenotes}\footnotesize
\item Coefficient on the author-female indicator, by section of the full text, full sample with field and year fixed effects. Robust standard errors (in parentheses); columns numbered (1)--(3). +$p<0.10$; *$p<0.05$; **$p<0.01$; ***$p<0.001$.
\end{tablenotes}
\end{threeparttable}
\end{adjustbox}
\end{table}
\clearpage
\begin{table}[tp]
\centering
\caption{Writing style by first-author gender across the full text of collaboratively authored papers}\label{tab:ft_collab}
\begin{adjustbox}{max width=\textwidth,center}
\begin{threeparttable}
\begin{tabular}{lccc}
\toprule
 & Involved--Informational Ratio & Involved rate & Informational rate \\
\cmidrule(lr){2-2}\cmidrule(lr){3-3}\cmidrule(lr){4-4}
 & (1) & (2) & (3) \\
\midrule
\multicolumn{4}{l}{\textbf{Collaborative, first author}} \\
Abstract & 0.0108*** & 0.0996*** & -0.2404*** \\
 & (0.0025) & (0.0256) & (0.0624) \\
$R^2$ & 0.0645 & 0.0493 & 0.0718 \\
Observations & 13,299 & 13,299 & 13,299 \\
Introduction & 0.0085*** & 0.0965*** & -0.1178* \\
 & (0.0019) & (0.0232) & (0.0550) \\
$R^2$ & 0.0387 & 0.0229 & 0.1003 \\
Observations & 8,868 & 8,868 & 8,868 \\
Methods & 0.0068*** & 0.1051*** & -0.2186*** \\
 & (0.0011) & (0.0178) & (0.0478) \\
$R^2$ & 0.2079 & 0.2129 & 0.2233 \\
Observations & 10,829 & 10,829 & 10,829 \\
Results & 0.0039** & 0.0512* & -0.1815** \\
 & (0.0013) & (0.0200) & (0.0629) \\
$R^2$ & 0.0689 & 0.0551 & 0.0796 \\
Observations & 10,652 & 10,652 & 10,652 \\
Discussion & 0.0086*** & 0.0824*** & -0.1820*** \\
 & (0.0020) & (0.0203) & (0.0503) \\
$R^2$ & 0.0424 & 0.0313 & 0.0777 \\
Observations & 10,623 & 10,623 & 10,623 \\
Full body & 0.0065*** & 0.0873*** & -0.1981*** \\
 & (0.0009) & (0.0132) & (0.0385) \\
$R^2$ & 0.1743 & 0.1367 & 0.1908 \\
Observations & 12,318 & 12,318 & 12,318 \\
\bottomrule
\end{tabular}
\begin{tablenotes}\footnotesize
\item Coefficient on the author-female indicator, by section of the full text, full sample with field and year fixed effects. Robust standard errors (in parentheses); columns numbered (1)--(3). +$p<0.10$; *$p<0.05$; **$p<0.01$; ***$p<0.001$.
\end{tablenotes}
\end{threeparttable}
\end{adjustbox}
\end{table}
\clearpage
\begin{table}[tp]
\centering
\caption{Writing style by last-author gender across the full text of collaboratively authored papers}\label{tab:ft_collab_last}
\begin{adjustbox}{max width=\textwidth,center}
\begin{threeparttable}
\begin{tabular}{lccc}
\toprule
 & Involved--Informational Ratio & Involved rate & Informational rate \\
\cmidrule(lr){2-2}\cmidrule(lr){3-3}\cmidrule(lr){4-4}
 & (1) & (2) & (3) \\
\midrule
\multicolumn{4}{l}{\textbf{Collaborative, last author}} \\
Abstract & 0.0106*** & 0.0783** & -0.2999*** \\
 & (0.0029) & (0.0298) & (0.0727) \\
$R^2$ & 0.0638 & 0.0487 & 0.0717 \\
Observations & 12,045 & 12,045 & 12,045 \\
Introduction & 0.0106*** & 0.0904** & -0.2732*** \\
 & (0.0024) & (0.0277) & (0.0645) \\
$R^2$ & 0.0421 & 0.0224 & 0.1030 \\
Observations & 7,995 & 7,995 & 7,995 \\
Methods & 0.0091*** & 0.1448*** & -0.2620*** \\
 & (0.0013) & (0.0209) & (0.0544) \\
$R^2$ & 0.2320 & 0.2173 & 0.2331 \\
Observations & 9,823 & 9,823 & 9,823 \\
Results & 0.0040** & 0.0538* & -0.1871* \\
 & (0.0015) & (0.0238) & (0.0729) \\
$R^2$ & 0.0696 & 0.0548 & 0.0811 \\
Observations & 9,668 & 9,668 & 9,668 \\
Discussion & 0.0114*** & 0.1363*** & -0.2111*** \\
 & (0.0022) & (0.0234) & (0.0573) \\
$R^2$ & 0.0437 & 0.0346 & 0.0791 \\
Observations & 9,626 & 9,626 & 9,626 \\
Full body & 0.0090*** & 0.1134*** & -0.2781*** \\
 & (0.0011) & (0.0159) & (0.0446) \\
$R^2$ & 0.1789 & 0.1398 & 0.1970 \\
Observations & 11,166 & 11,166 & 11,166 \\
\bottomrule
\end{tabular}
\begin{tablenotes}\footnotesize
\item Coefficient on the author-female indicator, by section of the full text, full sample with field and year fixed effects. Robust standard errors (in parentheses); columns numbered (1)--(3). +$p<0.10$; *$p<0.05$; **$p<0.01$; ***$p<0.001$.
\end{tablenotes}
\end{threeparttable}
\end{adjustbox}
\end{table}
\clearpage

\clearpage
\begin{table}[tp]
\centering
\caption{Gender differences in writing style by field, academic papers}\label{detailed_papers}
\begin{adjustbox}{max width=\textwidth,max totalheight=0.92\textheight,center}
\begin{threeparttable}
\begin{tabular}{lcccccc}
\toprule
 & \multicolumn{2}{c}{Involved--Informational Ratio} & \multicolumn{2}{c}{Involved rate} & \multicolumn{2}{c}{Informational rate} \\
\cmidrule(lr){2-3}\cmidrule(lr){4-5}\cmidrule(lr){6-7}
 & (1) & (2) & (3) & (4) & (5) & (6) \\
\midrule
\multicolumn{7}{l}{\textbf{Physical Sciences}} \\
Author is female (1 = Yes) & 0.0049*** & 0.0046*** & 0.1170*** & 0.1136*** & 0.2085*** & 0.2145*** \\
& (0.0012) & (0.0011) & (0.0136) & (0.0134) & (0.0311) & (0.0303) \\
Constant & 0.2314*** & 0.1668*** & 3.6371*** & 3.2202*** & 17.1514*** & 19.0286*** \\
& (0.0008) & (0.0279) & (0.0096) & (0.4059) & (0.0218) & (0.7060) \\
$R^2$ & 0.0003 & 0.0544 & 0.0012 & 0.0484 & 0.0007 & 0.0576 \\
Observations & 62,358 & 62,358 & 62,358 & 62,358 & 62,358 & 62,358 \\
\midrule
\multicolumn{7}{l}{\textbf{Technology}} \\
Author is female (1 = Yes) & 0.0096*** & 0.0101*** & 0.1404*** & 0.1444*** & 0.0162 & 0.0018 \\
& (0.0016) & (0.0015) & (0.0179) & (0.0169) & (0.0384) & (0.0371) \\
Constant & 0.2491*** & 0.1602*** & 3.8901*** & 2.8286*** & 17.3202*** & 19.2278*** \\
& (0.0011) & (0.0112) & (0.0126) & (0.1319) & (0.0269) & (0.3529) \\
$R^2$ & 0.0008 & 0.1160 & 0.0014 & 0.1176 & 0.0000 & 0.0811 \\
Observations & 43,675 & 43,675 & 43,675 & 43,675 & 43,675 & 43,675 \\
\midrule
\multicolumn{7}{l}{\textbf{Life Sciences \& Biomedicine}} \\
Author is female (1 = Yes) & 0.0185*** & 0.0186*** & 0.2046*** & 0.2052*** & -0.1583*** & -0.1571*** \\
& (0.0010) & (0.0010) & (0.0096) & (0.0090) & (0.0220) & (0.0211) \\
Constant & 0.2893*** & 0.1160*** & 4.1138*** & 2.2907*** & 15.9629*** & 19.3331*** \\
& (0.0007) & (0.0072) & (0.0067) & (0.0984) & (0.0156) & (0.2703) \\
$R^2$ & 0.0023 & 0.1007 & 0.0030 & 0.1230 & 0.0003 & 0.0892 \\
Observations & 149,541 & 149,541 & 149,541 & 149,541 & 149,541 & 149,541 \\
\midrule
\multicolumn{7}{l}{\textbf{Arts \& Humanities}} \\
Author is female (1 = Yes) & 0.0275*** & 0.0276*** & 0.2787*** & 0.2779*** & -0.3846*** & -0.3874*** \\
& (0.0018) & (0.0017) & (0.0177) & (0.0176) & (0.0302) & (0.0285) \\
Constant & 0.3860*** & 0.3551*** & 5.7840*** & 4.3604*** & 16.4391*** & 14.6273*** \\
& (0.0012) & (0.0624) & (0.0125) & (0.6235) & (0.0216) & (1.5462) \\
$R^2$ & 0.0037 & 0.0556 & 0.0038 & 0.0232 & 0.0025 & 0.1148 \\
Observations & 64,317 & 64,317 & 64,317 & 64,317 & 64,317 & 64,317 \\
\midrule
\multicolumn{7}{l}{\textbf{Social Sciences}} \\
Author is female (1 = Yes) & 0.0283*** & 0.0282*** & 0.2959*** & 0.2949*** & -0.2911*** & -0.2899*** \\
& (0.0010) & (0.0010) & (0.0097) & (0.0095) & (0.0173) & (0.0169) \\
Constant & 0.3482*** & 0.2189*** & 4.9205*** & 3.2420*** & 15.5635*** & 16.0783*** \\
& (0.0007) & (0.0342) & (0.0068) & (0.3761) & (0.0123) & (1.0835) \\
$R^2$ & 0.0042 & 0.0453 & 0.0048 & 0.0587 & 0.0015 & 0.0499 \\
Observations & 192,629 & 192,629 & 192,629 & 192,629 & 192,629 & 192,629 \\
\bottomrule
\end{tabular}
\begin{tablenotes}\footnotesize
\item Each field's models replicate Table~\ref{table:coefficients_academic papers_rate}, estimated separately by Web of Science research area. Within each field, columns (1), (3), and (5) include no fixed effects; columns (2), (4), and (6) add field and year fixed effects. Robust standard errors in parentheses. +$p<0.10$; *$p<0.05$; **$p<0.01$; ***$p<0.001$.
\end{tablenotes}
\end{threeparttable}
\end{adjustbox}
\end{table}
\clearpage

\pagebreak

\clearpage
\begin{table}[tp]
\centering
\caption{Gender differences in writing style by field, patents}\label{detailed_patents}
\begin{adjustbox}{max width=\textwidth,max totalheight=0.92\textheight,center}
\begin{threeparttable}
\begin{tabular}{lcccccc}
\toprule
 & \multicolumn{2}{c}{Involved--Informational Ratio} & \multicolumn{2}{c}{Involved rate} & \multicolumn{2}{c}{Informational rate} \\
\cmidrule(lr){2-3}\cmidrule(lr){4-5}\cmidrule(lr){6-7}
 & (1) & (2) & (3) & (4) & (5) & (6) \\
\midrule
\multicolumn{7}{l}{\textbf{Computers \& Comm}} \\
Inventor is female (1 = Yes) & 0.0004 & 0.0004 & 0.0578 & 0.0578 & 0.3038** & 0.3034** \\
& (0.0026) & (0.0026) & (0.0412) & (0.0411) & (0.1064) & (0.1054) \\
Lawyer is female (1 = Yes) & 0.0102+ & 0.0093+ & 0.0373 & 0.0356 & -0.7366*** & -0.6249** \\
& (0.0056) & (0.0055) & (0.0796) & (0.0793) & (0.2230) & (0.2208) \\
Constant & 0.1363*** & 0.1314*** & 2.6626*** & 2.6891*** & 20.9118*** & 20.9487*** \\
& (0.0019) & (0.0153) & (0.0293) & (0.2859) & (0.0764) & (0.6317) \\
$R^2$ & 0.0007 & 0.0156 & 0.0004 & 0.0121 & 0.0037 & 0.0311 \\
Observations & 5,580 & 5,580 & 5,580 & 5,580 & 5,580 & 5,580 \\
\midrule
\multicolumn{7}{l}{\textbf{Electrical \& Electronic}} \\
Inventor is female (1 = Yes) & 0.0076** & 0.0076** & 0.2715*** & 0.2716*** & 0.8669*** & 0.8668*** \\
& (0.0026) & (0.0026) & (0.0436) & (0.0434) & (0.1176) & (0.1156) \\
Lawyer is female (1 = Yes) & 0.0053 & 0.0048 & 0.0183 & 0.0034 & -0.6147* & -0.5997* \\
& (0.0058) & (0.0058) & (0.0956) & (0.0951) & (0.2460) & (0.2457) \\
Constant & 0.1340*** & 0.1366*** & 2.7003*** & 2.9902*** & 21.2449*** & 22.2131*** \\
& (0.0018) & (0.0101) & (0.0302) & (0.1956) & (0.0858) & (0.4293) \\
$R^2$ & 0.0021 & 0.0186 & 0.0083 & 0.0235 & 0.0128 & 0.0558 \\
Observations & 4,632 & 4,632 & 4,632 & 4,632 & 4,632 & 4,632 \\
\midrule
\multicolumn{7}{l}{\textbf{Mechanical}} \\
Inventor is female (1 = Yes) & -0.0012 & -0.0012 & 0.1130* & 0.1129* & 0.9326*** & 0.9312*** \\
& (0.0025) & (0.0025) & (0.0456) & (0.0454) & (0.1150) & (0.1133) \\
Lawyer is female (1 = Yes) & 0.0104+ & 0.0100 & 0.0477 & 0.0711 & -1.1050*** & -0.8692** \\
& (0.0060) & (0.0061) & (0.1066) & (0.1080) & (0.2791) & (0.2773) \\
Constant & 0.1445*** & 0.1516*** & 2.9765*** & 3.2065*** & 21.5424*** & 22.2138*** \\
& (0.0018) & (0.0107) & (0.0325) & (0.1994) & (0.0836) & (0.3680) \\
$R^2$ & 0.0008 & 0.0129 & 0.0014 & 0.0195 & 0.0180 & 0.0552 \\
Observations & 4,536 & 4,536 & 4,536 & 4,536 & 4,536 & 4,536 \\
\midrule
\multicolumn{7}{l}{\textbf{Drugs \& Medical}} \\
Inventor is female (1 = Yes) & 0.0034 & 0.0040 & 0.1690** & 0.1702** & 0.0993 & 0.0615 \\
& (0.0069) & (0.0068) & (0.0587) & (0.0583) & (0.1653) & (0.1466) \\
Lawyer is female (1 = Yes) & 0.0333** & 0.0047 & 0.1021 & 0.0420 & -2.0064*** & -0.2036 \\
& (0.0106) & (0.0113) & (0.0982) & (0.0984) & (0.2805) & (0.2719) \\
Constant & 0.1689*** & 0.2024*** & 3.0063*** & 3.2853*** & 20.3865*** & 18.1602*** \\
& (0.0062) & (0.0262) & (0.0416) & (0.5460) & (0.1189) & (0.8544) \\
$R^2$ & 0.0030 & 0.0405 & 0.0031 & 0.0332 & 0.0182 & 0.2395 \\
Observations & 3,056 & 3,056 & 3,056 & 3,056 & 3,056 & 3,056 \\
\midrule
\multicolumn{7}{l}{\textbf{Chemical}} \\
Inventor is female (1 = Yes) & -0.0004 & -0.0004 & 0.0743 & 0.0739 & -0.0344 & -0.0371 \\
& (0.0070) & (0.0069) & (0.0607) & (0.0607) & (0.1777) & (0.1659) \\
Lawyer is female (1 = Yes) & 0.0125 & 0.0101 & -0.0121 & 0.0143 & -1.4362*** & -1.2652*** \\
& (0.0088) & (0.0097) & (0.0998) & (0.1004) & (0.2872) & (0.2792) \\
Constant & 0.1725*** & 0.2024*** & 3.0471*** & 2.8248*** & 19.8252*** & 16.3100*** \\
& (0.0066) & (0.0294) & (0.0446) & (0.3505) & (0.1262) & (0.9924) \\
$R^2$ & 0.0004 & 0.0464 & 0.0006 & 0.0185 & 0.0092 & 0.1501 \\
Observations & 2,524 & 2,524 & 2,524 & 2,524 & 2,524 & 2,524 \\
\midrule
\multicolumn{7}{l}{\textbf{Others}} \\
Inventor is female (1 = Yes) & 0.0056*** & 0.0056*** & 0.1542*** & 0.1537*** & 0.2651*** & 0.2603*** \\
& (0.0017) & (0.0017) & (0.0292) & (0.0292) & (0.0700) & (0.0691) \\
Lawyer is female (1 = Yes) & 0.0084+ & 0.0079+ & -0.0074 & 0.0198 & -0.7232*** & -0.4835** \\
& (0.0045) & (0.0046) & (0.0649) & (0.0655) & (0.1569) & (0.1569) \\
Constant & 0.1495*** & 0.1559*** & 3.1071*** & 3.1456*** & 21.8632*** & 21.8371*** \\
& (0.0012) & (0.0091) & (0.0203) & (0.1543) & (0.0504) & (0.3421) \\
$R^2$ & 0.0014 & 0.0065 & 0.0024 & 0.0087 & 0.0031 & 0.0312 \\
Observations & 11,778 & 11,778 & 11,778 & 11,778 & 11,778 & 11,778 \\
\bottomrule
\end{tabular}
\begin{tablenotes}\footnotesize
\item Each category's models replicate Table~\ref{table:coefficients_patents_rate}, estimated separately by NBER patent category. Within each category, columns (1), (3), and (5) include no fixed effects; columns (2), (4), and (6) add field and year fixed effects. Robust standard errors in parentheses. +$p<0.10$; *$p<0.05$; **$p<0.01$; ***$p<0.001$.
\end{tablenotes}
\end{threeparttable}
\end{adjustbox}
\end{table}
\clearpage


 
\begin{landscape}
 
\thispagestyle{empty}

{
 
\scriptsize
 
\setlength{\tabcolsep}{2pt}
 
\renewcommand{\arraystretch}{1}
 
\begin{table}[tp]\centering
 
\scalebox{0.8}{%
 
\begin{threeparttable}
\def\sym#1{\ifmmode^{#1}\else\(^{#1}\)\fi}
\caption{Citations to sample papers as a function of author gender and writing style, standardized coefficients}
\label{table:RegressionsCitationsByGenderOfCitingAuthorPapersSeparateMeasuresV4Standardized}
\begin{tabular}{l*{8}{c}}
\toprule
 
 
                                             &\multicolumn{2}{c}{\shortstack{Citations from female first author}}&\multicolumn{2}{c}{\shortstack{Citations from male first author}}&\multicolumn{2}{c}{\shortstack{Citations from female last author}}&\multicolumn{2}{c}{\shortstack{Citations from male last author}}\\\cmidrule(lr){2-3}\cmidrule(lr){4-5}\cmidrule(lr){6-7}\cmidrule(lr){8-9}
                                             &\multicolumn{1}{c}{(1)}   &\multicolumn{1}{c}{(2)}   &\multicolumn{1}{c}{(3)}   &\multicolumn{1}{c}{(4)}   &\multicolumn{1}{c}{(5)}   &\multicolumn{1}{c}{(6)}   &\multicolumn{1}{c}{(7)}   &\multicolumn{1}{c}{(8)}   \\
\midrule
Author is female (1 = Yes)                   & 2.2000*** & 2.1644*** & -2.1896*** & -2.1658*** & 1.8241*** & 1.7858*** & -1.8543*** & -1.8202*** \\
                                             &       (0.0344)   &       (0.0344)   &       (0.0380)   &       (0.0380)   &       (0.0329)   &       (0.0329)   &       (0.0375)   &       (0.0376)   \\
Paper is cited 0 times (1 = Yes)             & -14.3178*** & -14.3005*** & -25.1422*** & -25.1459*** & -12.6816*** & -12.6724*** & -27.2020*** & -27.1917*** \\
                                             &       (0.0377)   &       (0.0379)   &       (0.0406)   &       (0.0409)   &       (0.0367)   &       (0.0369)   &       (0.0404)   &       (0.0406)   \\
Involved rate                                &  & 0.5196*** &  & -0.3898*** &  & 0.6102*** &  & -0.6426*** \\
                                             &                  &       (0.0394)   &                  &       (0.0427)   &                  &       (0.0382)   &                  &       (0.0423)   \\
Informational rate                           &  & -0.3266*** &  & 0.1108** &  & -0.2230*** &  & -0.0520 \\
                                             &                  &       (0.0376)   &                  &       (0.0418)   &                  &       (0.0357)   &                  &       (0.0412)   \\
Constant                                     & 17.3251*** & 17.3250*** & 29.9761*** & 29.9762*** & 14.8693*** & 14.8692*** & 33.0348*** & 33.0349*** \\
                                             &       (0.0344)   &       (0.0343)   &       (0.0379)   &       (0.0379)   &       (0.0329)   &       (0.0329)   &       (0.0375)   &       (0.0375)   \\\midrule
Field fixed effects                          &            \checkmark   &            \checkmark   &            \checkmark   &            \checkmark   &            \checkmark   &            \checkmark   &            \checkmark   &            \checkmark   \\
Year fixed effects                           &            \checkmark   &            \checkmark   &            \checkmark   &            \checkmark   &            \checkmark   &            \checkmark   &            \checkmark   &            \checkmark   \\
\midrule
$R^2$                                    &         0.3080   &         0.3084   &         0.4718   &         0.4720   &         0.2743   &         0.2748   &         0.5242   &         0.5244   \\
Observations                                            &         512,520   &         512,520   &         512,520   &         512,520   &         512,520   &         512,520   &         512,520   &         512,520   \\
\midrule Wald tests for linguistic predictors&                  &                  &                  &                  &                  &                  &                  &                  \\
F                                            &                  &       166.5310   &                  &        55.5259   &                  &       186.4656   &                  &       119.5908   \\
d.f.                                         &                  &         2.0000   &                  &         2.0000   &                  &         2.0000   &                  &         2.0000   \\
p-value                                      &                  &         0.0000   &                  &         0.0000   &                  &         0.0000   &                  &         0.0000   \\
 
\\
 
\bottomrule
 
\end{tabular}
 
\begin{tablenotes}
 
\item \emph{Notes:} Regression results for predicted outcomes if author is female, with the following dependent variables: rate of citations from papers with a female, male first author and with a female, male last author. These results represent data across all fields. Robust standard errors in parentheses; p-values are two-tailed. \checkmark\ indicates the fixed effect is included. 
\item +$p<0.10$; *$p<0.05$; **$p<0.01$; ***$p<0.001$
 
\end{tablenotes}
 
\end{threeparttable}
}
\end{table}
 
}%

\end{landscape}
 


\begin{landscape}

\thispagestyle{empty}

{

\scriptsize

\setlength{\tabcolsep}{2pt}

\renewcommand{\arraystretch}{1}

\begin{table}[tp]\centering

\scalebox{0.8}{%

\begin{threeparttable}
\def\sym#1{\ifmmode^{#1}\else\(^{#1}\)\fi}
\caption{Citations to sample patents as a function of inventor gender and writing style, standardized coefficients}
\label{table:RegressionsCitationsByGenderOfCitingInventorPatentsSeparateMeasuresV4Standardized}
\begin{tabular}{l*{8}{c}}
\toprule

 
                                             &\multicolumn{2}{c}{\shortstack{Citations from female first inventor}}&\multicolumn{2}{c}{\shortstack{Citations from male first inventor}}&\multicolumn{2}{c}{\shortstack{Citations from female last inventor}}&\multicolumn{2}{c}{\shortstack{Citations from male last inventor}}\\\cmidrule(lr){2-3}\cmidrule(lr){4-5}\cmidrule(lr){6-7}\cmidrule(lr){8-9}
                                             &\multicolumn{1}{c}{(1)}   &\multicolumn{1}{c}{(2)}   &\multicolumn{1}{c}{(3)}   &\multicolumn{1}{c}{(4)}   &\multicolumn{1}{c}{(5)}   &\multicolumn{1}{c}{(6)}   &\multicolumn{1}{c}{(7)}   &\multicolumn{1}{c}{(8)}   \\
\midrule
Inventor is female (1 = Yes) &         1.3748***&         1.3883***&        -1.5209***&        -1.5433***&         1.3625***&         1.3737***&        -1.5832***&        -1.5998***\\
                                             &       (0.0779)   &       (0.0786)   &       (0.1718)   &       (0.1725)   &       (0.0802)   &       (0.0806)   &       (0.1705)   &       (0.1710)   \\
Lawyer is female (1 = Yes)                   & 0.0904 & 0.0792 & 0.0485 & 0.0536 & 0.0172 & 0.0066 & 0.1194 & 0.1250 \\
                                             &       (0.0834)   &       (0.0838)   &       (0.1730)   &       (0.1731)   &       (0.0837)   &       (0.0839)   &       (0.1717)   &       (0.1719)   \\
Patent is cited 0 times (1 = Yes)            & -2.3336*** & -2.3281*** & -19.4657*** & -19.4661*** & -2.5636*** & -2.5582*** & -19.2030*** & -19.2044*** \\
                                             &       (0.0604)   &       (0.0603)   &       (0.1341)   &       (0.1343)   &       (0.0639)   &       (0.0638)   &       (0.1327)   &       (0.1328)   \\
Involved rate                                &  & 0.0255 &  & 0.3072+ &  & 0.0556 &  & 0.1805 \\
                                             &                  &       (0.0824)   &                  &       (0.1757)   &                  &       (0.0861)   &                  &       (0.1731)   \\
Informational rate                           &  & -0.2948*** &  & 0.1665 &  & -0.2756*** &  & 0.1673 \\
                                             &                  &       (0.0811)   &                  &       (0.1826)   &                  &       (0.0822)   &                  &       (0.1802)   \\
Constant                                     & 3.4763*** & 3.4763*** & 29.1863*** & 29.1863*** & 3.7251*** & 3.7251*** & 28.9506*** & 28.9506*** \\
                                             &       (0.0775)   &       (0.0774)   &       (0.1717)   &       (0.1717)   &       (0.0798)   &       (0.0798)   &       (0.1704)   &       (0.1704)   \\\midrule
Field fixed effects                          &            \checkmark   &            \checkmark   &            \checkmark   &            \checkmark   &            \checkmark   &            \checkmark   &            \checkmark   &            \checkmark   \\
Year fixed effects                           &            \checkmark   &            \checkmark   &            \checkmark   &            \checkmark   &            \checkmark   &            \checkmark   &            \checkmark   &            \checkmark   \\
\midrule
$R^2$                                    &         0.0585   &         0.0589   &         0.2962   &         0.2963   &         0.0572   &         0.0575   &         0.2964   &         0.2964   \\
Observations                                            &          32,106   &          32,106   &          32,106   &          32,106   &          32,106   &          32,106   &          32,106   &          32,106   \\
\midrule Wald tests for linguistic predictors&                  &                  &                  &                  &                  &                  &                  &                  \\
F                                            &                  &         6.7987   &                  &         1.7288   &                  &         6.3657   &                  &         0.8430   \\
d.f.                                         &                  &         2.0000   &                  &         2.0000   &                  &         2.0000   &                  &         2.0000   \\
p-value                                      &                  &         0.0011   &                  &         0.1775   &                  &         0.0017   &                  &         0.4304   \\
 
\\

\bottomrule

\end{tabular}

\begin{tablenotes}

\item \emph{Notes:} Regression results for predicted outcomes if inventor or patent lawyer is female, with the following dependent variables: rate of citations from patents with a female or male first inventor and with a female or male last inventor. Robust standard errors in parentheses; p-values are two-tailed. \checkmark\ indicates the fixed effect is included. 
\item +$p<0.10$; *$p<0.05$; **$p<0.01$; ***$p<0.001$

\end{tablenotes}

\end{threeparttable}
}
\end{table}

}%

\end{landscape}


\clearpage
\begin{table}[tp]
\centering
\caption{Involved and informational language predict the gender of citing authors, net of content similarity}\label{table:risk_papers_citations}
\begin{adjustbox}{max width=\textwidth,center}
\begin{threeparttable}
\begin{tabular}{lcccc}
\toprule
 & \multicolumn{4}{c}{DV: Citations from papers with a\ldots} \\
\cmidrule(lr){2-5}
 & (1) & (2) & (3) & (4) \\
 & Female & Male & Female & Male \\
 & first author & first author & last author & last author \\
\midrule
Involved rate & 0.0015*** & -0.0045*** & 0.0019*** & -0.0053*** \\
    & (0.0002) & (0.0003) & (0.0002) & (0.0003) \\
Informational rate & -0.0031*** & -0.0039*** & -0.0024*** & -0.0050*** \\
    & (0.0001) & (0.0002) & (0.0001) & (0.0002) \\
Author is female (1 = Yes) & 0.0243*** & -0.0344*** & 0.0214*** & -0.0326*** \\
    & (0.0007) & (0.0011) & (0.0006) & (0.0012) \\
Constant & 0.1842*** & 0.3510*** & 0.1404*** & 0.4146*** \\
    & (0.0020) & (0.0033) & (0.0016) & (0.0037) \\
\midrule
Field fixed effects & \checkmark & \checkmark & \checkmark & \checkmark \\
Year fixed effects & \checkmark & \checkmark & \checkmark & \checkmark \\
\midrule
$R^2$ & 0.0026 & 0.0029 & 0.0022 & 0.0033 \\
Observations & 8,509,506 & 8,509,506 & 8,509,506 & 8,509,506 \\
\bottomrule
\end{tabular}
\begin{tablenotes}\footnotesize
\item Linear probability models estimating whether an observed or counterfactual citation comes from a paper with a female or male first or last author, over a content-matched risk set of the $k=100$ nearest neighbors by SBERT cosine similarity. All models include field and year fixed effects; \checkmark\ indicates the fixed effect is included. Robust standard errors clustered by focal paper in parentheses. +$p<0.10$; *$p<0.05$; **$p<0.01$; ***$p<0.001$.
\end{tablenotes}
\end{threeparttable}
\end{adjustbox}
\end{table}
\clearpage

\clearpage
\begin{table}[tp]
\centering
\caption{Involved and informational language predict the gender of citing inventors, net of content similarity}\label{table:risk_patents_citations}
\begin{adjustbox}{max width=\textwidth,center}
\begin{threeparttable}
\begin{tabular}{lcccc}
\toprule
 & \multicolumn{4}{c}{DV: Citations from patents with a\ldots} \\
\cmidrule(lr){2-5}
 & (1) & (2) & (3) & (4) \\
 & Female & Male & Female & Male \\
 & first inventor & first inventor & last inventor & last inventor \\
\midrule
Involved rate & 0.0006 & 0.0019 & 0.0007+ & 0.0023 \\
    & (0.0004) & (0.0017) & (0.0004) & (0.0017) \\
Informational rate & -0.0005** & 0.0012+ & -0.0006*** & 0.0014* \\
    & (0.0002) & (0.0006) & (0.0002) & (0.0006) \\
Inventor is female (1 = Yes) & 0.0276*** & -0.0593*** & 0.0252*** & -0.0570*** \\
    & (0.0013) & (0.0050) & (0.0012) & (0.0049) \\
Lawyer is female (1 = Yes) & 0.0017 & -0.0110 & 0.0048+ & -0.0125 \\
    & (0.0026) & (0.0088) & (0.0027) & (0.0086) \\
Constant & 0.0389*** & 0.3738*** & 0.0462*** & 0.3633*** \\
    & (0.0037) & (0.0156) & (0.0036) & (0.0152) \\
\midrule
Field fixed effects & \checkmark & \checkmark & \checkmark & \checkmark \\
Year fixed effects & \checkmark & \checkmark & \checkmark & \checkmark \\
\midrule
$R^2$ & 0.0047 & 0.0038 & 0.0037 & 0.0036 \\
Observations & 1,049,463 & 1,049,463 & 1,049,463 & 1,049,463 \\
\bottomrule
\end{tabular}
\begin{tablenotes}\footnotesize
\item Linear probability models estimating whether an observed or counterfactual citation comes from a patent with a female or male first or last inventor, over a content-matched risk set of the $k=100$ nearest neighbors by SBERT cosine similarity. All models include field and year fixed effects; \checkmark\ indicates the fixed effect is included. Robust standard errors clustered by focal patent in parentheses. +$p<0.10$; *$p<0.05$; **$p<0.01$; ***$p<0.001$.
\end{tablenotes}
\end{threeparttable}
\end{adjustbox}
\end{table}
\clearpage

\clearpage
\begin{table}[tp]
\centering
\caption{Involved and informational language predict the gender of citing authors, net of content similarity, using 50 nearest neighbors}\label{table:risk_papers_citations_k50}
\begin{adjustbox}{max width=\textwidth,center}
\begin{threeparttable}
\begin{tabular}{lcccc}
\toprule
 & \multicolumn{4}{c}{DV: Citations from papers with a\ldots} \\
\cmidrule(lr){2-5}
 & (1) & (2) & (3) & (4) \\
 & Female & Male & Female & Male \\
 & first author & first author & last author & last author \\
\midrule
Involved rate & 0.0016*** & -0.0044*** & 0.0020*** & -0.0052*** \\
    & (0.0002) & (0.0003) & (0.0002) & (0.0003) \\
Informational rate & -0.0032*** & -0.0039*** & -0.0024*** & -0.0051*** \\
    & (0.0001) & (0.0002) & (0.0001) & (0.0002) \\
Author is female (1 = Yes) & 0.0248*** & -0.0356*** & 0.0218*** & -0.0337*** \\
    & (0.0007) & (0.0011) & (0.0006) & (0.0012) \\
Constant & 0.1883*** & 0.3586*** & 0.1436*** & 0.4236*** \\
    & (0.0020) & (0.0033) & (0.0016) & (0.0037) \\
\midrule
Field fixed effects & \checkmark & \checkmark & \checkmark & \checkmark \\
Year fixed effects & \checkmark & \checkmark & \checkmark & \checkmark \\
\midrule
$R^2$ & 0.0027 & 0.0030 & 0.0023 & 0.0033 \\
Observations & 8,288,361 & 8,288,361 & 8,288,361 & 8,288,361 \\
\bottomrule
\end{tabular}
\begin{tablenotes}\footnotesize
\item Linear probability models estimating whether an observed or counterfactual citation comes from a paper with a female or male first or last author, over a content-matched risk set of the $k=50$ nearest neighbors by SBERT cosine similarity. All models include field and year fixed effects; \checkmark\ indicates the fixed effect is included. Robust standard errors clustered by focal paper in parentheses. +$p<0.10$; *$p<0.05$; **$p<0.01$; ***$p<0.001$.
\end{tablenotes}
\end{threeparttable}
\end{adjustbox}
\end{table}
\clearpage

\clearpage
\begin{table}[tp]
\centering
\caption{Involved and informational language predict the gender of citing inventors, net of content similarity, using 50 nearest neighbors}\label{table:risk_patents_citations_k50}
\begin{adjustbox}{max width=\textwidth,center}
\begin{threeparttable}
\begin{tabular}{lcccc}
\toprule
 & \multicolumn{4}{c}{DV: Citations from patents with a\ldots} \\
\cmidrule(lr){2-5}
 & (1) & (2) & (3) & (4) \\
 & Female & Male & Female & Male \\
 & first inventor & first inventor & last inventor & last inventor \\
\midrule
Involved rate & 0.0006 & 0.0019 & 0.0007+ & 0.0023 \\
    & (0.0004) & (0.0017) & (0.0004) & (0.0017) \\
Informational rate & -0.0005** & 0.0011+ & -0.0006*** & 0.0013* \\
    & (0.0002) & (0.0006) & (0.0002) & (0.0006) \\
Inventor is female (1 = Yes) & 0.0279*** & -0.0595*** & 0.0255*** & -0.0573*** \\
    & (0.0013) & (0.0050) & (0.0013) & (0.0050) \\
Lawyer is female (1 = Yes) & 0.0021 & -0.0090 & 0.0052+ & -0.0106 \\
    & (0.0026) & (0.0088) & (0.0027) & (0.0086) \\
Constant & 0.0395*** & 0.3793*** & 0.0469*** & 0.3686*** \\
    & (0.0037) & (0.0157) & (0.0036) & (0.0154) \\
\midrule
Field fixed effects & \checkmark & \checkmark & \checkmark & \checkmark \\
Year fixed effects & \checkmark & \checkmark & \checkmark & \checkmark \\
\midrule
$R^2$ & 0.0048 & 0.0038 & 0.0037 & 0.0036 \\
Observations & 1,038,649 & 1,038,649 & 1,038,649 & 1,038,649 \\
\bottomrule
\end{tabular}
\begin{tablenotes}\footnotesize
\item Linear probability models estimating whether an observed or counterfactual citation comes from a patent with a female or male first or last inventor, over a content-matched risk set of the $k=50$ nearest neighbors by SBERT cosine similarity. All models include field and year fixed effects; \checkmark\ indicates the fixed effect is included. Robust standard errors clustered by focal patent in parentheses. +$p<0.10$; *$p<0.05$; **$p<0.01$; ***$p<0.001$.
\end{tablenotes}
\end{threeparttable}
\end{adjustbox}
\end{table}
\clearpage

\clearpage
\begin{table}[tp]
\centering
\caption{Involved and informational language predict the gender of citing authors, net of content similarity, using 150 nearest neighbors}\label{table:risk_papers_citations_k150}
\begin{adjustbox}{max width=\textwidth,center}
\begin{threeparttable}
\begin{tabular}{lcccc}
\toprule
 & \multicolumn{4}{c}{DV: Citations from papers with a\ldots} \\
\cmidrule(lr){2-5}
 & (1) & (2) & (3) & (4) \\
 & Female & Male & Female & Male \\
 & first author & first author & last author & last author \\
\midrule
Involved rate & 0.0014*** & -0.0045*** & 0.0018*** & -0.0054*** \\
    & (0.0002) & (0.0003) & (0.0002) & (0.0003) \\
Informational rate & -0.0031*** & -0.0039*** & -0.0023*** & -0.0050*** \\
    & (0.0001) & (0.0002) & (0.0001) & (0.0002) \\
Author is female (1 = Yes) & 0.0243*** & -0.0341*** & 0.0213*** & -0.0322*** \\
    & (0.0007) & (0.0011) & (0.0006) & (0.0012) \\
Constant & 0.1833*** & 0.3493*** & 0.1397*** & 0.4126*** \\
    & (0.0020) & (0.0033) & (0.0016) & (0.0037) \\
\midrule
Field fixed effects & \checkmark & \checkmark & \checkmark & \checkmark \\
Year fixed effects & \checkmark & \checkmark & \checkmark & \checkmark \\
\midrule
$R^2$ & 0.0026 & 0.0029 & 0.0022 & 0.0033 \\
Observations & 8,562,395 & 8,562,395 & 8,562,395 & 8,562,395 \\
\bottomrule
\end{tabular}
\begin{tablenotes}\footnotesize
\item Linear probability models estimating whether an observed or counterfactual citation comes from a paper with a female or male first or last author, over a content-matched risk set of the $k=150$ nearest neighbors by SBERT cosine similarity. All models include field and year fixed effects; \checkmark\ indicates the fixed effect is included. Robust standard errors clustered by focal paper in parentheses. +$p<0.10$; *$p<0.05$; **$p<0.01$; ***$p<0.001$.
\end{tablenotes}
\end{threeparttable}
\end{adjustbox}
\end{table}
\clearpage

\clearpage
\begin{table}[tp]
\centering
\caption{Involved and informational language predict the gender of citing inventors, net of content similarity, using 150 nearest neighbors}\label{table:risk_patents_citations_k150}
\begin{adjustbox}{max width=\textwidth,center}
\begin{threeparttable}
\begin{tabular}{lcccc}
\toprule
 & \multicolumn{4}{c}{DV: Citations from patents with a\ldots} \\
\cmidrule(lr){2-5}
 & (1) & (2) & (3) & (4) \\
 & Female & Male & Female & Male \\
 & first inventor & first inventor & last inventor & last inventor \\
\midrule
Involved rate & 0.0006 & 0.0019 & 0.0007+ & 0.0023 \\
    & (0.0004) & (0.0017) & (0.0004) & (0.0017) \\
Informational rate & -0.0005** & 0.0012+ & -0.0006*** & 0.0014* \\
    & (0.0002) & (0.0006) & (0.0002) & (0.0006) \\
Inventor is female (1 = Yes) & 0.0275*** & -0.0593*** & 0.0251*** & -0.0570*** \\
    & (0.0013) & (0.0050) & (0.0012) & (0.0049) \\
Lawyer is female (1 = Yes) & 0.0016 & -0.0115 & 0.0047+ & -0.0131 \\
    & (0.0025) & (0.0087) & (0.0027) & (0.0085) \\
Constant & 0.0388*** & 0.3728*** & 0.0461*** & 0.3623*** \\
    & (0.0037) & (0.0155) & (0.0036) & (0.0152) \\
\midrule
Field fixed effects & \checkmark & \checkmark & \checkmark & \checkmark \\
Year fixed effects & \checkmark & \checkmark & \checkmark & \checkmark \\
\midrule
$R^2$ & 0.0047 & 0.0038 & 0.0037 & 0.0036 \\
Observations & 1,051,472 & 1,051,472 & 1,051,472 & 1,051,472 \\
\bottomrule
\end{tabular}
\begin{tablenotes}\footnotesize
\item Linear probability models estimating whether an observed or counterfactual citation comes from a patent with a female or male first or last inventor, over a content-matched risk set of the $k=150$ nearest neighbors by SBERT cosine similarity. All models include field and year fixed effects; \checkmark\ indicates the fixed effect is included. Robust standard errors clustered by focal patent in parentheses. +$p<0.10$; *$p<0.05$; **$p<0.01$; ***$p<0.001$.
\end{tablenotes}
\end{threeparttable}
\end{adjustbox}
\end{table}
\clearpage

\end{document}